\documentclass[a4paper]{article} 

\usepackage{amsmath}
\usepackage{array}
\usepackage{color}
\usepackage{enumerate}
\usepackage{float}
\usepackage{geometry}
\usepackage{graphicx,amsfonts} 
\usepackage{hyperref}
\usepackage{multirow}
\usepackage{mathptmx}  
\usepackage{ulem}

\usepackage{caption}
\usepackage{subcaption}

\begin{document}


\title{Stress-constrained continuum topology optimization: a new approach based on elasto-plasticity}
\author{Oded Amir \\
Faculty of Civil and Environmental Engineering \\
Technion -- Israel Institute of Technology \\
odedamir@technion.ac.il }
 
\maketitle

\begin{abstract}
A new approach for generating stress-constrained topological designs in continua is presented.
The main novelty is in the use of elasto-plastic modeling and in optimizing the design such that it will exhibit a linear-elastic response.
This is achieved by imposing a single global constraint on the total sum of equivalent plastic strains, providing accurate control over all local stress violations.
The single constraint essentially replaces a large number of local stress constraints or an approximate aggregation of them--two common approaches in the literature.  
A classical rate-independent plasticity model is utilized, for which analytical adjoint sensitivity analysis is derived and verified.
Several examples demonstrate the capability of the computational procedure to generate designs that challenge results from the literature, in terms of the obtained stiffness-strength-weight trade-offs. 
A full elasto-plastic analysis of the optimized designs shows that prior to the initial yielding, these designs can sustain significantly higher loads than  minimum compliance topological layouts, with only a minor compromise on stiffness. 

\paragraph{keywords:} Topology Optimization, Stress Constraints, Elasto-plasticity

\end{abstract}


\section{Introduction} \label{sec:intro}

Topology optimization of continua is a computational method aimed at optimizing the distribution of one or several materials in a given design domain.
The purpose is typically to achieve a minimum-weight structural design with a constraint on displacements, or vice-versa:
minimize compliance (i.e.\ maximize stiffness) using a given amount of available material.
For extensive reviews see for example \cite{eschenauer2001topology}, \cite{BendsoeSigmund2003} and recently \cite{sigmund2013topology} and \cite{deaton2014survey}.
One of the most challenging aspects in developing computational topology optimization procedures is the consideration of stress constraints.
From an engineering standpoint, limiting the stresses of an optimized design is a fundamental requirement:
load-bearing components are typically designed to remain in the linear-elastic regime throughout their service life, meaning that the yield stress should not be exceeded.  
In this article, a new approach to satisfying stress constraints is proposed. 
The central idea is to consider the nonlinear, inelastic material behavior and via optimization to drive the design towards a linear-elastic response.

The incorporation of stress constraints imposes several challenges.
First and foremost is the local nature of stress constraints. 
Most applications of topology optimization involve an objective functional of global nature and only a few global constraints that control volume, weight, displacements or compliance.
If stress constraints are incorporated it is considerably more difficult to tackle the corresponding optimization problem. 
In principle, stress constraints should be imposed on every material point in the design domain, meaning that the number of constraints is comparable to the number of design variables--both related to the resolution of the underlying finite element mesh.
Therefore it is expected that the solution time will be significantly longer than for standard topology optimization problems, and the vulnerability of numerical algorithms to arrive at local minima will be aggravated.
Review of the existing literature highlights two dominating strategies for formulating and solving the optimization problem:
1) All local stress constraints are considered in the problem formulation, whereas in the actual solution only a subset of ``active'' constraints are included;  
and 2) Local stress constraints are aggregated into a single or into a few global constraints.  

The former strategy was implemented in one of the earliest publications on stress-constrained topology optimization of continua by \cite{Duysinx1998}.
Throughout most of the optimization process, roughly one third of the local constraints are considered for sensitivity analysis and optimization. 
In the final optimization steps, 180 local constraints (corresponding to 15\% of the design variables) are actually active.
Local constraints were also considered by \cite{bruggi2008mixed} and similar tendencies are reported in a later, detailed study of various problem formulations with local stress constraints \cite{bruggi2012topology}.
Therefore the efficiency of imposing local constraints in large-scale problems is questionable.
A similar problem formulation but with a different numerical treatment was presented by \cite{pereira2004topology}.
All stress constraints are considered but an Augmented Lagrangian technique is utilized for solving the optimization problem, facilitating a reduction in computational effort invested in sensitivity analysis.
Results appear promising as they exhibit layouts that circumvent regions with potentially high stress concentrations. 
On the other hand, the authors report computational times of up to 10 times higher than for standard minimum compliance formulations.   
The Augmented Lagrangian approach was followed also by \cite{fancello2006topology} who presented layouts that avoid stress concentrations.
The author reported difficulties regarding the numerical implementation and the number of function evaluations indicates that the procedure may not be suitable for large-scale applications.

The second, widely adopted strategy for dealing with the large number of stress constraints involves various forms of constraint aggregation, i.e.\ collecting the constraints into a global stress function.  
In an early study, \cite{yang1996stress} examine the use of both Kreisselmeier-Steinhauser (KS) functions and $p$-norm functions (referring to \cite{park1995extensions}). 
Their problem formulations aim at minimizing either the global stress or a weighted combination of global stress and compliance, subject to a volume constraint.
A step further in the direction of utilizing global stress measures was proposed by \cite{duysinx1998new}. 
Two global stress functions were suggested, namely the $p$-mean and the $p$-norm. 
It was shown that for any given $p$, the maximum local stress is bounded from above by the $p$-norm and from below by the $p$-mean.
Due to ill-conditioning and oscillatory behavior, the maximum value of $p$ is limited to 4 which is not large enough for identifying the actual peak stress.

Several recent studies demonstrate that constraint aggregation can in fact lead to satisfactory results--in particular for the classical L-bracket case.
\cite{le2010stress} provide an extensive critical review and propose regional stress measures where local stress constraints are grouped in interlacing regions according to their stress level.
A similar approach of ``block aggregation'' was suggested also by \cite{paris2007block,paris2010block}.  
The regional stress measures are based on $p$-norms and a normalization with respect to the actual maximum stress is proposed in order to improve the approximation of the maximum stress.
In the optimization of the L-bracket, a layout that avoids the re-entrant corner is generated, demonstrating the potential of the approach.
On the other hand, the numerical implementation suffers from several drawbacks: 
First, the normalization is non-differentiable as it changes discontinuously between optimization cycles; 
second, also the regional constraints can change every optimization cycle according to the sorting of local stresses, thus introducing some inconsistency in the optimization process;
and third, it is shown that increasing the number of regions does not always improve the optimized design as one might expect due to the tighter control over local stresses.       

Level-set methods combined with topological derivatives have also succeeded in generating designs free of stress concentrations. 
\cite{allaire2008minimum} minimize an integral measure of a power-law penalty of stress. 
This approach is shown to provide smooth designs in re-entrant corners but there is no direct control over the actual stress nor the compliance. 
Another set of positive results was presented by \cite{amstutz2010topological}.
The large number of stress constraints is replaced by an external penalty functional that mimics the point-wise constraint.
The layouts avoid stress concentrations at re-entrant corners.
Only slight violations of the stress constraints occur in the final designs, e.g.\ 1\%-2\%  above the target stress.
The main drawbacks of this approach are the reliance on penalization parameters that may be problem-specific, and the lack of direct control over local stresses.
The significance of the latter depends on the degree of localization of high stresses--which may differ considerably from one problem to another. 

An interesting novel approach was proposed recently by \cite{verbart2013new}.
Instead of imposing a large number of constraints, material is penalized if the stress exceeds the allowable stress.
Numerical results demonstrate the potential of this efficient approach. 
However, because penalization is utilized it is hard to satisfy the admissible stress criterion accurately.
Furthermore, the penalized material law is somewhat artificial so it may be difficult to generalize the method.

Another major difficulty in computational stress-constrained topology optimization is the so-called ``singularity'' problem, originally demonstrated in the context of truss topological design.  
It was shown that the optimal topology might correspond to a singular point in the design space, therefore making it difficult or in some cases impossible to arrive at the true optimum by numerical search algorithms \cite{sved1968structural,kirsch1990singular,cheng1992study}.
Such singular points are encountered in cases where removal of a certain truss bar (or a material point in the continuum case) results in a feasible design space with better optimum due to the removal of the corresponding constraint.
This article does not target the difficulties related to the singularity phenomenon.
In fact, an appropriate relaxation scheme is an essential ingredient in the suggested computational approach.
Possibly the most widespread remedy for dealing with the singularity problem is the so-called $\epsilon$-relaxation \cite{cheng1997varepsilon}, where the actual stress constraints are relaxed so that the resulting feasible domain does not possess degenerate branches.
A similar relaxation scheme involving smooth envelope functions was suggested in the context of local buckling constraints by \cite{rozvany1996difficulties}.
The $\epsilon$-relaxation approach was first integrated into continuum topology optimization by \cite{Duysinx1998} and by \cite{duysinx1998new}, who implemented a continuation scheme for gradually reducing $\epsilon$ hence approaching the actual constraints.
The $\epsilon$-relaxation approach was successfully applied to various test cases in later studies, see for example  \cite{pereira2004topology,fancello2006topology,le2010stress}.
An alternative relaxation for avoiding the singularity phenomenon was introduced by \cite{bruggi2008alternative}.
In the SIMP rule, the penalization power $q$ for the yield stress was chosen to be lower than the penalization power $p$ for the stiffness. 
It is noted that separate penalization exponents for stiffness and for yield stress have been suggested much earlier in an extension of SIMP-based topology optimization to elasto-plastic structures \cite{maute1998adaptive}.   
In practice, the so-called $qp$-relaxation appears to provide similar results to those of the $\epsilon$-relaxation.
Both are highly dependent on the continuation scheme and it was shown that a sequence of solutions to the relaxed problem may not converge to the global optimum \cite{stolpe2001trajectories}.
Despite these shortcomings it seems necessary to apply some form of continuous relaxation in order to arrive at practical structural designs that satisfy stress requirements. 

The central idea of the approach proposed in this article is to optimize the inelastic structural response for the particular purpose of satisfying stress constraints in linear elasticity.
Up to date, applications of topology optimization that considered inelastic response were concerned with objectives other than the one pursued herein.
Material nonlinearities in topology optimization were initially considered by \cite{yuge1995optimization}.
Layout optimization of frame structures undergoing plastic deformation was presented, based on homogenization of porous material.
\cite{swan1997voigt} suggested a framework for topology optimization of structures with material nonlinearity based on Voigt and Reuss mixing rules.
The SIMP (Solid Isotropic Material with Penalization) interpolation scheme, originally proposed for linear elastic material \cite{bendsoe1989optimal}, was extended for elasto-plastic behavior by \cite{maute1998adaptive}.
Although several other articles on the subject were published over the last two decades, topology optimization involving elasto-plasticity is still not well established.
One difficulty lies in obtaining accurate design sensitivities.
In some cases, several derivative terms are neglected \cite{maute1998adaptive,schwarz2001topology}.
Apparently this has a minor effect on the outcome of the optimization but in general these terms are not negligible.
Moreover, when comparing analytical design sensitivities to finite difference calculations, errors in the order of $10^{-2}$ are observed \cite{swan1997voigt,yoon2007topology}.
Recent studies have incorporated analytical adjoint sensitivity analysis for rate-independent elasto-plasticity, based on the framework by \cite{michaleris1994tangent}.
Accurate sensitivities have been reported for problems involving reinforced concrete design \cite{bogomolny2012conceptual} and effective energy management under dynamic loading \cite{nakshatrala2015topology}.
Another analytical sensitivity analysis scheme for topology optimization of elasto-plastic structures was recently presented by \cite{kato2015analytical}.
Highly accurate derivatives are obtained, however the formulation is limited to cases in which the load is applied only to the nodes whose displacements are controlled. 
Adjoint sensitivity analysis has been applied also for topology optimization with viscoelastic material \cite{james2015topology};
viscoplastic micro-heterogeneous materials in a multiscale approach \cite{fritzen2015topology};
and continuum damage models \cite{amir2013reinforcement,amir2013topology,james2014failure}.
The latter study in fact targeted a similar goal as in this article--mitigating failure, i.e.\ imposing stress constraints--but was based on a different constitutive model and on a different problem formulation which involved constraint aggregation.  

As mentioned in the beginning of the introduction, the proposed approach relies on modeling the inelastic behavior and driving the design towards a linear-elastic response.
This is achieved by constraining the total sum of equivalent plastic strains.
A single global constraint is added to the standard stiffness vs. volume problem, inherently providing accurate control over all local stress violations.  
Consequently stress limits can be implicitly satisfied, without imposing a large number of local constraints.
The corresponding computational procedure can alleviate one of the major obstacles in stress-constrained topology optimization--the need to solve a nonlinear, non-convex optimization problem with a large number of design variables and an equally large number of constraints.

The remainder of the article is organized as follows.
In Section \ref{sec:epmodel} we briefly review the elasto-plastic material model and the nonlinear finite element analysis formulation.
The topology optimization problem formulation and the design parametrization are then introduced in Section \ref{sec:topopt},
followed by a derivation and verification of the adjoint sensitivity analysis in Section \ref{sec:sa}.
Several examples demonstrate the applicability of the proposed approach in Section \ref{sec:examples}.
Finally, a discussion of the results and of necessary future investigations is given in Section \ref{sec:conclusion}.  


\section{Elasto-plastic model and finite element analysis} \label{sec:epmodel}
In this section we briefly review the material model and the subsequent nonlinear finite element analysis.
The purpose is to provide the necessary background for the optimization problem formulation presented in Section \ref{sec:topopt}, which involves state variables related to the elasto-plastic material model, as well as for the adjoint sensitivity analysis, presented in Section \ref{sec:sa}.
   
\subsection{Classical rate-independent plasticity} \label{sec:epmodel_rateindplast}
The derivation of the governing equations herein follows the textbooks by \cite{simo2006computational} and by \cite{zienkiewicz2000finite}.
The model is essentially composed of the following assumptions and rules:
elastic stress-strain relationships;
a yield condition, defining the elastic domain;
a flow rule and hardening law;
Kuhn-Tucker complementarity conditions;
and a consistency condition. We first assume that the total strain tensor $\boldsymbol\epsilon$ can be split into its elastic and plastic parts, $\boldsymbol\epsilon^{el}$ and $\boldsymbol\epsilon^{pl}$ respectively, 
\begin{equation}
\boldsymbol\epsilon = \boldsymbol\epsilon^{el} + \boldsymbol\epsilon^{pl} .
\end{equation}
Furthermore, we relate the stress tensor $\boldsymbol\sigma$ to the elastic strains using the elastic constitutive tensor $\mathbf{D}$,
\begin{equation} \label{eq:stress=Dstrain}
{\boldsymbol\sigma} = \mathbf{D} {\boldsymbol\epsilon}^{el}.
\end{equation}
The yield criterion $f$ is a function that defines the admissible stress states
\begin{equation}
f(\boldsymbol{\sigma},\mathbf{q}) \leq 0
\end{equation}
where $\mathbf{q}$ are internal variables related to the plastic strains and to the hardening parameters.
The elastic domain is defined by the interior of the yield criterion where $f < 0$; 
the yield surface is defined by $f = 0$; 
and the stress state corresponding to $f > 0$ is considered non-admissible.   

The irreversible plastic flow is governed by the evolution of plastic strains and internal variables
\begin{eqnarray} 
\dot{\boldsymbol\epsilon}^{pl} & = & \dot{\lambda} \mathbf{r}(\boldsymbol{\sigma},\mathbf{q}) \label{plstrainflow}\\
\dot{\mathbf{q}} & = & - \dot{\lambda} \mathbf{h}(\boldsymbol{\sigma},\mathbf{q}) \label{intvariableflow}
\end{eqnarray}
where $\mathbf{r}$ and $\mathbf{h}$ are functions defining the direction of plastic flow and the hardening of the material.
The parameter $\dot{\lambda}$ is typically called the \textit{consistency parameter} or \textit{plastic multiplier}.
Together with the yield criterion, $\dot{\lambda}$ must satisfy the Kuhn-Tucker complementarity conditions
\begin{eqnarray}
\dot{\lambda} & \geq & 0 \nonumber\\
f(\boldsymbol{\sigma},\mathbf{q}) & \leq & 0 \nonumber \\
\dot{\lambda} f(\boldsymbol{\sigma},\mathbf{q}) & = & 0 \label{KTconditions}
\end{eqnarray}
as well as the consistency requirement 
\begin{equation}
\dot{\lambda} \dot{f}(\boldsymbol{\sigma},\mathbf{q}) = 0 .
\end{equation}
The consistency requirement means that during plastic loading, the stress state must remain on the yield surface, meaning $\dot{f} = 0$ if $\dot{\lambda} > 0$. 

A widely accepted model of rate-independent plasticity in metals is usually known as $J_2$ flow theory or simply $J_2$-plasticity.
It is based on the von Mises yield criterion \cite{vonMises1928} that relates the yielding of the material to the deviatoric stresses, measured by the second deviatoric stress invariant $J_2$.
The model is hereby presented as a particular case of rate-independent plasticity.

The yield criterion is the von Mises yield function expressed as  
\begin{equation}\label{vonmisesyieldsurface}
f(\boldsymbol{\sigma},\kappa) = \sqrt{3 J_2}  - \sigma_{y}(\kappa) \leq 0
\end{equation}
where the expression $\sqrt{3 J_2}$ is usually named the \emph{von Mises stress} or \emph{equivalent stress}.
$\sigma_{y}$ is the yield stress in uniaxial tension, which depends on a single internal parameter $\kappa$ according to an isotropic hardening function.
Kinematic hardening is not considered in the current work.
A popular choice for the hardening rule is the bi-linear function
\begin{equation} \label{sigmayhardening}
\sigma_{y}(\kappa) =  \sigma^0_{y} + H E \kappa 
\end{equation}
where $\sigma^0_{y}$ is the initial yield stress, $H$ is a scalar (usually in the order of $10^{-2}$) and $E$ is Young's modulus.
An associative flow rule is assumed, meaning that the flow of plastic strains is in a direction normal to the yield surface
\begin{equation} \label{epsilonplflow}
\dot{\boldsymbol\epsilon}^{pl} = \dot{\lambda} \frac{\partial{f}}{\partial{\boldsymbol\sigma}}.
\end{equation}
Finally, the internal variable governing the hardening is the equivalent plastic strain, evolving according to the rule
\begin{equation} \label{kappaflow}
\dot{\kappa} = \sqrt{\frac{2}{3}} {\left\| \dot{\boldsymbol\epsilon}^{pl} \right\|}_2. 
\end{equation}
The factor $\sqrt{\frac{2}{3}}$ is introduced so that for the particular one-dimensional case (involving uniaxial plastic deformation), the obvious relation will be obtained, i.e. $\dot{\kappa} = \dot{\epsilon^{pl}}$.

\subsection{Finite element implementation} \label{sec:epmodel_fe}
For finite element analysis, the process of rate-independent plasticity is conveniently represented as a flow evolving in time, where each time step corresponds to an increment of load or displacement. 
In the current work, a standard Newton-Raphson incremental-iterative scheme with displacement control is employed.
For the purpose of sensitivity analysis in optimal design, the finite element equations are cast into the framework for transient, coupled and nonlinear systems suggested by \cite{michaleris1994tangent}.

In the coupled approach, for every `time' increment $n$ in the transient analysis, we determine the unknowns $\mathbf{u}_n$, $\mathbf{v}_n$ and $\theta_n$ that satisfy the residual equations
\begin{eqnarray}
\mathbf{R}_n(\mathbf{u}_n,\mathbf{u}_{n-1},\mathbf{v}_n,\mathbf{v}_{n-1},\theta_n) = \mathbf{0} \label{eq:epmodel_globalresidual} \\ 
\mathbf{H}_n(\mathbf{u}_n,\mathbf{u}_{n-1},\mathbf{v}_n,\mathbf{v}_{n-1}) = \mathbf{0} \label{eq:epmodel_localresidual}
\end{eqnarray}
where $\mathbf{u}_n$ is the displacements vector, $\theta_n$ is the load factor and $\mathbf{v}_n$ are the internal variables--all corresponding to the time $t_n$.
$\mathbf{R}_n=\mathbf{0}$ is satisfied at the global level and $\mathbf{H}_n=\mathbf{0}$ is satisfied at each Gauss integration point.
The transient, coupled and nonlinear system of equations is uncoupled by treating the response $\mathbf{v}$ as a function of the response $\mathbf{u}$.
When solving the residual equations for the $n$-th increment, the responses $\mathbf{u}_{n-1}$ and $\mathbf{v}_{n-1}$ are known from the previous converged increment.
The independent response $\mathbf{u}_n$ is found by an iterative prediction-correction procedure in the global level, while for each iterative step the dependent response $\mathbf{v}_n(\mathbf{u}_n)$ is found by an inner iterative loop.
The responses $\mathbf{u}_n$ and its dependant $\mathbf{v}_n$ are corrected until Eqs.~\eqref{eq:epmodel_globalresidual} and \eqref{eq:epmodel_localresidual} are satisfied to sufficient accuracy.
This procedure is repeated for all $N$ increments.

Neglecting body forces, $\mathbf{R}_n$ is defined in the current study as the difference between external and internal forces and depends explicitly on $\mathbf{v}_n$ and $\theta_n$ only 
\begin{equation} 
\mathbf{R}_n(\mathbf{v}_n,\theta_n) = \theta_n \mathbf{\hat{f}} - \int_V {\mathbf{B}^T} {\boldsymbol\sigma_n} dV \label{eq:epmodel_R}
\end{equation}
where $\mathbf{\hat{f}}$ is a constant reference load vector with non-zero entries only at loaded degrees of freedom and $\mathbf{B}$ is the standard strain-displacement matrix in the context of finite element procedures. 
For the particular material model used in this study, the vector $\mathbf{v}_n$ is given by
\begin{equation}
\mathbf{v}_n = \left[
\begin{array}{c}
\boldsymbol\epsilon^{pl}_n \\
\kappa_n \\
\boldsymbol\sigma_n \\
\lambda_n
\end{array} \right].
\end{equation} 

For solving the local nonlinear constitutive problem, an implicit backward-Euler scheme is employed.
The central feature of this scheme is the introduction of a trial elastic state.
For any given incremental displacement field, it is first assumed that there is no plastic flow between time $t_n$ and the next time step $t_{n+1}$, meaning the incremental elastic strains are the incremental total strains.
It can be shown that the loading/unloading situation which is governed by the Kuhn-Tucker conditions can be identified using the trial elastic state \cite{simo2006computational}.
Once a plastic increment occurs, the new state variables can be found by solving a nonlinear equation system resulting from the time discretization of the governing equations.
This results in the nonlinear system $\mathbf{H}_n$ which is derived specifically for any given elasto-plastic model. 
For the particular model used in the current study, $\mathbf{H}_n$ is defined as the collection of four incremental residuals, resulting from the time linearization of the governing constitutive equations

\begin{eqnarray} \label{eq:epmodel_H}
^1\mathbf{H}_{n} & = & {\boldsymbol\epsilon^{pl}_{n-1}} + ({\lambda_n} - {\lambda_{n-1}}) (\frac{\partial f}{\partial {\boldsymbol\sigma}_n})^T - {\boldsymbol\epsilon^{pl}_n} \nonumber \\
^2\mathbf{H}_{n} & = & {\kappa_{n-1}} + ({\lambda_n} - {\lambda_{n-1}}) \sqrt{\frac{2}{3} (\frac{\partial f}{\partial {\boldsymbol\sigma}_n})^T (\frac{\partial f}{\partial {\boldsymbol\sigma}_n})} - {\kappa_n} \nonumber  \\
^3\mathbf{H}_{n} & = & {\boldsymbol\sigma_{n-1}} + \mathbf{D} \left[\mathbf{B} {\mathbf{u}_n} - \mathbf{B} {\mathbf{u}_{n-1}} - ({\boldsymbol\epsilon^{pl}_n} - {\boldsymbol\epsilon^{pl}_{n-1}}) \right]  - {\boldsymbol\sigma_n} \nonumber  \\
^4\mathbf{H}_{n} & = & J_{2n}  - \frac{1}{3} (\sigma_{y}(\kappa_n))^2.
\end{eqnarray}
The equation $^1\mathbf{H}_{n}$ represents the associative flow rule; 
$^2\mathbf{H}_{n}$ represents the evolution of the isotropic hardening parameter;
$^3\mathbf{H}_{n}$ relates stresses to elastic strains;;
and $^4\mathbf{H}_{n}$ is the yield criterion in squared form, with $J_{2n}$ representing the second deviatoric stress invariant evaluated using $\boldsymbol\sigma_n$.
It is worth noting that the local nonlinear problem of Eq.~\eqref{eq:epmodel_H} can be solved efficiently as a scalar equation by a return-mapping algorithm, for example as derived by \cite{Simo1985} for plane stress situations.
Nevertheless, for the purpose of sensitivity analysis we find it convenient to use the full representation as suggested by \cite{michaleris1994tangent}. 


\section{Topology optimization approach} \label{sec:topopt}

The central goal of the proposed formulation is to generate optimized structural layouts that can sustain a certain load under a prescribed range of displacements, while not exceeding the allowable stress limitations.
In the majority of studies so far, this design problem was formulated as an optimization of a linear-elastic structure, aimed at minimizing either compliance or volume.
Stress limitations were imposed as constraints, either locally on each material point or in a global, aggregated manner.
In the suggested formulation, we approach the same design goal in a completely different way.
Essentially, we seek the best trade-off between three quantities: 
1) The weight of the structure, coinciding with the volume for single-material layouts;
2) The load-bearing capacity, represented by the end-compliance--the product of loads and displacements at the final (time-wise) equilibrium point;
and 3) The overall sum of plastic strains, representing the violation of allowable stress limits.
In the numerical experiments, two variants of the optimization problem are examined.
These arise from assigning each of the above quantities 1 or 2 as an objective, while constraining the other one and quantity 3. 
It will be shown that both variants lead to satisfactory results.
In the remainder of this section, quantity 2 is considered in the objective.
Other variants can be derived in a very similar manner.    

\subsection{Problem formulation}
For the purpose of optimizing the topological layout of a continuum, we follow the material distribution approach \cite{bendsoe1988generating} together with the SIMP interpolation scheme \cite{bendsoe1989optimal} and its extension to multiple phases, usually known as Modified SIMP \cite{sigmund1997design}.
This implies that the design variables $\mathbf{x}$ are densities at discrete material points, assigned at the centroid of each finite element in the design domain and varying between zero (void) and 1 (solid).
The optimization problem can be stated as follows 
\small{
\begin{eqnarray} \label{eq:problemformulation}
\min_{\mathbf{x}} &         & g_0 = - \theta_{N} \hat{\mathbf{f}}^T {\mathbf{u}_{N}} \nonumber \\
\text{s.t.:} &              & g_1 = \sum_{e=1}^{N_e} v_e \overline{\mathbf{x}}_e - g_1^{\star} \leq 0 \nonumber \\
             &              & g_2 = \sum_{e=1}^{N_e} \sum_{k=1}^{N_{GP}} {\kappa^{ek}_N} - g_2^{\star} \leq 0 \nonumber \\
						 &							& 0 \leq x_e \leq 1, \qquad e = 1,...,N_{e} \nonumber \\
\text{with:} 	          		& & \mathbf{R}_n(\mathbf{v}_n,\theta_n,\mathbf{x}) = \mathbf{0} \qquad n = 1,...,N \nonumber \\
														& & \mathbf{H}_n(\mathbf{u}_n,\mathbf{u}_{n-1},\mathbf{v}_n,\mathbf{v}_{n-1},\mathbf{x}) = \mathbf{0} \qquad n = 1,...,N .
\end{eqnarray}
}
\normalsize
The objective is to maximize the end-compliance for a given prescribed displacement, i.e.~maximize the load-bearing capacity for a given magnitude of deformation.
This quantity is evaluated using the terminal values of the load factor $\theta_{N}$ and of the displacements ${\mathbf{u}_{N}}$.
The constraint $g_1$ ensures that no more than a certain prescribed volume $g_1^{\star}$ is utilized. 
The design volume is measured according to the physical material density $\overline{{x}}_e$ of each finite element.
The physical densities $\overline{\mathbf{x}}$ are related to the mathematical variables $\mathbf{x}$ via widely used filtering and projection techniques which will be presented explicitly in the next section. 
The constraint $g_2$ ensures that the overall spatial sum of the plastic strains does not exceed a certain small threshold $g_2^{\star}$, which can in theory be zero.
The sum of plastic strains is evaluated based on the quantity $\kappa_N$, i.e.~ the plastic strain measured at each Gauss point in the finite element mesh, at the terminal equilibrium point.
Finally, the nonlinear residuals ${\mathbf{R}_n}$ and ${\mathbf{H}_n}$ are as defined in Section \ref{sec:epmodel_fe} according to the respective elasto-plastic model.

\subsection{Design parametrization}
The correspondence between the mathematical optimization variables $\mathbf{x}$ and the nonlinear finite element analysis is as follows. 
First a standard density filter is applied \cite{bruns2001topology,bourdin2001filters} with a simple linear weighting function to obtain $\widetilde{\mathbf{x}}$.
The purpose of applying a density filter is to overcome the well-known difficulty of artificial checkerboard patterns as well as to introduce a length scale in the design, thus avoiding results with very thin features that are difficult to manufacture.
Then, a Heaviside projection function \cite{guest2004achieving,xu2010volume} is utilized in order to `push' the design towards a distinct 0-1 (or void-solid) layout.
This yields the physical density distribution,  
\small{
\begin{equation}\label{Xuprojection}
\overline{x}_e = \left\{ \begin{array}{ll}
\eta \left[ e^{-\beta_{HS} (1 - \widetilde{x}_e / \eta)} - (1 - \widetilde{x}_e / \eta) e^{-\beta_{HS}} \right] & 0 \leq {\widetilde{x}}_e \leq \eta \\
(1-\eta) \left[1 - e^{-\beta_{HS} (\widetilde{x}_e-\eta)/(1-\eta)} + \right. &  \\
\qquad \left. (\widetilde{x}_e-\eta)/(1-\eta) e^{-\beta_{HS}} \right] + \eta & \eta < {\widetilde{x}}_e \leq 1
\end{array}
\right.
\end{equation}
}
\normalsize
where $\eta$ is a threshold value and $\beta_{HS}$ is a parameter determining the `sharpness' of the smooth projection function.
In the current study we use $\eta = 0.5$, meaning that any filtered density above 0.5 is projected to 1 and any value below 0.5 is projected to 0.
The initial value of $\beta_{HS}$ is usually set to 1 and it is increased gradually as the optimization progresses.
Heaviside projections are typically introduced in order to achieve crisp void-solid layouts which are necessary in some design problems due to manufacturing requirements.
For the cases addressed in this article, it is not absolutely necessary to utilize such projections, which increase the degree of nonlinearity and may cause difficulties in convergence.
Nevertheless, it is useful to apply the Heaviside projection, even with rather mild $\beta_{HS}$ values, in order to minimize material transition regions.
In regions where the density is between zero and one, also known as “gray” regions in topology optimization, the elasto-plastic material law is artificial.
This is due to the choice of penalization scheme as will be explained below.
Therefore, the true stress in the actual (manufactured) structure may differ from the computed stress within the optimization. 
This motivates the minimization of “gray” transition regions. 

The constitutive model corresponding to $J_2$ flow theory involves three material parameters: Young's modulus $E$, the hardening fraction $H$ and the initial yield stress $\sigma^0_y$.
As mentioned above, an extension to the SIMP approach for interpolating the three parameters was originally presented by \cite{maute1998adaptive}.
For evaluating the tangent stiffness matrix and the internal forces vector, Young's modulus is interpolated in each finite element as follows
\begin{equation} \label{eq:E_ModSIMP}
E(\overline{x}_e) = E_{min} + (E_{max}-E_{min}) {\overline{x}_e}^{p_E}.
\end{equation}
In general, $E_{min}$ and $E_{max}$ are the values of Young's modulus of two candidate materials which are distributed in the design domain.
For the case of distributing a single material and void, $E_{min}$ is set to be several orders of magnitude smaller than $E_{max}$.
Finally, $p_E$ is a penalization factor required to drive the design toward a 0-1 layout.
The initial yield stress is penalized similarly, 
\begin{equation} \label{eq:sigmay_ModSIMP}
\sigma^0_{y}(\overline{x}_e) = \sigma_{y,min}^0 + (\sigma_{y,max}^0 - \sigma_{y,min}^0) {\overline{x}_e}^{p_{\sigma_y}} 
\end{equation}
where $\sigma_{y,min}^0$ and $\sigma_{y,max}^0$ are the initial yield stresses for the two candidate materials, corresponding to $\overline{x} = 0$ and $\overline{x} = 1$ respectively.
From a physical point of view, the penalization factor ${p_{\sigma_y}}$ should be equal to $p_E$ so that the yield strain does not depend on the density. 
However, in many cases it is necessary to set ${p_{\sigma_y}} < p_E$ in order to avoid numerical difficulties arising when low density elements reach their yield limit.
The physical consequence is that in intermediate densities, the yield strain is artificially higher than that of the full material, as shown in Figure \ref{fig:SigmaEpsilonRhoPenal.}
Separate exponents in elasto-plastic topology optimization were already introduced by \cite{maute1998adaptive}.
This approach is used also with stress constraints (namely the $qp$-relaxation, \cite{bruggi2008alternative}) and is similar to $\epsilon$-relaxation approaches \cite{cheng1997varepsilon}.

\begin{figure}%
	\centering
	
	\begin{subfigure}[c]{0.4\textwidth}
		\includegraphics[width=\textwidth]{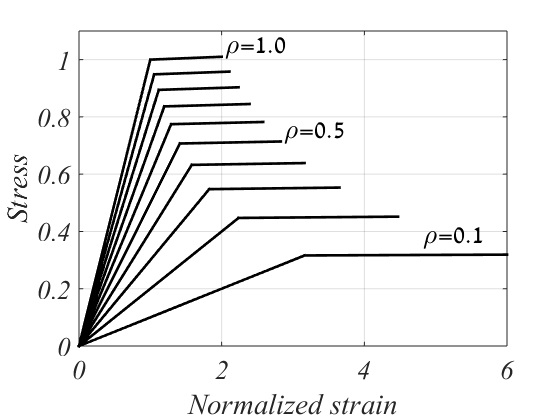}
		\caption{}
		\label{fig:SigmaEpsilonRhoPenal_1_05}
	\end{subfigure}
	~
	\begin{subfigure}[c]{0.4\textwidth}
		\includegraphics[width=\textwidth]{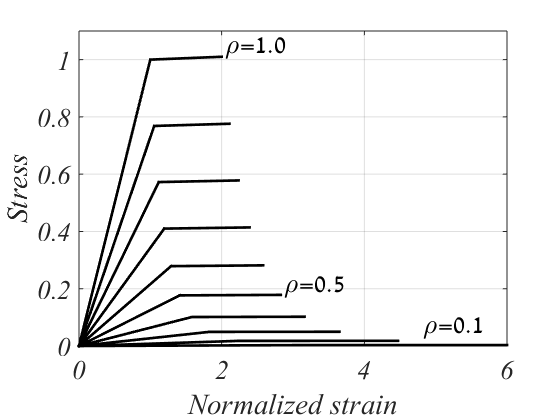}
		\caption{}
		\label{fig:SigmaEpsilonRhoPenal_3_25}
	\end{subfigure}
	
	\caption{Normalized stress-strain curves for various densities $\rho$ and separate penalty exponents $p_E$ and ${p_{\sigma_y}}$. Left: $p_E=1.0, {p_{\sigma_y}}=0.5$; Right: $p_E=3.0, {p_{\sigma_y}}=2.5$. For densities smaller than 1, the yield strain is relatively delayed.}
	\label{fig:SigmaEpsilonRhoPenal}

\end{figure}

In this study, we keep $H$ independent of the design variables because the post-yield stiffness is already penalized via Eq.~\eqref{eq:E_ModSIMP}.
Furthermore, the essence of the design problem is to find designs that do not yield or that have a very short post-yield response.
For such cases, it is not necessary to consider an accurate post-yield response, especially for intermediate material densities.
For solid material, a constant $H$ is the same as having a SIMP-type interpolation for $H$. 


\section{Sensitivity Analysis} \label{sec:sa}
Considering the optimization problem in Eq.~\eqref{eq:problemformulation}, the derivatives of the volume constraint $g_1$ are straightforward.
The objective $g_0$ and the constraint $g_2$ involve state variables, therefore an adjoint sensitivity analysis procedure is necessary.
As mentioned earlier, the design sensitivities are computed following the framework for transient, nonlinear coupled problems described by \cite{michaleris1994tangent}. 
In the following we focus on computing the derivatives of a general functional with respect to the physical densities $\overline{\mathbf{x}}$ whereas the derivatives with respect to ${\mathbf{x}}$ can then be computed by the chain rule.

\subsection{Backwards-incremental adjoint procedure}
We begin by forming the augmented functional $\hat{g}(\overline{\mathbf{x}})$
\small{
\begin{eqnarray}
\hat{g}(\overline{\mathbf{x}}) 	& = & g - \sum_{n=1}^{N} \boldsymbol\lambda_{n}^T \mathbf{R}_n(\mathbf{v}_n(\overline{\mathbf{x}}), \theta_n(\overline{\mathbf{x}})) \nonumber \\
															&   & - \sum_{n=1}^{N} \boldsymbol\gamma_n^T \mathbf{H}_n (\mathbf{u}_n(\overline{\mathbf{x}}),\mathbf{u}_{n-1}(\overline{\mathbf{x}}),\mathbf{v}_n(\overline{\mathbf{x}}), \mathbf{v}_{n-1}(\overline{\mathbf{x}}),\overline{\mathbf{x}})   
\end{eqnarray}
}
\normalsize
where for clarity, the dependency of $g$ on state and design variables was omitted.
From here on, $\boldsymbol\lambda_{n}$ represents an adjoint vector corresponding to increment $n$, not to be confused with the scalar $\lambda_n$ which is used for the time discretization of the plastic multiplier $\dot{\lambda}$. 
Furthermore, $\boldsymbol\lambda_{n}$ is a global adjoint vector whereas $\boldsymbol\gamma_n$ is a local (Gauss-point) adjoint vector.
In principle, $g$ can be a function of all state variables throughout all time steps, in addition to its dependency on design variables.
For the particular functionals in Eq.~\eqref{eq:problemformulation}, we see that $g_0(\overline{\mathbf{x}}) = g_0(\mathbf{u}_N(\overline{\mathbf{x}}),\theta_N(\overline{\mathbf{x}}))$ and $g_2(\overline{\mathbf{x}}) = g_2(\mathbf{v}_N(\overline{\mathbf{x}}))$.
These relations are utilized in the particular implementation of the adjoint procedure for each functional.

The purpose of the adjoint procedure is to eliminate all terms involving derivatives of state variables with respect to design variables, which cannot be computed explicitly.
It can be seen that the only explicit dependency upon design variables is contained in $\mathbf{H}_n$, yielding the expression for the explicit sensitivity with respect to an element physical density 
\begin{equation}
\frac{\partial\hat{g}_{exp}}{\partial \overline{x}_e} = - \sum_{n=1}^{N} \boldsymbol\gamma_n^T \frac{\partial(\mathbf{H}_n)}{\partial \overline{x}_e}. 
\end{equation} 
The adjoint vectors $\boldsymbol\gamma_n$ ($n = 1,...,N$) are computed in each Gauss integration point by a backwards-incremental procedure, which is required due to path dependency of the elasto-plastic response.
The backwards procedure consists of the collection of equation systems resulting from the requirement that all implicit derivatives with respect to the design variables will vanish.
Complete details regarding the derivation of the adjoint procedure can be found in \cite{michaleris1994tangent}, whereas specific implementations are described by  \cite{AmirPhDThesis}, \cite{bogomolny2012conceptual} and \cite{nakshatrala2015topology}.
Implementations with other nonlinear material models were mentioned in the introduction.

The adjoint procedure begins with a coupled system to be solved for ${\boldsymbol\lambda_N}$, 
\begin{eqnarray} \label{eq:lambdaN}
\left[- \frac{\partial({\mathbf{R}_N})}{\partial({\mathbf{v}_N})} \frac{\partial({\mathbf{H}_N})}{\partial({\mathbf{v}_N})}^{-1} \frac{\partial({\mathbf{H}_N})}{\partial({\mathbf{u}_N})} \right]^T {\boldsymbol\lambda_N} & = & 
\frac{\partial g}{\partial({\mathbf{u}_N})}^T - \left[ \frac{\partial g}{\partial({\mathbf{v}_N})}  \frac{\partial({\mathbf{H}_N})}{\partial({\mathbf{v}_N})}^{-1} \frac{\partial({\mathbf{H}_N})}{\partial({\mathbf{u}_N})} \right]^T \nonumber \\
\frac{\partial({\mathbf{R}_N})}{\partial({\theta}_N)}^T {\boldsymbol\lambda_N} & = & \frac{\partial g}{\partial({\theta_N})}
\end{eqnarray}
where $\left[\frac{\partial({\mathbf{R}_N})}{\partial({\mathbf{v}_N})} \frac{\partial({\mathbf{H}_N})}{\partial({\mathbf{v}_N})}^{-1} \frac{\partial({\mathbf{H}_N})}{\partial({\mathbf{u}_N})} \right]$ is the tangent stiffness matrix corresponding to the converged state at increment $N$ \cite{michaleris1994tangent}.
${\boldsymbol\gamma_N}$ is then determined on a Gauss-point level by solving
\begin{equation} \label{eq:gammaN}
\frac{\partial({\mathbf{H}_N})}{\partial({\mathbf{v}_N})}^T {\boldsymbol\gamma_N} = - \frac{\partial({\mathbf{R}_N})}{\partial({\mathbf{v}_N})}^T {\boldsymbol\lambda_N} + \frac{\partial g}{\partial({\mathbf{v}_N})}^T.
\end{equation}
Proceeding incrementally backwards in time, in the $n$-th increment the coupled adjoint equations are solved to determine ${\boldsymbol\lambda_n}$
\begin{eqnarray} \label{eq:lambdan}
\left[- \frac{\partial({\mathbf{R}_n})}{\partial({\mathbf{v}_{n}})} \frac{\partial({\mathbf{H}_n})}{\partial({\mathbf{v}_{n}})}^{-1} \frac{\partial({\mathbf{H}_n})}{\partial({\mathbf{u}_n})} \right]^T {\boldsymbol\lambda_n} & = & 
\frac{\partial g}{\partial({\mathbf{u}_n})}^T 
- \left[ \frac{\partial g}{\partial({\mathbf{v}_{n}})}  \frac{\partial({\mathbf{H}_n})}{\partial({\mathbf{v}_{n}})}^{-1} \frac{\partial({\mathbf{H}_n})}{\partial({\mathbf{u}_n})} \right] ^T \nonumber \\
& - & \left[\frac{\partial({\mathbf{H}_{n+1}})}{\partial({\mathbf{u}_n})}  
- \frac{\partial({\mathbf{H}_{n+1}})}{\partial({\mathbf{v}_{n}})} \frac{\partial({\mathbf{H}_n})}{\partial({\mathbf{v}_{n}})}^{-1} \frac{\partial({\mathbf{H}_n})}{\partial({\mathbf{u}_n})} \right]^T {\boldsymbol\gamma_{n+1}} \nonumber \\
\frac{\partial({\mathbf{R}_n})}{\partial({\theta_n})}^T {\boldsymbol\lambda_n} & = & \frac{\partial g}{\partial({\theta_n})}
\end{eqnarray}
followed by the solution of the local adjoint vector ${\boldsymbol\gamma_n}$ on a Gauss-point level
\begin{equation} \label{gammaNm1}
\frac{\partial({\mathbf{H}_{n}})}{\partial({\mathbf{v}_{n}})}^T {\boldsymbol\gamma_n} = - \frac{\partial({\mathbf{R}_n})}{\partial({\mathbf{v}_{n}})}^T {\boldsymbol\lambda_n} 
- \frac{\partial({\mathbf{H}_{n+1}})}{\partial({\mathbf{v}_{n}})}^T {\boldsymbol\gamma_{n+1}} + \frac{\partial g}{\partial({\mathbf{v}_{n}})}^T.
\end{equation}
Once ${\boldsymbol\gamma_n}$ is determined, its contribution to the design sensitivities is computed.
Then the procedure continues to the previous increment denoted by $n-1$.
This is repeated until all contributions are collected to obtain the required design sensitivities.

The partial derivatives of the objective, the constraints, the global residuals and the local residuals with respect to the state variables are required for implementing the adjoint procedure.
The derivatives $\frac{\partial({\mathbf{R}_n})}{\partial({\mathbf{v}_n})}$ and $\frac{\partial({\mathbf{R}_n})}{\partial({\theta_n})}$ can be easily obtained from Eq.~\eqref{eq:epmodel_R} whereas the derivatives
$\frac{\partial({\mathbf{H}_n})}{\partial({\mathbf{u}_n})}$,
$\frac{\partial({\mathbf{H}_{n+1}})}{\partial({\mathbf{u}_n})}$,
$\frac{\partial({\mathbf{H}_n})}{\partial({\mathbf{v}_n})}$,
$\frac{\partial({\mathbf{H}_{n+1}})}{\partial({\mathbf{v}_n})}$
and $\frac{\partial({\mathbf{H}_n})}{\partial \overline{x}_e}$
are related to the particular elasto-plastic model and to the choice of the internal variables $\mathbf{v}$.
For the model considered herein based on classical $J_2$ flow theory, they can be derived by differentiation of Eq.~\eqref{eq:epmodel_H}.  
An explicit example of these derivatives was given in \cite{AmirPhDThesis}. 
Finally, the partial derivatives 
$\frac{\partial g}{\partial({\mathbf{u}_N})},
\frac{\partial g}{\partial({\mathbf{v}_N})}, 
\frac{\partial g}{\partial({\theta_N})}$,
$\frac{\partial g}{\partial({\mathbf{u}_n})},
\frac{\partial g}{\partial({\mathbf{v}_n})}$ and 
$\frac{\partial g}{\partial({\theta_n})}$ can be derived explicitly for each functional to be considered in the problem formulation of Eq.~\eqref{eq:problemformulation}. 

It should be noted that when implementing the adjoint procedure, the derivatives of the local residuals ${\mathbf{H}_n}$ and ${\mathbf{H}_{n+1}}$ should maintain consistency with respect to the analysis. 
In essence, four situations are possible at a certain sequence of increments $\left\{n, n+1\right\}$:
1) Continuous elastic response; 
2) Elastic-plastic transition;
3) Continuous plastic response;
and 4) Plastic-elastic transition (during unloading).
The actual situation encountered affects the computation of the derivatives of the respective residuals.
In general, the derivatives of the local residual are matrices of varying sizes, depending on the situation which is determined exclusively by the elastic trial state. 

The final component required for performing the sensitivity analysis is the derivative of the residual ${\mathbf{H}_n}$ with respect to the physical material density. 
Combining Eqs.~\eqref{eq:epmodel_H}, \eqref{eq:E_ModSIMP} and \eqref{eq:sigmay_ModSIMP}, we obtain
\begin{equation}
\frac{\partial({\mathbf{H}_n})}{\partial{\overline{x}_e}} = \left[ \begin{array}{c}
																										\mathbf{0} \\
																										0 \\
																										{\frac{\partial E}{\partial \overline{x}_e} \mathbf{D}_0 ({\boldsymbol\epsilon^{el}_n} - {\boldsymbol\epsilon^{el}_{n-1}})} \\
																										{-\frac{2}{3} (\sigma^0_{y} + H E \kappa_n) ( \frac{\partial \sigma^0_y}{\partial \overline{x}_e} + H \frac{\partial E}{\partial \overline{x}_e} \kappa_n)} \end{array} \right]									
\end{equation}
where $\frac{\partial E}{\partial \overline{x}_e}$ and $\frac{\partial \sigma^0_y}{\partial \overline{x}_e}$ are derived from Eqs.~\eqref{eq:E_ModSIMP} and \eqref{eq:sigmay_ModSIMP}, and $\mathbf{D}_0$ is the elastic constitutive tensor for Young's modulus equal to 1.


\section{Examples} \label{sec:examples}

In this section we present several numerical examples that demonstrate the applicability of the proposed approach for solving stress-constrained topology optimization.
Different variants of the optimization problem of Eq.~\eqref{eq:problemformulation} are considered.
All optimization problems are solved by the method of moving asymptotes - MMA \cite{svanberg1987method}.
Specific parameters required for reproducing the results are given within the text. 

\subsection{Example 1: Stress-constrained L-bracket design}
\begin{figure}
	\centering
	\includegraphics[width=0.5\textwidth]{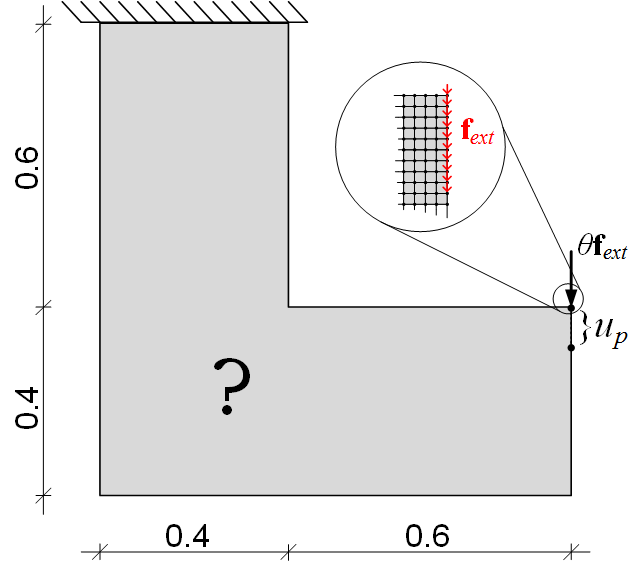}
	\caption{Problem setup for topology optimization of an L-bracket. The load is distributed over the top 10 nodes in order to avoid artificial stress concentrations at the loading point.}
	\label{fig:lbracketsetup}
\end{figure}

In the first example we consider the classical case of an L-bracket, which is often used to evaluate new procedures for topology optimization under stress constraints (e.g.~\cite{le2010stress} and references therein), see Figure \ref{fig:lbracketsetup} for the problem setup.
In particular, the case of the L-shaped domain with a point load at the position $\left\{ 1.0,0.4\right\}$ is thoroughly examined, as by \cite{le2010stress} and \cite{verbart2013new}.
A similar problem, often appearing in articles about stress constraints, is the case in which the point load is applied at the position $\left\{ 1.0,0.2\right\}$.
In the author's opinion, the latter is somewhat easier to deal with because the optimized layout for compliance only has a wider angle in the re-entrant corner, see Figures \ref{fig:MaxComp150layout} and \ref{fig:maxcomp150midlayout}. 
Therefore the former case is in the center of the following examination, in order to fairly evaluate the proposed approach.
The latter case is presented subsequently for the sake of completeness.
 
\subsubsection{Reference design: no stress limitation}
First, a maximization of end-compliance subject to a volume constraint only is performed, given a certain prescribed displacement.
This is necessary in order to identify the ``stiffest design'' achievable without any limitation on stresses.
The load is distributed over the top 10 nodes in order to avoid artificial stress concentrations at the loading point.
Assuming the 10 adjacent nodes will have almost identical vertical displacements, it is sufficient to measure the end-compliance based on a single DOF where the displacement is prescribed, instead of measuring the complete end-compliance.  
This somewhat simplifies the computational implementation of the adjoint equations, though the derivation above is general and applicable to any loading situation.
With reference to the formulation in Eq.~\eqref{eq:problemformulation}, the objective now includes only the product of force and displacement at the prescribed DOF and the constraint $g_2$ is omitted.   
The resulting layout and performance coincide with those that can be obtained by a linear-elastic minimum compliance topology optimization procedure.
This is expected for an elasto-plastic single-material optimization with strain hardening.
If multiple materials are considered, with distinct yield stresses and hardening behaviors, the optimized layouts may differ from linear-elastic minimum compliance layouts, see for example \cite{kato2015analytical}. 
The model is discretized with a 150$\times$150 mesh resolution consisting of 14,400 square, bi-linear elements.
The available volume is set to 35\% of the total volume of the L-shaped domain and the filter radius is 0.02.
The prescribed displacement the position $\left\{ 1.0,0.4\right\}$ is set to $u_p = 0.01$ and automatic displacement incrementation is applied, where the increment size is adapted based on the convergence of Newton-Raphson iterations in the previous increment. 
The material parameters are given in Table \ref{tab:matparams} and are essentially constant for all examples.

\begin{table}%
\centering
\begin{tabular}{|c|c|}
\hline 
$E_{min}$ & $1.0\cdot10^{-3}$ \\
\hline 
$E_{max}$ & $1.0\cdot10^3$ \\
\hline 
$\nu$ & 0.3\\
\hline 
$\sigma_{y,min}^0$ & 0.0 \\ 
\hline 
$\sigma_{y,max}^0$ & 2.0 \\
\hline 
$H$ & 0.01 \\
\hline
\end{tabular}
\caption{Material parameters used for all examples.}
\label{tab:matparams}
\end{table}

According to the numerical experiments, a continuation scheme involving both the penalty exponents and the Heaviside sharpness yields the best results.
The parameters $p_E$ and $p_{\sigma_y}$ are increased gradually throughout the optimization process.
The initial values are set to $p_E=1.0$ and $p_{\sigma_y}=0.5$ and they are increased by 0.1 every 10 design cycles, up to the values of 5.0 and 4.5 respectively.
The parameter $\beta_{HS}$ is initialized at 1.0 and multiplied by 1.1 every 10 design cycles, but only when $p_E \geq 3.0$. 
The upper limit for $\beta_{HS}$ is set to 10.0 in order to avoid highly nonlinear projection functions. 
In the call to MMA, the derivatives of the compliance objective are multiplied by $10^5$ in order to obtain good scaling and consequently fast convergence of the MMA sub-problems.
It is known that the performance of MMA can be affected by this scaling parameter, which should be chosen according to the values of the actual quantities--in this case, the magnitude of the end-compliance is in the order of $10^{-5}$.
According to the author's experience, if the problem is badly scaled then it can slow down convergence and in some cases lead the overall optimization process to inferior local minima.
An external move limit of 0.2 on the MMA update is enforced. 
In all examples presented in this section, the optimization is terminated after 500 design cycles.
The stopping criterion that was imposed, requiring that the maximum change in an element's density is below $10^{-3}$, was never achieved.

The optimized topology is presented in Figure \ref{fig:MaxComp150layout}.
The end-compliance, the sum of equivalent plastic strains and the volume are presented in the second column of Table \ref{tab:lbracketresults}.
The stress distribution in terms of von Mises stresses is presented in Figure \ref{fig:MaxComp150vmstress} and the distribution of equivalent plastic strains in Figure \ref{fig:MaxComp150equivps}.
It can be seen that there is a significant stress concentration in the vicinity of the re-entrant corner.

\begin{table*}
	\footnotesize
	\centering
		\begin{tabular}{|c||c||c||c|}
		\hline
			\multirow{2}{*}{} & Max. end-comp.  & Max. end-comp. s.t. vol. & Min. vol. s.t. end-comp. \\
							                  & s.t. vol.       & and plastic strains      & and plastic strains \\
		\hline
			End-compliance \hfill $\theta_{N} \hat{{f}^p} {{u}_{N}^p}$                                          & $4.3013\cdot10^{-5}$ & $3.8163\cdot10^{-5}$ & $3.7908\cdot10^{-5}$  \\
			Plastic strains \hfill $\sum_{e=1}^{N_e} \sum_{k=1}^{N_{GP}} {\kappa^{ek}_N}$                       & $2.0455\cdot10^{-1}$ & $1.3353\cdot10^{-3}$ & $1.1369\cdot10^{-2}$  \\
			Volume \hfill $\frac{\sum_{e=1}^{N_e} v_e \overline{\mathbf{x}}_e}{0.35\cdot{N_e}\cdot{v_e}} - 1$ & $-5.6769\cdot10^{-7}$ & $-2.7461\cdot10^{-4}$ & $-2.2727\cdot10^{-2}$ \\
			Figures \hfill { } & \ref{fig:MaxComp150layout}, \ref{fig:MaxComp150vmstress}, \ref{fig:MaxComp150equivps} & \ref{fig:plstr150layout}, \ref{fig:plstr150vmstress}, \ref{fig:plstr150equivps} & \ref{fig:minvol150layout}, \ref{fig:minvol150vmstress}, \ref{fig:minvol150equivps} \\ 
		\hline
		\end{tabular}
	\caption{Results of the topology optimization of an L-bracket with a load at the top right corner. For the same volume and under the same prescribed displacement, constraining the sum of equivalent plastic strains leads to nearly zero plastic strains while compromising the end-compliance by 13\%.}
	\label{tab:lbracketresults}
\end{table*}

\subsubsection{Constraining the plastic straining}
\begin{figure*}%
	\centering
	
	\begin{subfigure}[c]{0.3\textwidth}
		\includegraphics[width=\textwidth,trim={0.5in 2in 1in 3in}]{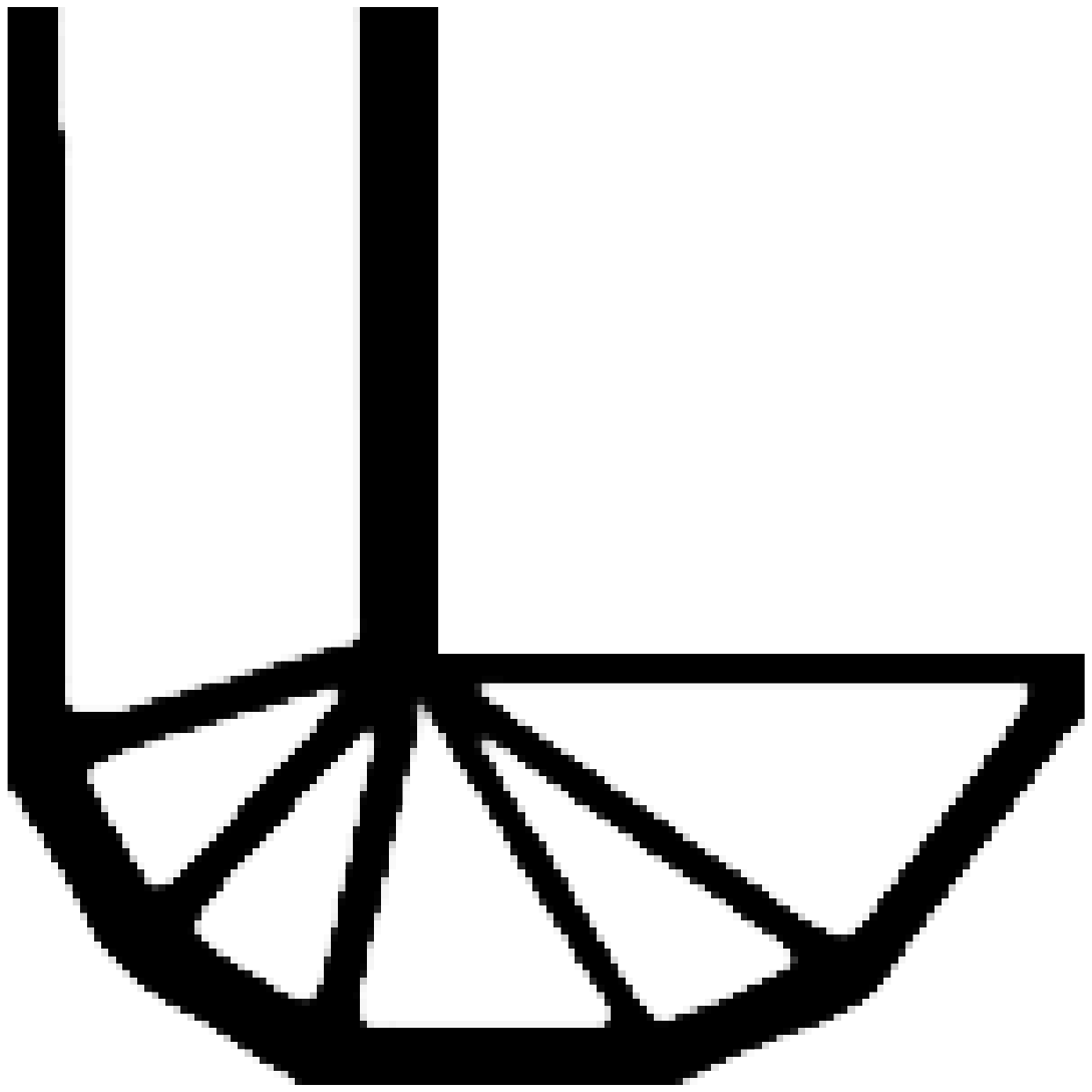}
		\caption{}
		\label{fig:MaxComp150layout}
	\end{subfigure}
	~
	\begin{subfigure}[c]{0.3\textwidth}
		\includegraphics[width=\textwidth,trim={0.5in 2in 1in 3in}]{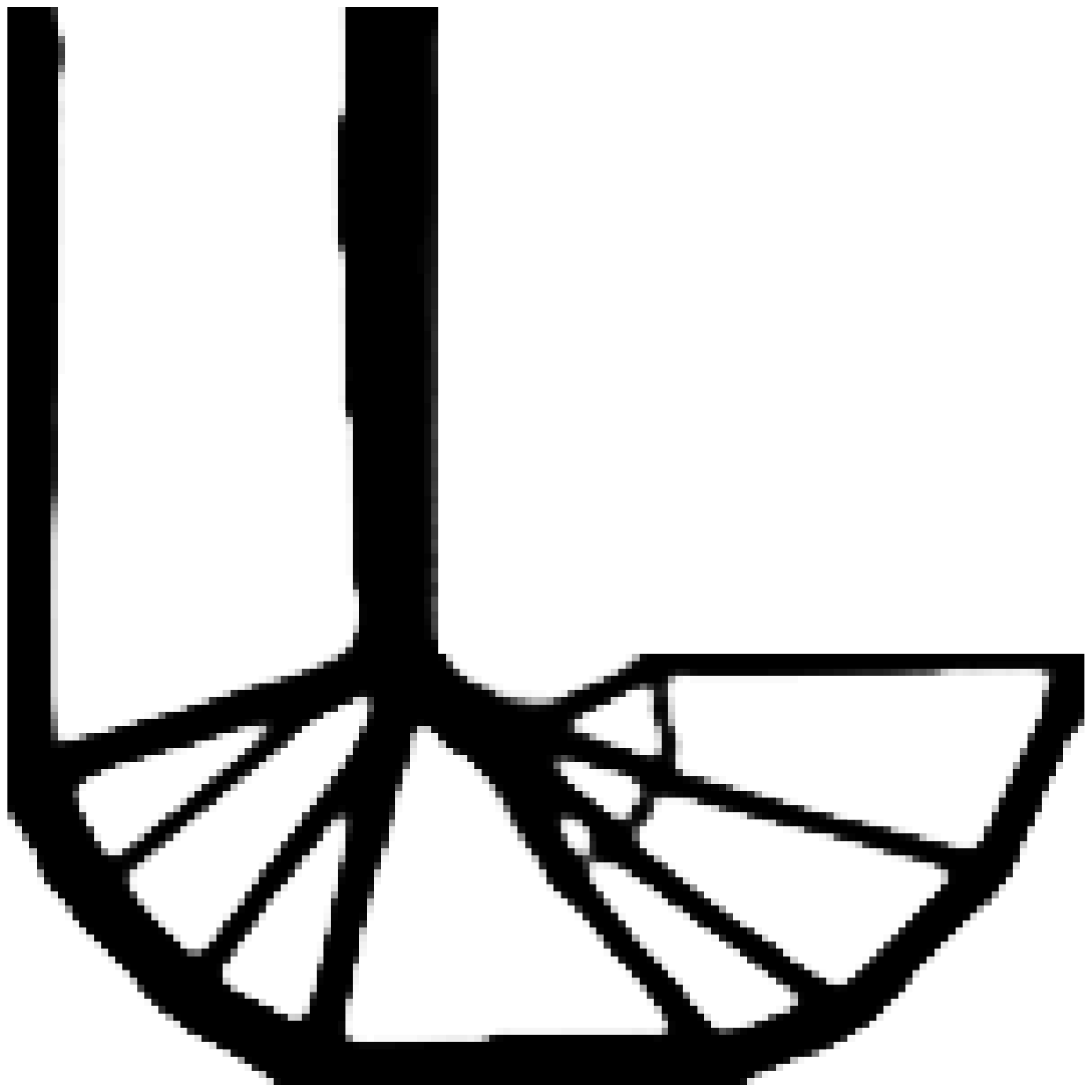}
		\caption{}
		\label{fig:plstr150layout}
	\end{subfigure}
	~
	\begin{subfigure}[c]{0.3\textwidth}
		\includegraphics[width=\textwidth,trim={0.5in 2in 1in 3in}]{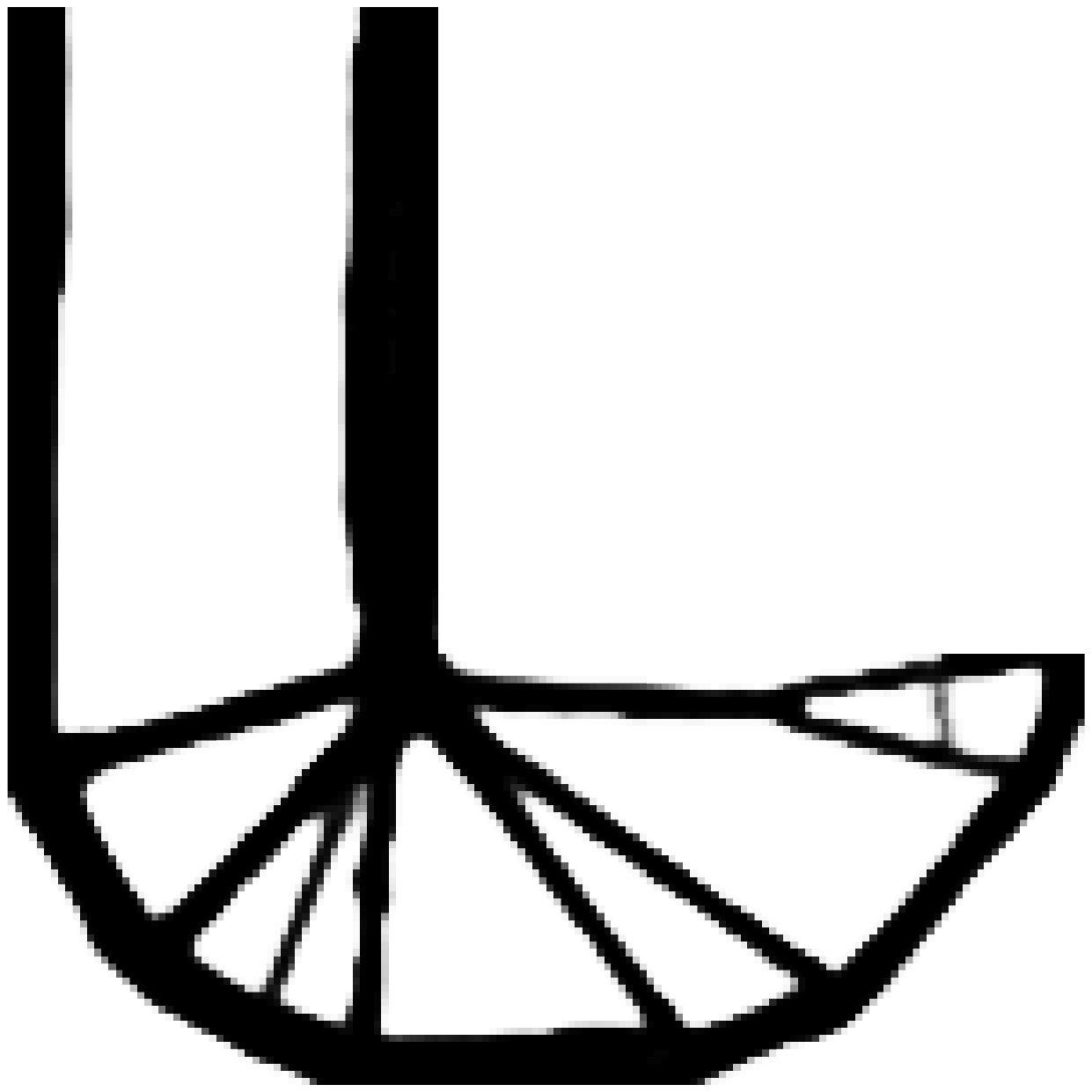}
		\caption{}
		\label{fig:minvol150layout}
	\end{subfigure}
	
	\begin{subfigure}[c]{0.3\textwidth}
		\includegraphics[width=\textwidth]{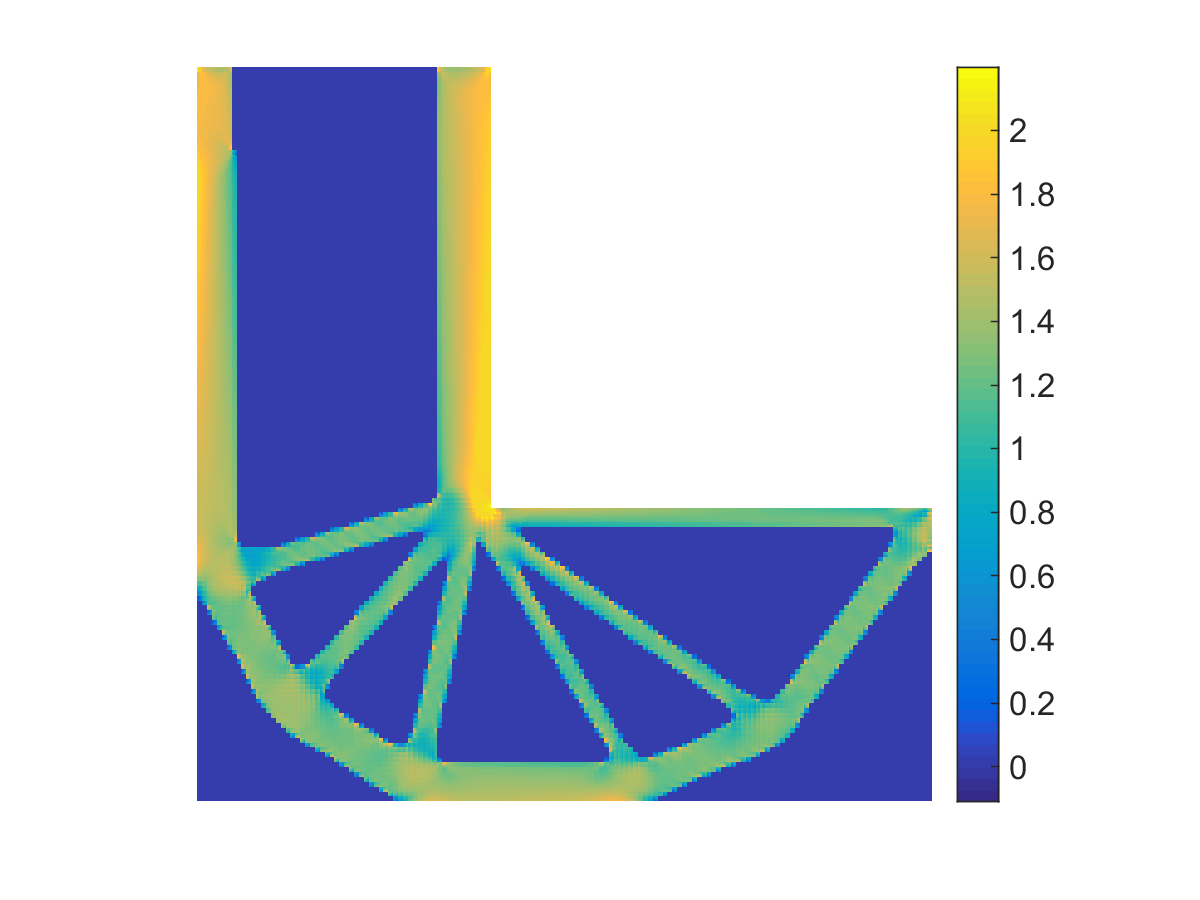}
		\caption{}
		\label{fig:MaxComp150vmstress}
	\end{subfigure}
	~
	\begin{subfigure}[c]{0.3\textwidth}
		\includegraphics[width=\textwidth]{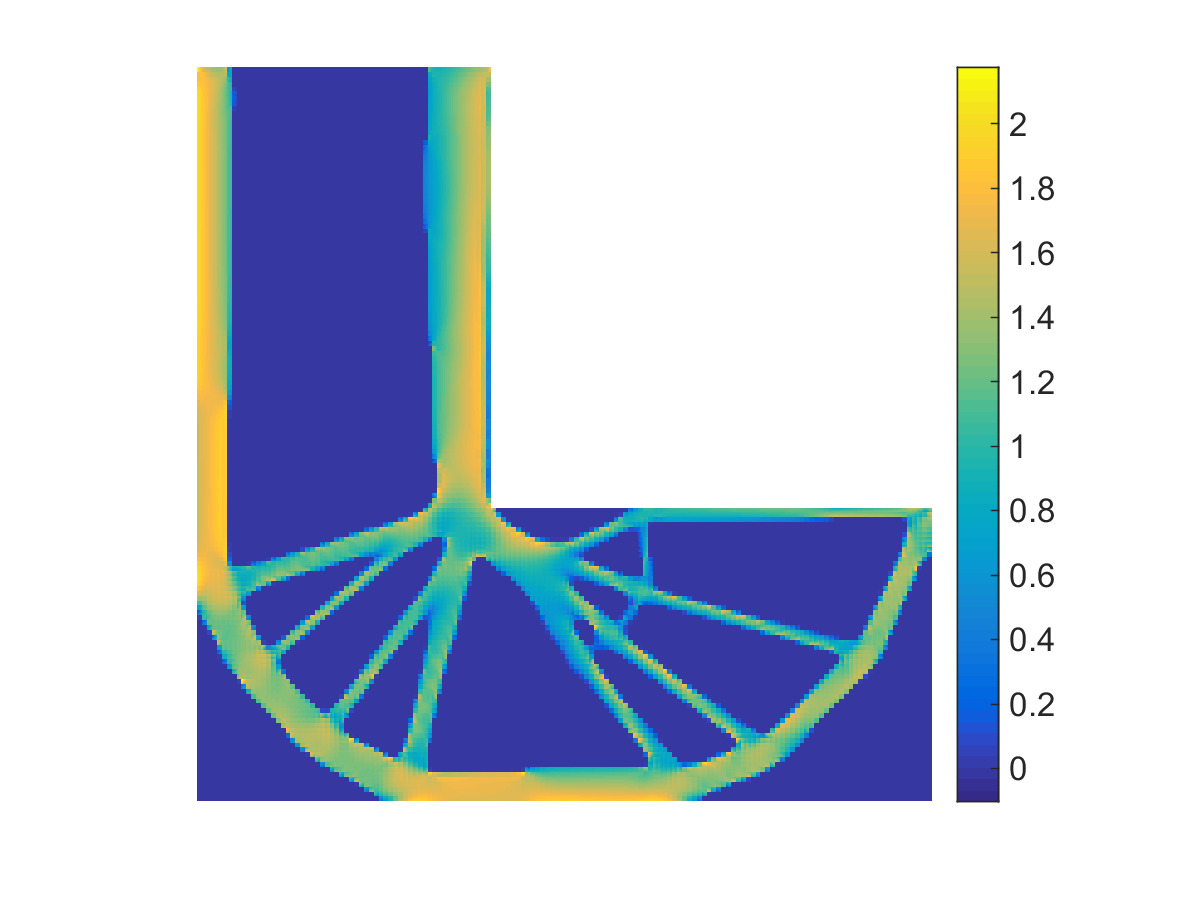}
		\caption{}
		\label{fig:plstr150vmstress}
	\end{subfigure}
	~
	\begin{subfigure}[c]{0.3\textwidth}
		\includegraphics[width=\textwidth]{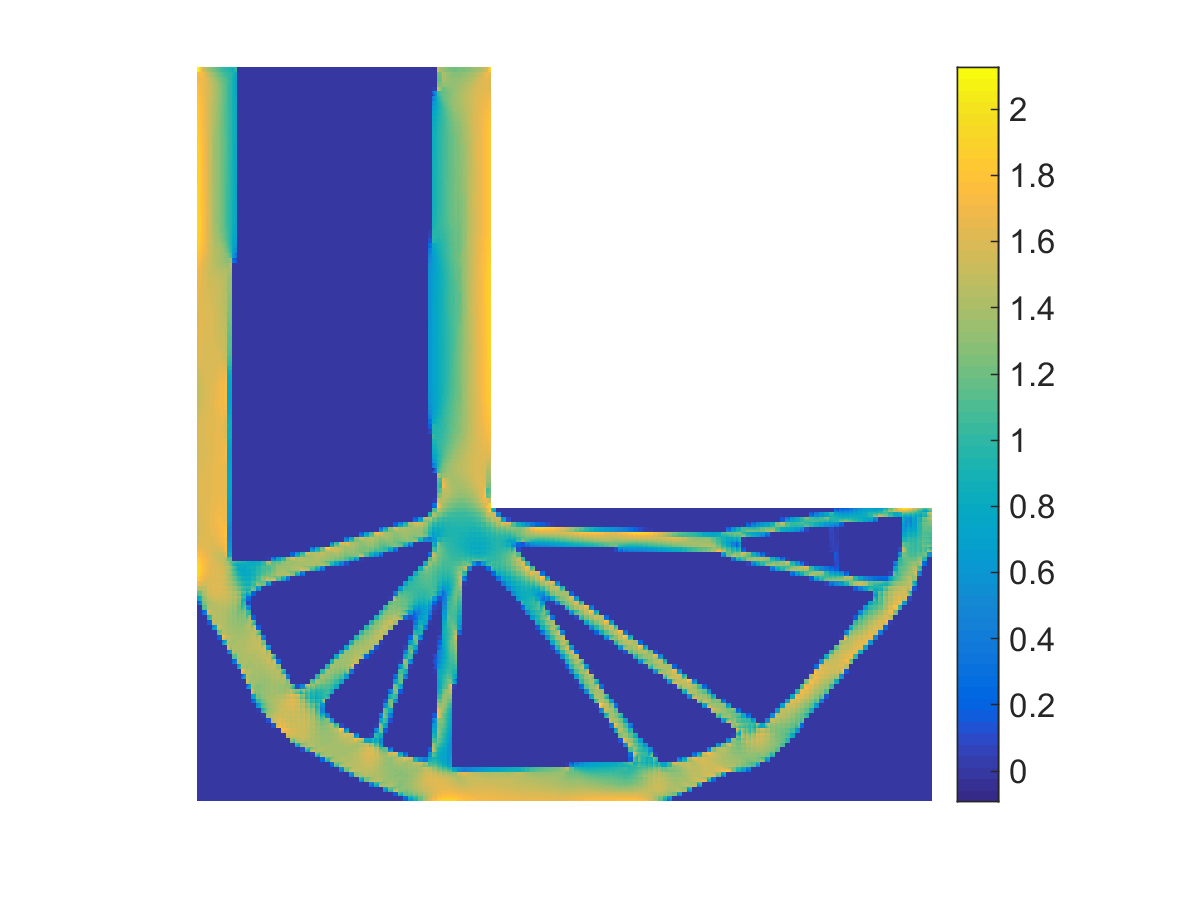}
		\caption{}
		\label{fig:minvol150vmstress}
	\end{subfigure}
	
	\begin{subfigure}[c]{0.3\textwidth}
		\includegraphics[width=\textwidth]{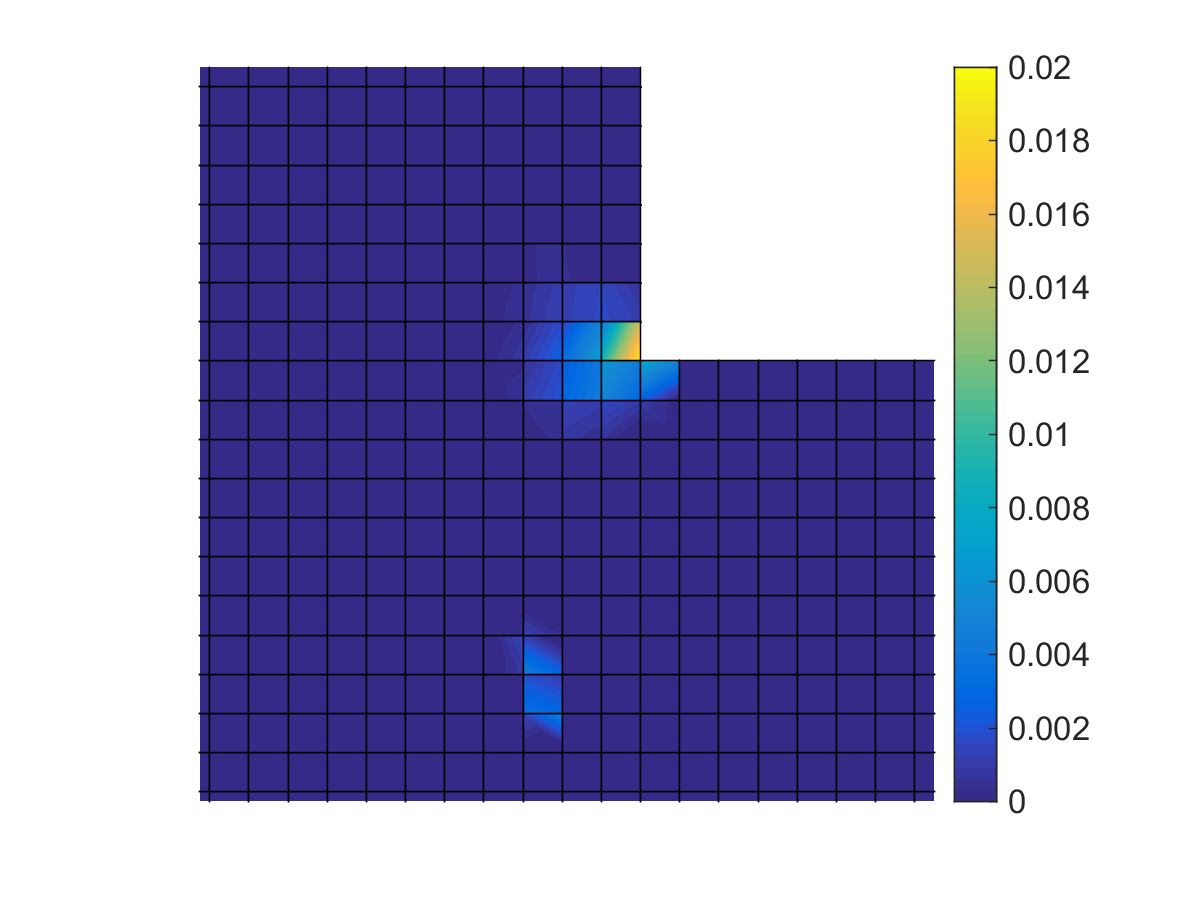}
		\caption{}
		\label{fig:MaxComp150equivps}
	\end{subfigure}
	~
	\begin{subfigure}[c]{0.3\textwidth}
		\includegraphics[width=\textwidth]{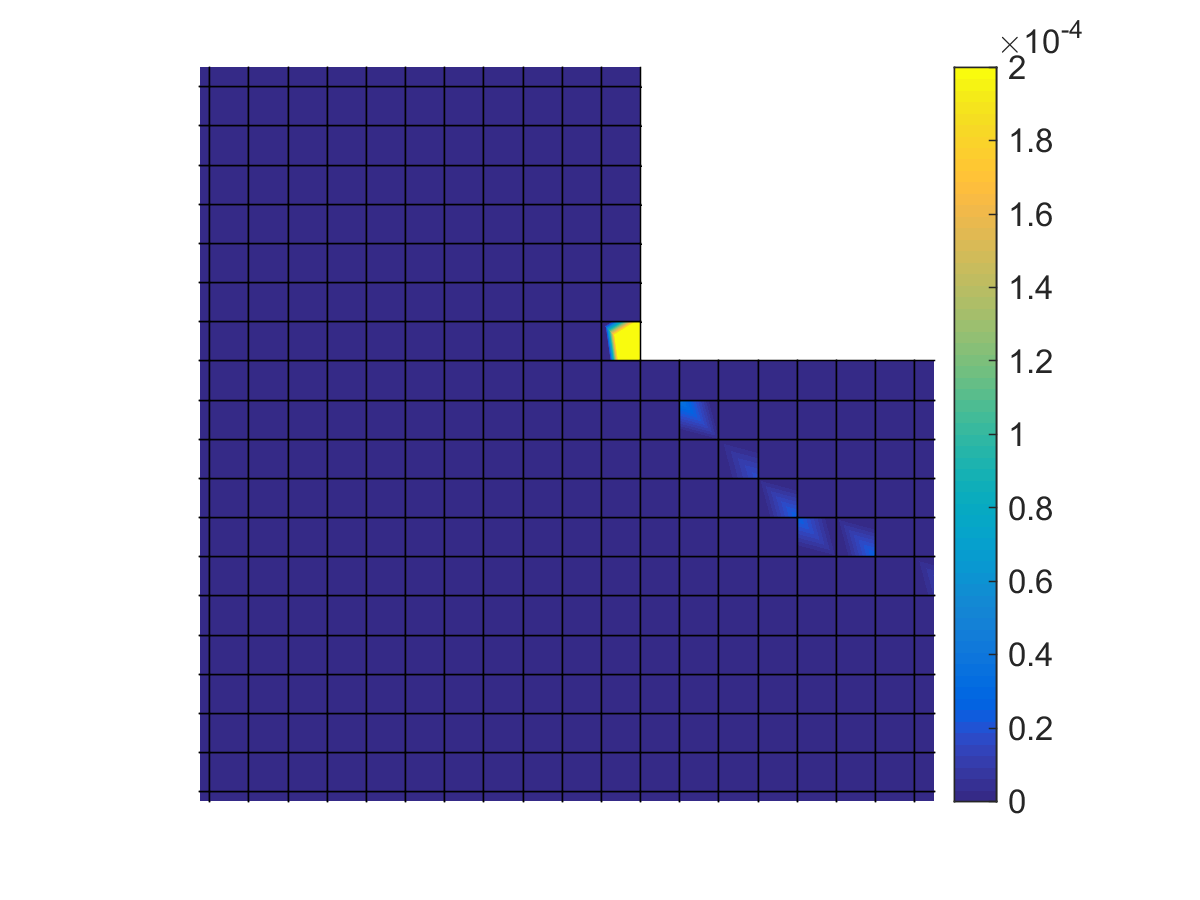}
		\caption{}
		\label{fig:plstr150equivps}
	\end{subfigure}
	~
	\begin{subfigure}[c]{0.3\textwidth}
		\includegraphics[width=\textwidth]{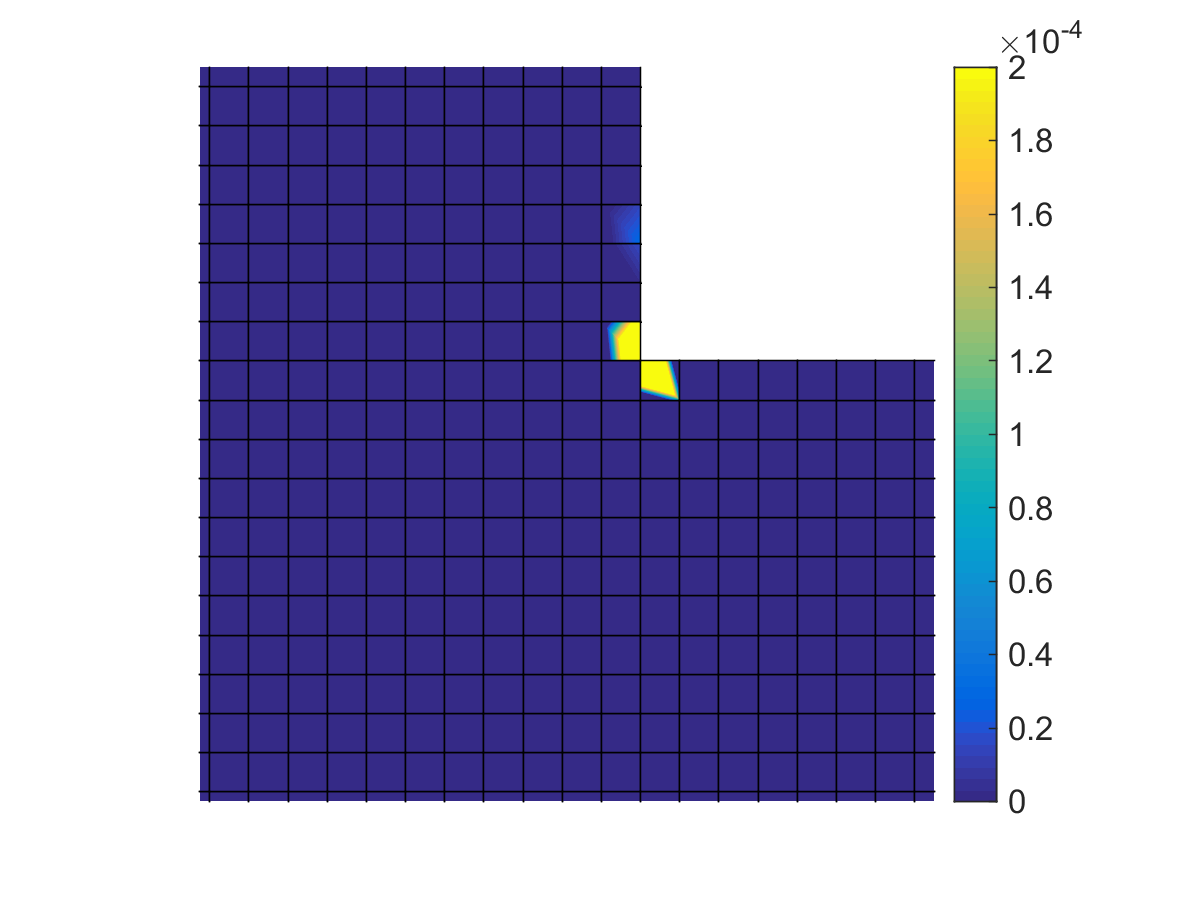}
		\caption{}
		\label{fig:minvol150equivps}
	\end{subfigure}
	
	\caption{Topology optimization of an L-bracket with a load and a prescribed displacement at the top right corner. From the left: maximizing the end-compliance s.t. a volume constraint; maximizing the end-compliance s.t. constraints on volume and on the total sum of equivalent plastic strains; minimizing volume s.t. constraints on end-compliance and on the total sum of equivalent plastic strains. From top: optimized layouts; von Mises stress distributions; equivalent plastic strains in the vicinity of the re-entrant corner. Note the two orders of magnitude difference between the scale of sub-figure (g) compared to (h) and (i).}%
	\label{fig:lbracketresults}%
\end{figure*}

The exact same problem setup is used to demonstrate the capability of the proposed approach to capture local stress concentrations and to consider them when seeking the optimized topology.
We add a single, global constraint on the total sum of the equivalent plastic strains at the final equilibrium state, as in Eq.~\eqref{eq:problemformulation}.
It will be shown that the stress distribution is improved and stress concentrations are avoided.
All material parameters remain the same, so does the continuation scheme for  $p_E$, $p_{\sigma_y}$ and $\beta_{HS}$ as well as the scaling and move limit for MMA.
A small threshold of $g_2^{\star}=10^{-4}$ is set for the constraint on plastic strains, providing some slack and improving the convergence to a design with almost zero plastic strains.
The optimized topology is presented in Figure \ref{fig:plstr150layout}.
The end-compliance, the sum of equivalent plastic strains and the volume are presented in the third column of Table \ref{tab:lbracketresults}.
The stress distribution in terms of von Mises stresses is presented in Figure \ref{fig:plstr150vmstress} and the distribution of equivalent plastic strains in Figure \ref{fig:plstr150equivps}.

Examining the optimized layout, it can be seen that the proposed approach can indeed generate designs that circumvent stress concentrations.
The modification of the design, compared to the one optimized for end-compliance only, is quite subtle--a rounding of the re-entrant corner and a stiffening of the bars meeting at the corner in order to compensate for the reduced stiffness of the rounded corner.  
This result is slightly different from those achieved in previous studies referenced above, and appears to be the closest one to the layout obtained without stress considerations. 
It resembles a result achieved by \cite{bruggi2008mixed} with a mixed-FEM approach and local stress constraints. 
It also resembles a result achieved by \cite{le2010stress} but for a much higher volume fraction (see Figure 7(b) in the referenced article).
It can be argued that the primary design change is a shape modification, that may have been generated by a shape optimization procedure following topology optimization without stress constraints.
Nevertheless, it can be seen that topological changes are indeed introduced, for example the stiffening bars orthogonal to the main bars. 
This means that performing topology optimization without stress considerations, followed by shape optimization with stress considerations, may not be sufficient for finding the best layout in terms of both topology and shape.

As for the stress distribution, it is evident that adding a constraint on plastic strains leads to a more uniform distribution of extreme stresses.
Hence stress constraint violation is implicitly avoided without actually imposing local stress constraints on each material point.
Nevertheless, a slight violation of the global constraint on plastic strains is observed--plastic straining is present in the first element near the re-entrant corner. 
This violation can be attributed to several factors: 1) The non-differentiability of plastic strains at the yield point, causing difficulties in satisfying the constraint precisely; 2) The inherent approximation due to the use of a sequential convex programming method for solving a non-convex problem; and 3) The quality of the optimization algorithm itself.

Another possibility to achieve the same design goal is by interchanging end-compliance and volume in Eq.~\eqref{eq:problemformulation}.
This corresponds to minimizing the volume of the optimized design while requiring a certain load-bearing capacity for a given prescribed displacement.
The end-compliance is constrained to $3.8\cdot10^{-5}$ in order to obtain a good comparison with the result of maximization of end-compliance s.t.~volume and plastic strains. 
The initial volume fraction is set to 100\% of the domain. 
All parameters used in the solution of the previous case retain the same values, except for the continuation on the penalization exponents.
For effectively constraining the end-compliance, it is necessary to begin the process with some penalty in Eqs.~\eqref{eq:E_ModSIMP} and \eqref{eq:sigmay_ModSIMP}.
Therefore the initial values are chosen as $p_E=3.0$ and $p_{\sigma_y}=2.5$ and they are kept constant for the first 200 iterations. 
Then, the same continuation scheme is applied as for the previous cases.  
The optimized topology is presented in Figure \ref{fig:minvol150layout}.
The end-compliance, the sum of equivalent plastic strains and the volume are presented in the fourth column of Table \ref{tab:lbracketresults}.
The stress distribution in terms of von Mises stresses is presented in Figure \ref{fig:minvol150vmstress} and the distribution of equivalent plastic strains in Figure \ref{fig:minvol150equivps}.

\subsubsection{Elasto-plastic performance of the optimized design}
For examining the actual benefit from the proposed formulation, it is interesting to examine the elasto-plastic response of the three designs obtained in the numerical experiments.
The responses are directly compared to a result by \cite{le2010stress}, in particular the layout in Fig.~7(e) of the referenced article whose setting is the closest to the current problem definition.

\begin{figure*}%
	\centering
	
	\begin{subfigure}[c]{0.4\textwidth}
		\includegraphics[width=\textwidth,trim={0.5in 2in 1in 3in}]{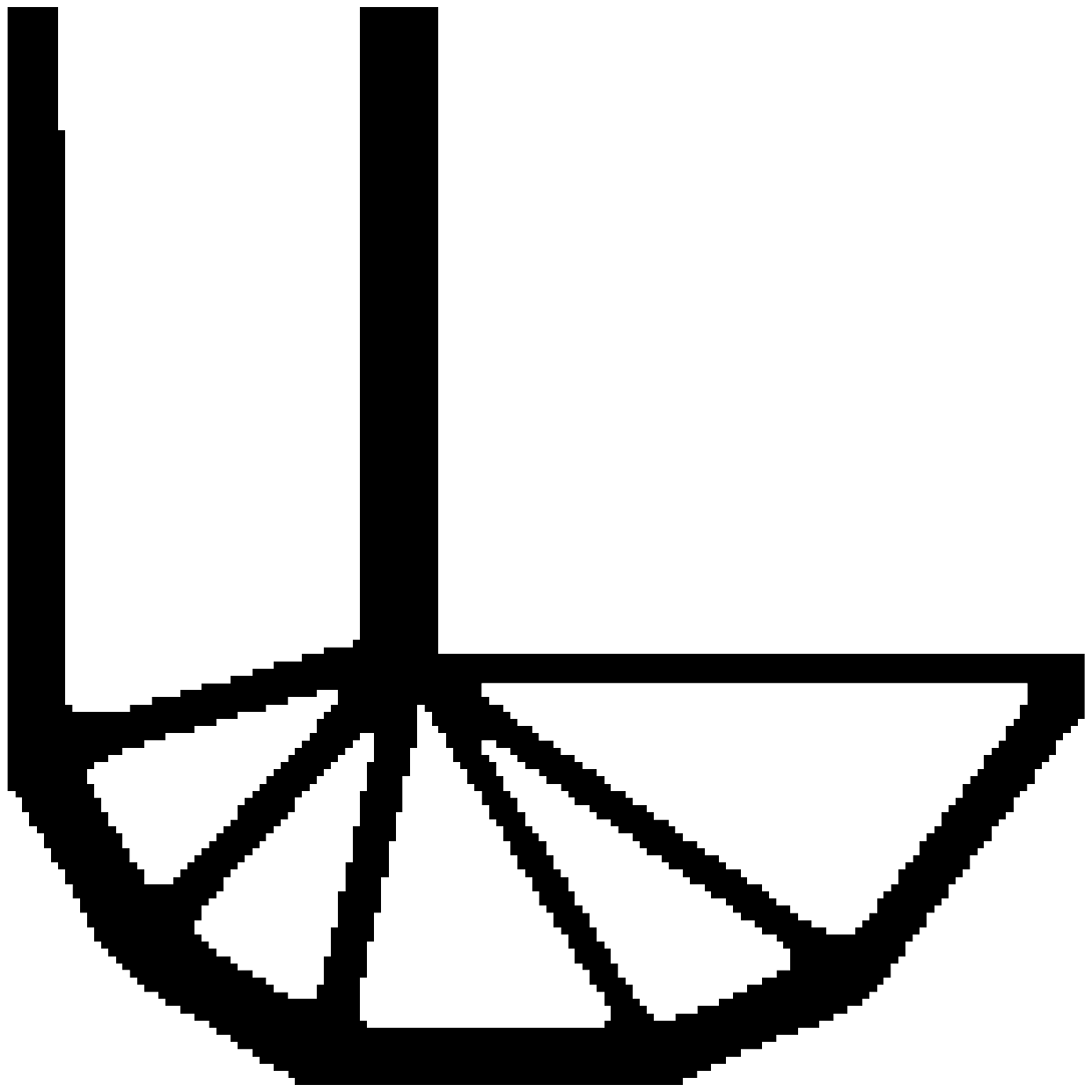}
		\caption{}
		\label{fig:maxcomp150layoutPP}
	\end{subfigure}
	~
	\begin{subfigure}[c]{0.4\textwidth}
		\includegraphics[width=\textwidth,trim={0.5in 2in 1in 3in}]{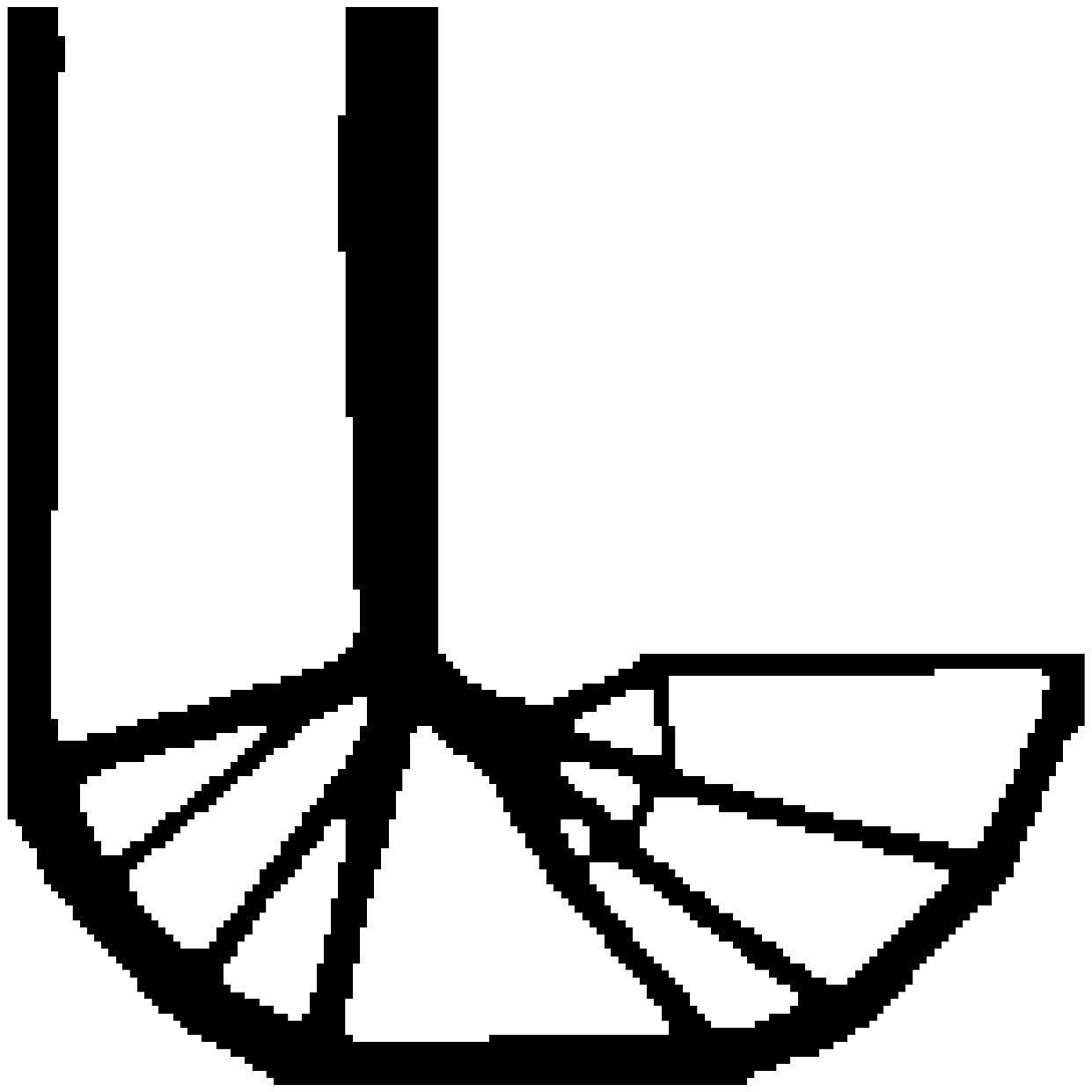}
		\caption{}
		\label{fig:plstr150layoutPP}
	\end{subfigure}
	~
	\begin{subfigure}[c]{0.4\textwidth}
		\includegraphics[width=\textwidth,trim={0.5in 2in 1in 3in}]{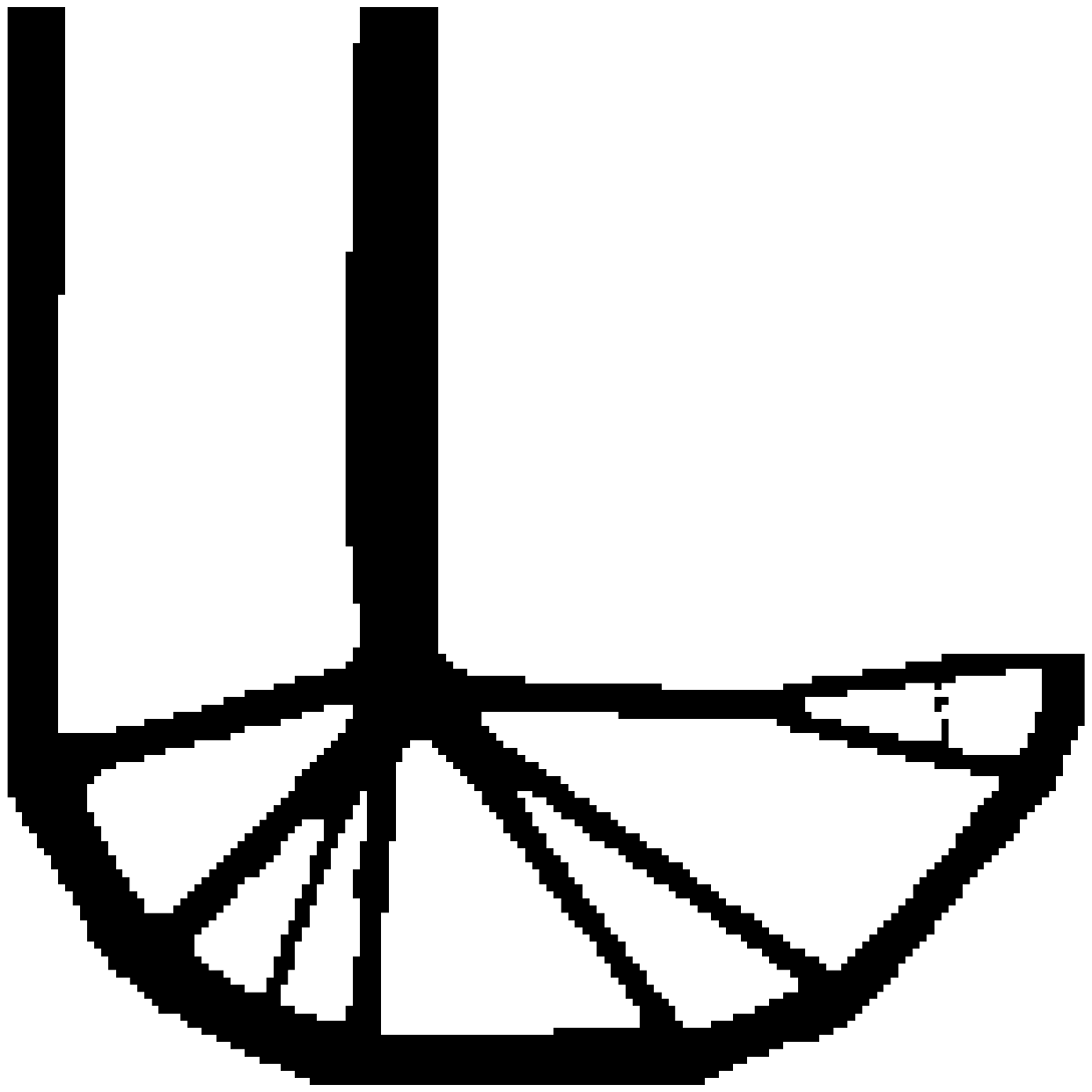}
		\caption{}
		\label{fig:minvol150layoutPP}
	\end{subfigure}
	~
	\begin{subfigure}[c]{0.4\textwidth}
		\includegraphics[width=\textwidth,trim={0.5in 2in 1in 3in}]{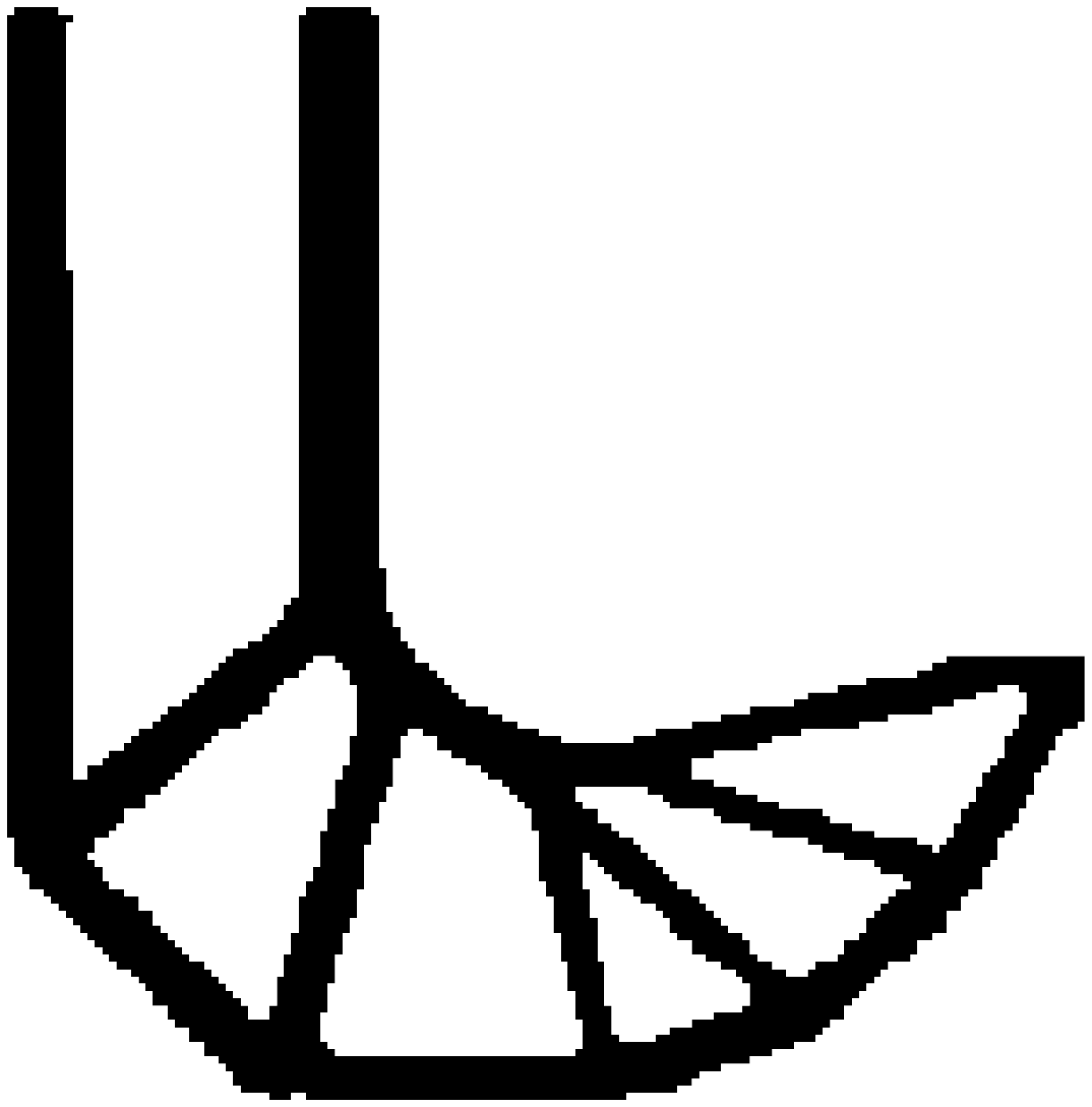}
		\caption{}
		\label{fig:Le150layoutPP}
	\end{subfigure}
	
	\caption{Post-processed 0/1 designs of the L-bracket for comparing the elasto-plastic responses of the optimized layouts. Only minor changes are observed compared to the results of optimization in Figure \ref{fig:lbracketresults}. The layout (d) is based on an image imported from \cite{le2010stress} for comparing the responses.}
	\label{fig:checkdesignslayouts}
	
\end{figure*}

A simple post-processing of the layouts obtained in the current study is performed, consisting of a rounding of all intermediate design variables to $10^{-6}$ and 1, with 0.5 as the threshold value.
This hardly affects the layout and volume (thanks to the Heaviside projection utilized within the optimization process) and facilitates a more accurate elasto-plastic analysis.
The only visible difference is that a thin bar in the minimum volume design is partially deleted, hence it will not contribute to the transfer of forces in the comparative case.
An image of the layout obtained by \cite{le2010stress} was imported and processed to obtain a design as similar as possible to the original, while adapting the mesh resolution to that of the current study.
The same projection scheme was applied as described above but with a threshold of 0.26 for achieving roughly the same volume fraction as the other designs.  
The four post-processed layouts are presented in Figure \ref{fig:checkdesignslayouts}.

For convenience, the layouts are tagged \#1, \#2, \#3 and \#4, corresponding to maximum end-compliance s.t.~volume, maximum end-compliance s.t.~volume and plastic strains, minimum volume s.t.~end-compliance and plastic strains and the design by a $p$-norm approach \cite{le2010stress}.
The computed volume fractions are 34.82\%, 35.11\%, 34.21\% and 34.84\% for designs \#1, \#2, \#3 and \#4, respectively.
The solution parameters are the same as for the optimization runs, except for the prescribed displacement:
it is increased to $0.02$ in order to guarantee that all designs enter the plastic regime so that their actual elasto-plastic response can be compared.
The displacement is applied within 40 equal increments.
Furthermore, $p_E=1.0$ and $p_{\sigma_y}=0$ so that for the solid parts the true elasto-plastic law is obtained, whereas for void there is effectively no yield. 

\begin{figure}%
\centering
	\includegraphics[width=0.6\textwidth,trim={3in 0in 3in 0in}]{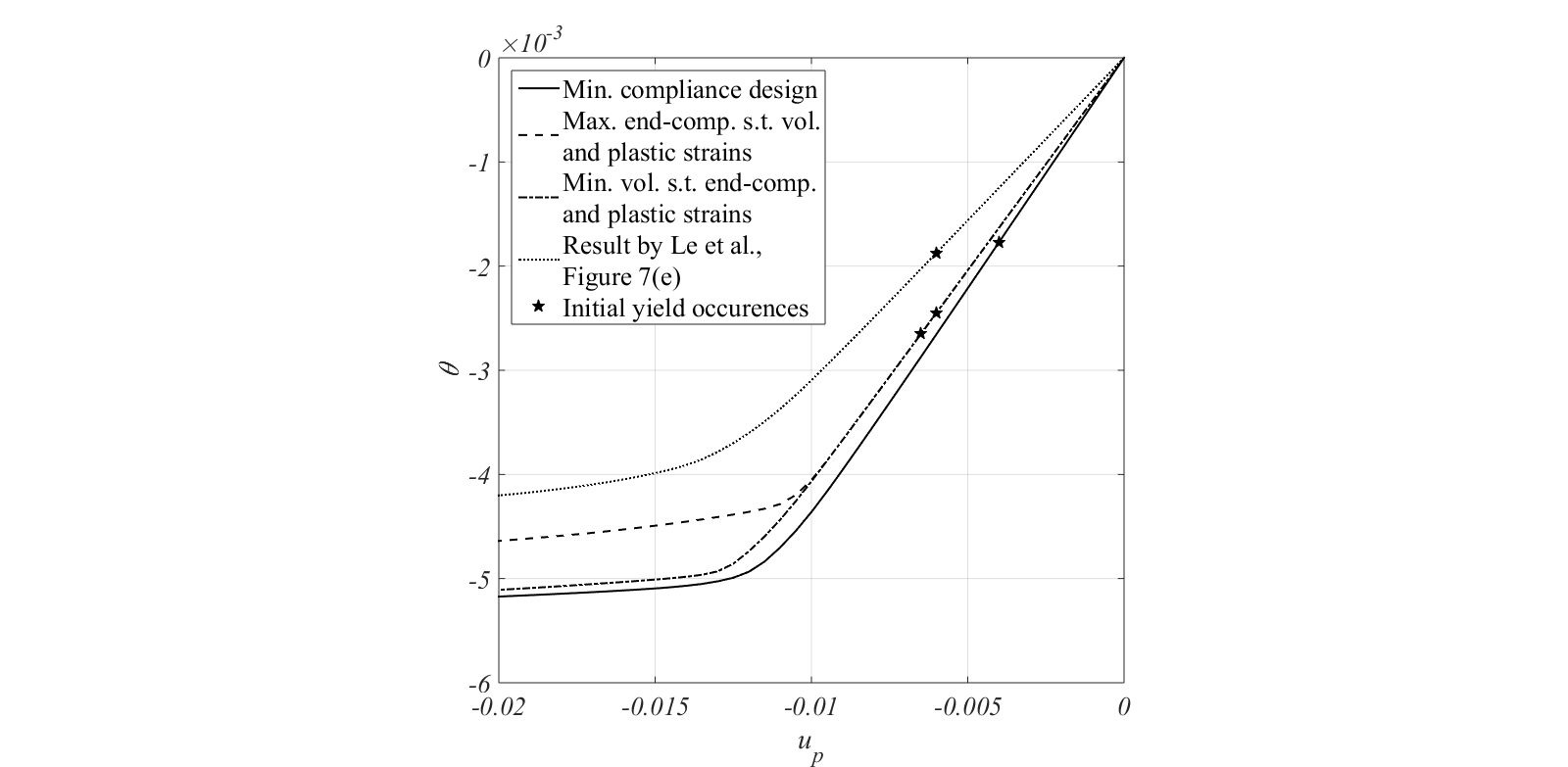}%
	\caption{Elasto-plastic response of the three optimized designs of an L-bracket loaded at the top of the right side. The designs optimized with a constraint on plastic strains exhibit an initial yield delayed by 42\% and 54\% in terms of forces, compared to the stiffest design without stress considerations. Performance is improved in comparison to the $p$-norm approach \cite{le2010stress}, in terms of stiffness and strength for a given volume.}%
	\label{fig:checkdesignsUF}%
\end{figure}

The load-displacement curves at the prescribed DOF are presented in Figure \ref{fig:checkdesignsUF}.
It can be seen that the responses of the designs obtained with a constraint on plastic strains exhibit a significant delay of the initial yield.
The magnitude of load at the initial yield is increased by approximately 42\% and 54\% compared to the reference design \#1, for designs \#2 and \#3, respectively.
This comes with a certain compromise on stiffness, meaning that the displacement level for a given load is slightly higher--this is expected because layout \#1 is ``the stiffest design''.
The elasto-plastic optimization also provides superior trade-offs of volume-stiffness-strength, compared to those obtained with a $p$-norm approach \cite{le2010stress}.
These results demonstrate the capability of the proposed approach to provide topological designs that account for stress constraints in an early design stage.
The topological layouts obtained provide a significant delay in the initial yield alongside a minor compromise on stiffness. 
This coincides with the common design goal of finding a stiff design that does not fail prematurely. 

\subsubsection{L-bracket with a load at the mid-point}
As mentioned above, many previous studies on stress-constrained topology optimization treated the L-bracket problem with a point load at the middle of the right side edge.
Therefore it is interesting to apply the proposed approach also for this setup.
A direct comparison of the final elasto-plastic response to a published result is sought also for this case.
For this purpose, a layout obtained using a level-set approach by \cite{amstutz2010topological} is chosen--in particular, the design in the center of Fig.~8 in the referenced article.
The volume fraction for the current optimization runs is modified accordingly, to 48.25\%.
All other parameters remain the same as for the L-bracket with a load at the top, except for the local distribution of the point load--5 nodes are sufficient in this case for avoiding a stress concentration.

\begin{figure*}%
	\centering
	
	\begin{subfigure}[c]{0.4\textwidth}
		\includegraphics[width=\textwidth,trim={0.5in 2in 1in 3in}]{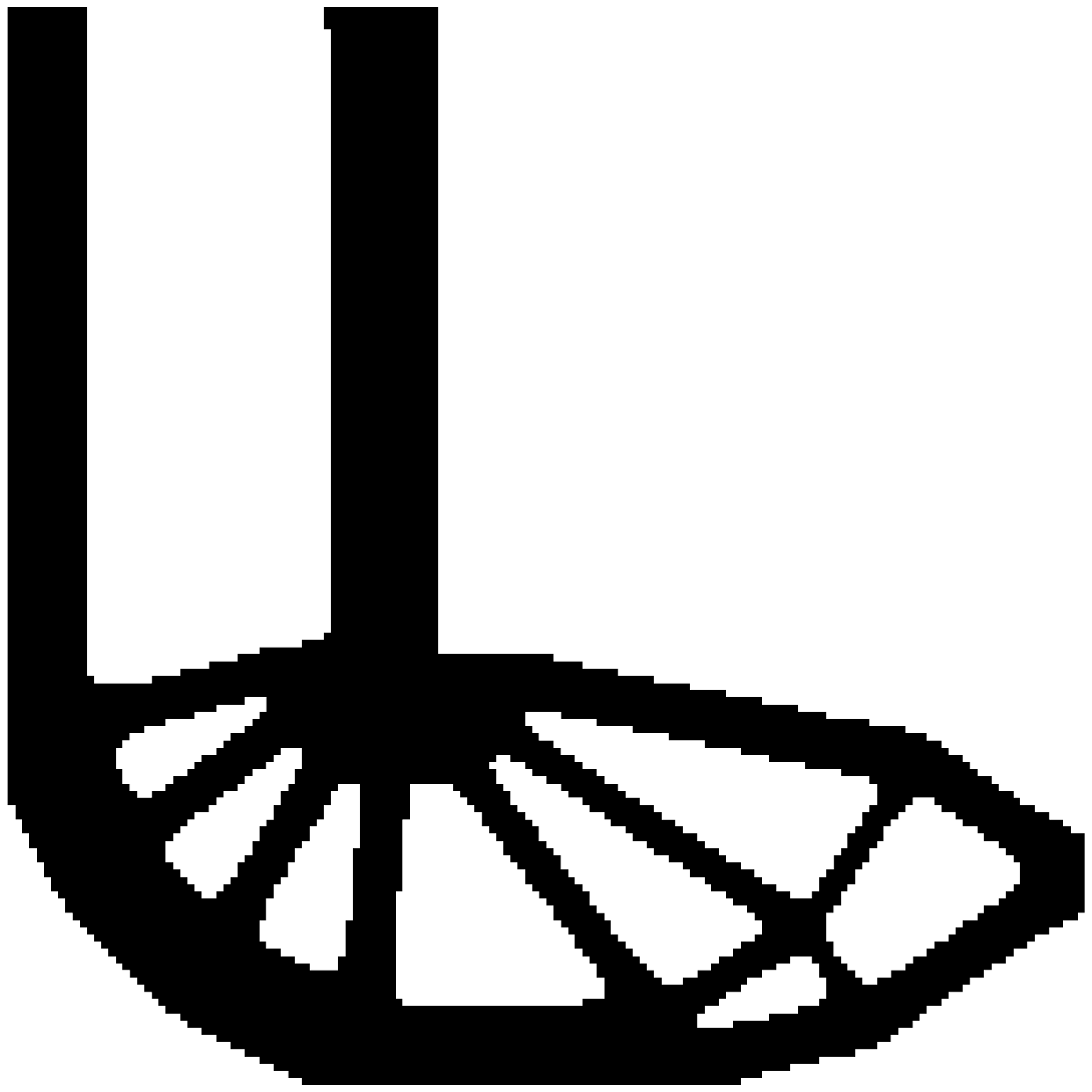}
		\caption{}
		\label{fig:maxcomp150midlayout}
	\end{subfigure}
	~
	\begin{subfigure}[c]{0.4\textwidth}
		\includegraphics[width=\textwidth,trim={0.5in 2in 1in 3in}]{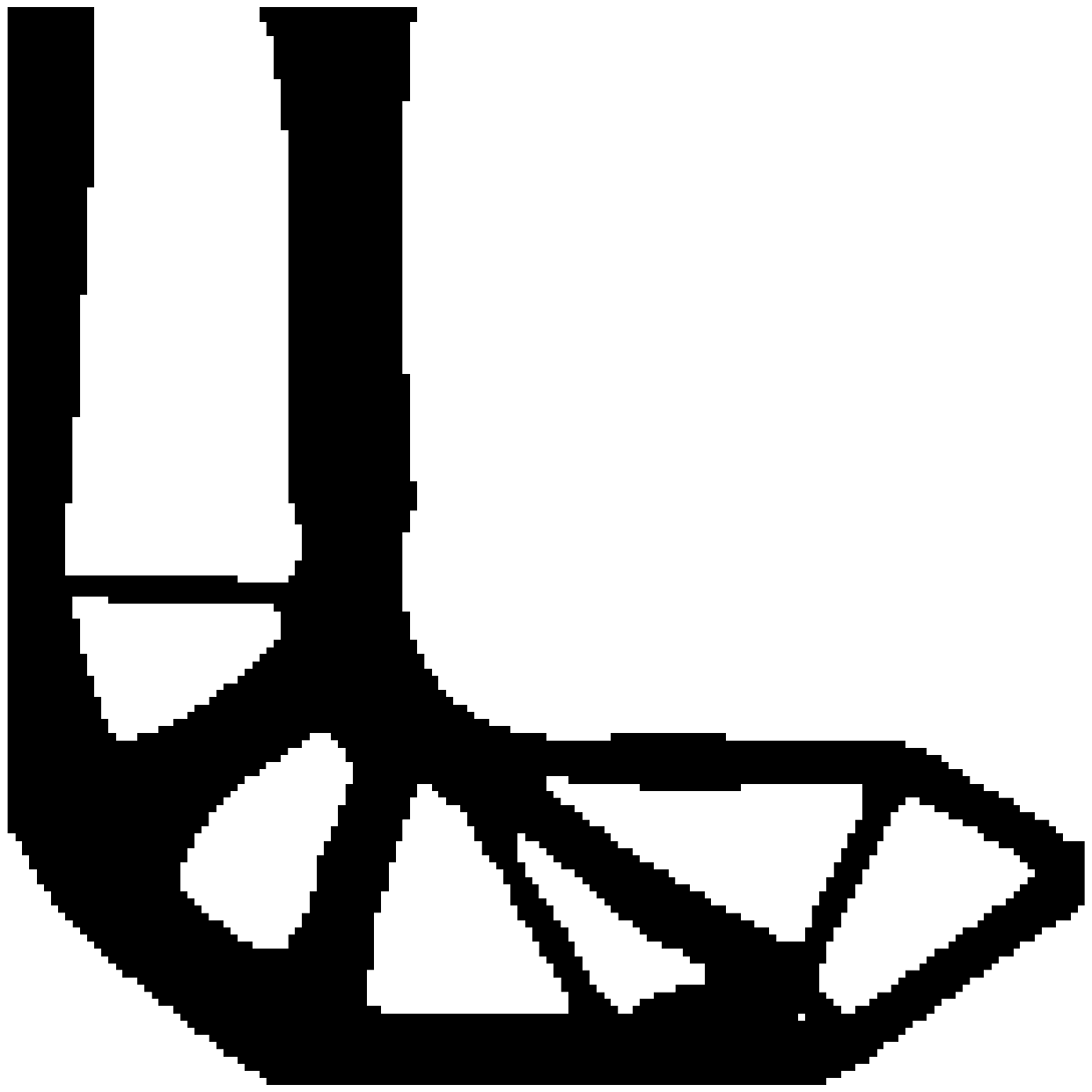}
		\caption{}
		\label{fig:plstr150midlayout}
	\end{subfigure}
	~
	\begin{subfigure}[c]{0.4\textwidth}
		\includegraphics[width=\textwidth,trim={0.5in 2in 1in 3in}]{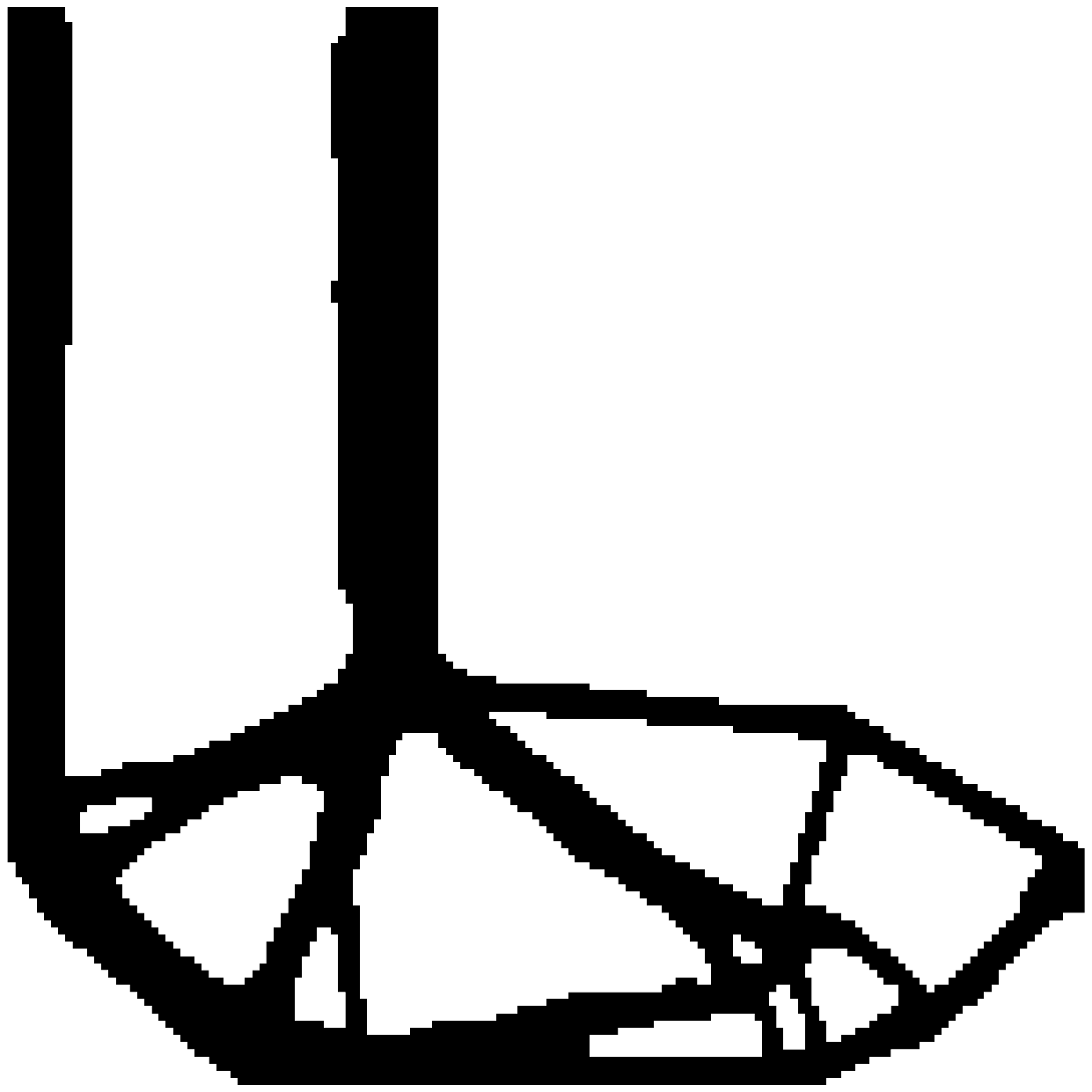}
		\caption{}
		\label{fig:minvol150midlayout}
	\end{subfigure}
	~
	\begin{subfigure}[c]{0.4\textwidth}
		\includegraphics[width=\textwidth,trim={0.5in 2in 1in 3in}]{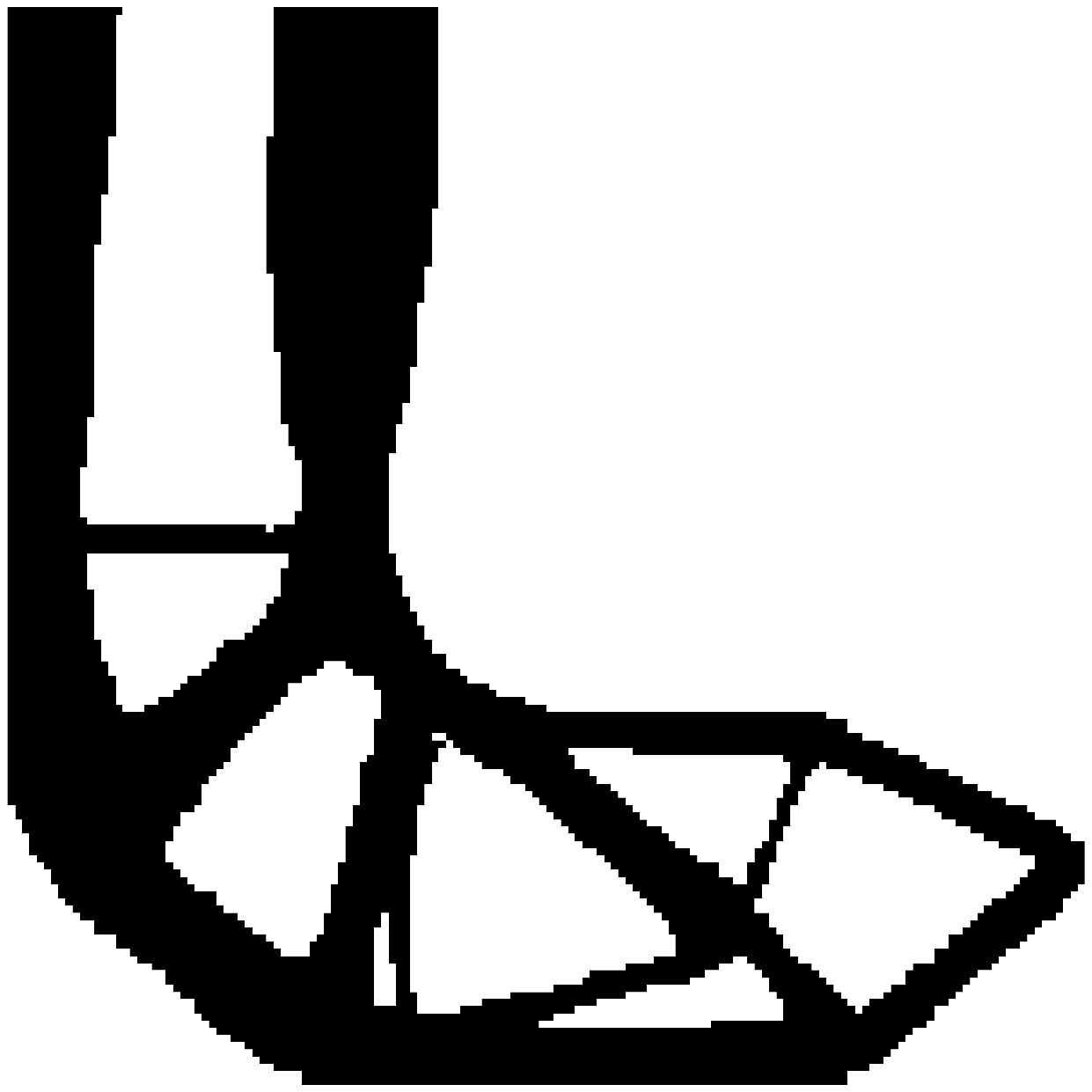}
		\caption{}
		\label{fig:Amst150midlayout}
	\end{subfigure}

\caption{Topology optimization of an L-bracket with a load and a prescribed displacement at the mid right. The presented layouts are obtained after a 0/1 projection as described in the text, for further examination of the elasto-plastic response.  (a) maximizing the end-compliance s.t.~a volume constraint; (b) maximizing the end-compliance s.t.~constraints on volume and on the total sum of equivalent plastic strains; (c) minimizing the volume s.t.~constraints on the end-compliance and on the total sum of equivalent plastic strains. (d) The layout based on an image imported from \cite{amstutz2010topological} for comparing the responses.}
	\label{fig:lbracketresultsmid}%

\end{figure*}

The results for the three formulations are presented in Table \ref{tab:lbracketresultsmid}.
The observed trends are the same as for the first case:
with the same volume fraction, nearly zero plastic strains can be achieved with a compromise of roughly 23\% on end-compliance.
The minimum volume procedure achieves a volume fraction far below 48.25\% but also exhibits slightly higher plastic strains.
The optimized layouts are presented in Figure \ref{fig:lbracketresultsmid}.
It can be seen again, that rather subtle shape and topological changes are introduced in order to avoid the stress concentrations.
This facilitates a relatively minor reduction in stiffness compared to the ``stiffest'' design. 
The generated layouts resemble those achieved by other approaches, e.g.~by \cite{allaire2008minimum}, by \cite{amstutz2010topological} and by \cite{james2014failure}.

\begin{table*}
	\footnotesize
	\centering
		\begin{tabular}{|c||c||c||c|}
		\hline
			\multirow{2}{*}{} & Max. end-comp.  & Max. end-comp. s.t. vol. & Min. vol. s.t. end-comp. \\
							                  & s.t. vol.       & and plastic strains      & and plastic strains \\
		\hline
			End-compliance \hfill $\theta_{N} \hat{{f}^p} {{u}_{N}^p}$                                          & $1.0979\times10^{-4}$ & $8.4328\times10^{-5}$ & $8.4309\times10^{-5}$  \\
			Plastic strains \hfill $\sum_{e=1}^{N_e} \sum_{k=1}^{N_{GP}} {\kappa^{ek}_N}$                       & $1.1400\times10^{0}$ & $5.9734\times10^{-4}$ & $2.1170\times10^{-3}$  \\
			Volume \hfill $\frac{\sum_{e=1}^{N_e} v_e \overline{\mathbf{x}}_e}{0.4825\times{N_e}\times{v_e}} - 1$ & $-5.1457\times10^{-7}$ & $-3.9286\times10^{-4}$ & $-2.0116\times10^{-1}$ \\
			Figures \hfill { } & \ref{fig:maxcomp150midlayout} & \ref{fig:plstr150midlayout} & \ref{fig:minvol150midlayout} \\ 
		\hline
		\end{tabular}
	\caption{Results of the topology optimization of an L-bracket with a load at the mid point of the right side. For the same volume and under the same prescribed displacement, constraining the sum of equivalent plastic strains leads to nearly zero plastic strains while compromising the end-compliance by 23\%. The minimum volume procedure reaches a volume fraction of 39.45\% but exhibits higher plastic strains.}
	\label{tab:lbracketresultsmid}
\end{table*}
								
\begin{figure}%
\centering
	\includegraphics[width=0.6\textwidth,trim={3in 0in 3in 0in}]{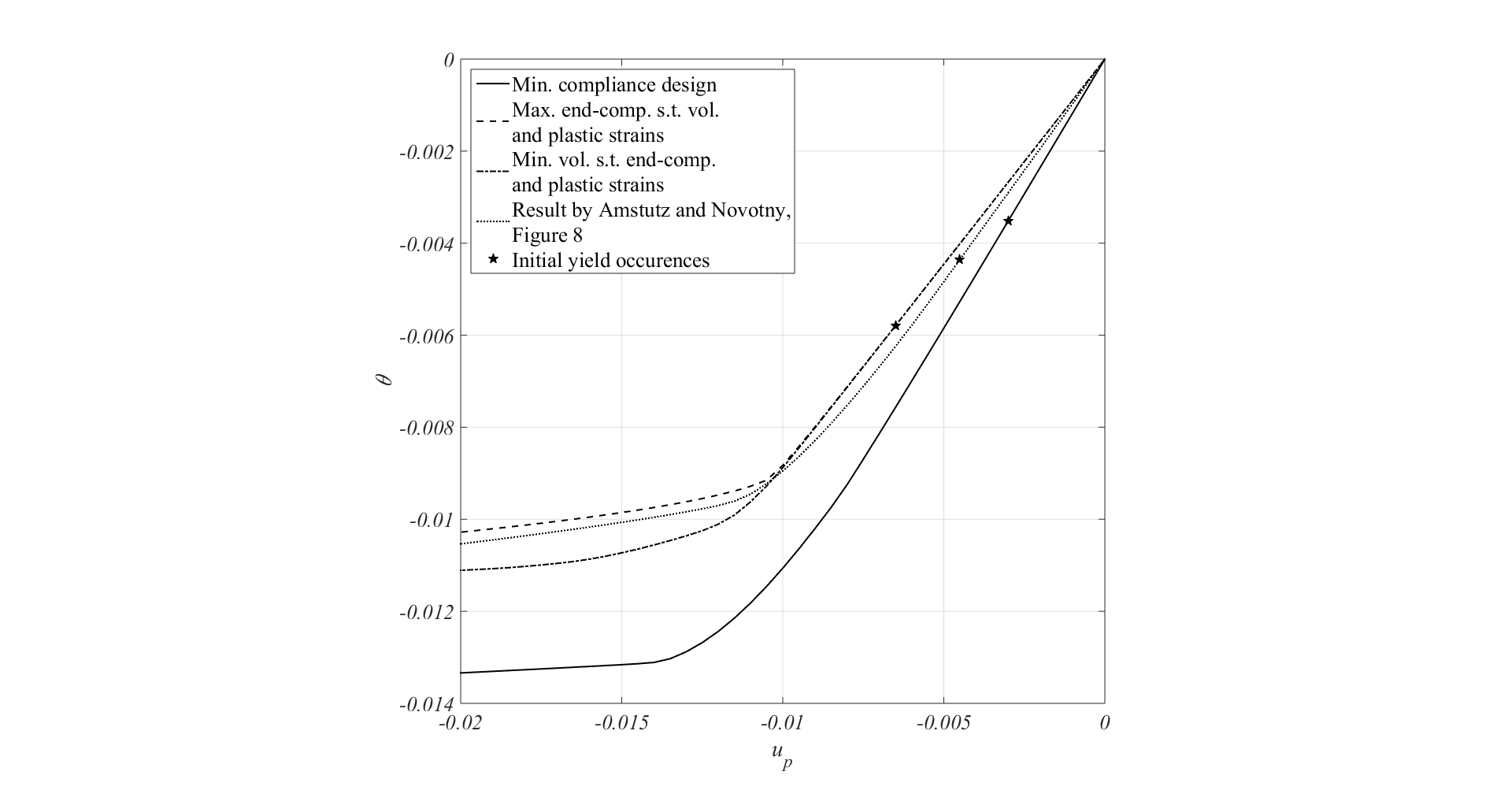}%
	\caption{Elasto-plastic response of the three optimized designs of an L-bracket loaded at the middle of the right side. The designs optimized with a constraint on plastic strains exhibit an initial yield delayed by 72\% in terms of forces, compared to the stiffest designs without stress considerations. The attained stiffness is slightly lower than in \cite{amstutz2010topological}, but the initial yield is postponed.}%
	\label{fig:checkdesignsUFmid}%
\end{figure}

For examining the elasto-plastic response, the post-processing described above is repeated.
The computed volume fractions are 48.05\%, 48.35\%, 38.54\% and 48.74\% for designs \#1, \#2,\#3 and \#4, respectively.
Design \#4 corresponds to the interpretation from \cite{amstutz2010topological}, Fig.~8.
The load-displacement curves at the prescribed DOF for a displacement incrementation up to 0.02 are presented in Figure \ref{fig:checkdesignsUFmid}.
Again, the responses of the designs obtained with a constraint on plastic strains exhibit a significant delay of the initial yield.
Quite remarkably, the magnitude of load at the initial yield is increased by approximately 72\% compared to the reference design \#1, for designs \#2 and \#3.
The yielding of designs \#2 and \#3 is postponed also in comparison with the layout taken from \cite{amstutz2010topological}.
This highlights the capability of the proposed approach to generate designs with high quality trade-offs of volume-stiffness-strength.

\subsection{Example 2: Stress-constrained U-bracket design}
\begin{figure*}
	\centering
	\includegraphics[width=0.8\textwidth]{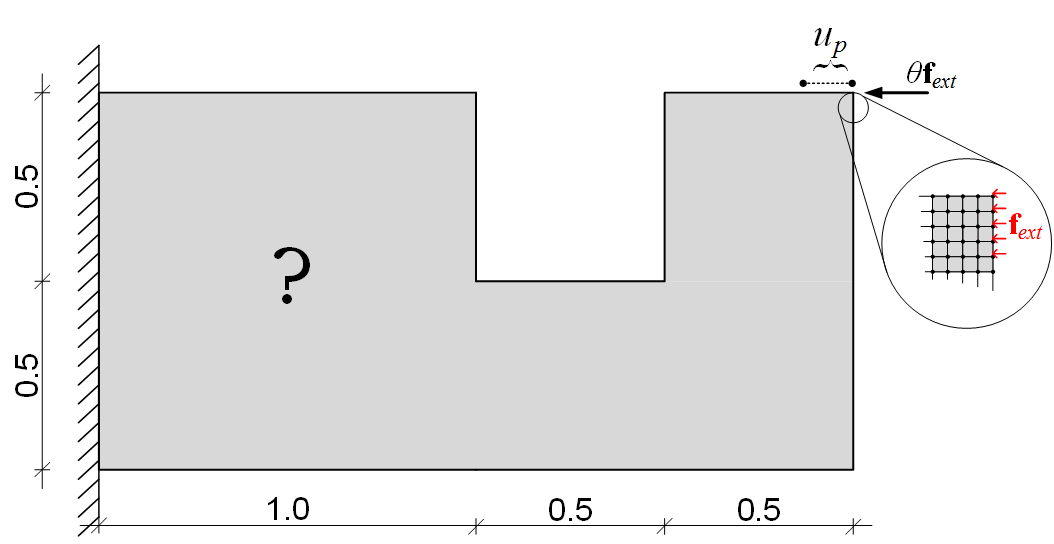}
	\caption{Problem setup for topology optimization of a U-bracket. The load is distributed over the top 10 nodes in order to avoid artificial stress concentrations at the loading point.}
	\label{fig:ubracketsetup}
\end{figure*}

The second example demonstrates the topological design of a U-bracket with a horizontal load, see Figure \ref{fig:ubracketsetup} for the problem setup.
Here, two regions of stress concentrations are expected because the load path should pass via both re-entrant corners.
The model is discretized with a 200$\times$100 mesh resolution consisting of 17,500 square, bi-linear elements.
The available volume of material is set to 40\% of the total volume and the filter radius is 0.03.
The prescribed displacement is set to $u_p = 0.01$ and automatic displacement incrementation is applied.
The material parameters are the same as for the previous examples. 
The continuation scheme is slightly modified in order to examine the capability of beginning the optimization with some penalization, namely $p_E=3.0$ and $p_{\sigma_y}=2.5$.
This is in contrast to initial values of $p_E=1.0$ and $p_{\sigma_y}=0.5$ that were used in the previous example.
Again, for the minimum volume case the penalty exponents $p_E=3.0$ and $p_{\sigma_y}=2.5$ are constant for the first 200 design cycles in order to enable feasibility of the compliance constraint.
Otherwise, the continuation scheme and MMA parameters are identical to the previous example.

\begin{figure*}%
	\centering
	
	\begin{subfigure}[c]{0.3\textwidth}
		\includegraphics[width=\textwidth,trim={0.5in 2in 1in 3in}]{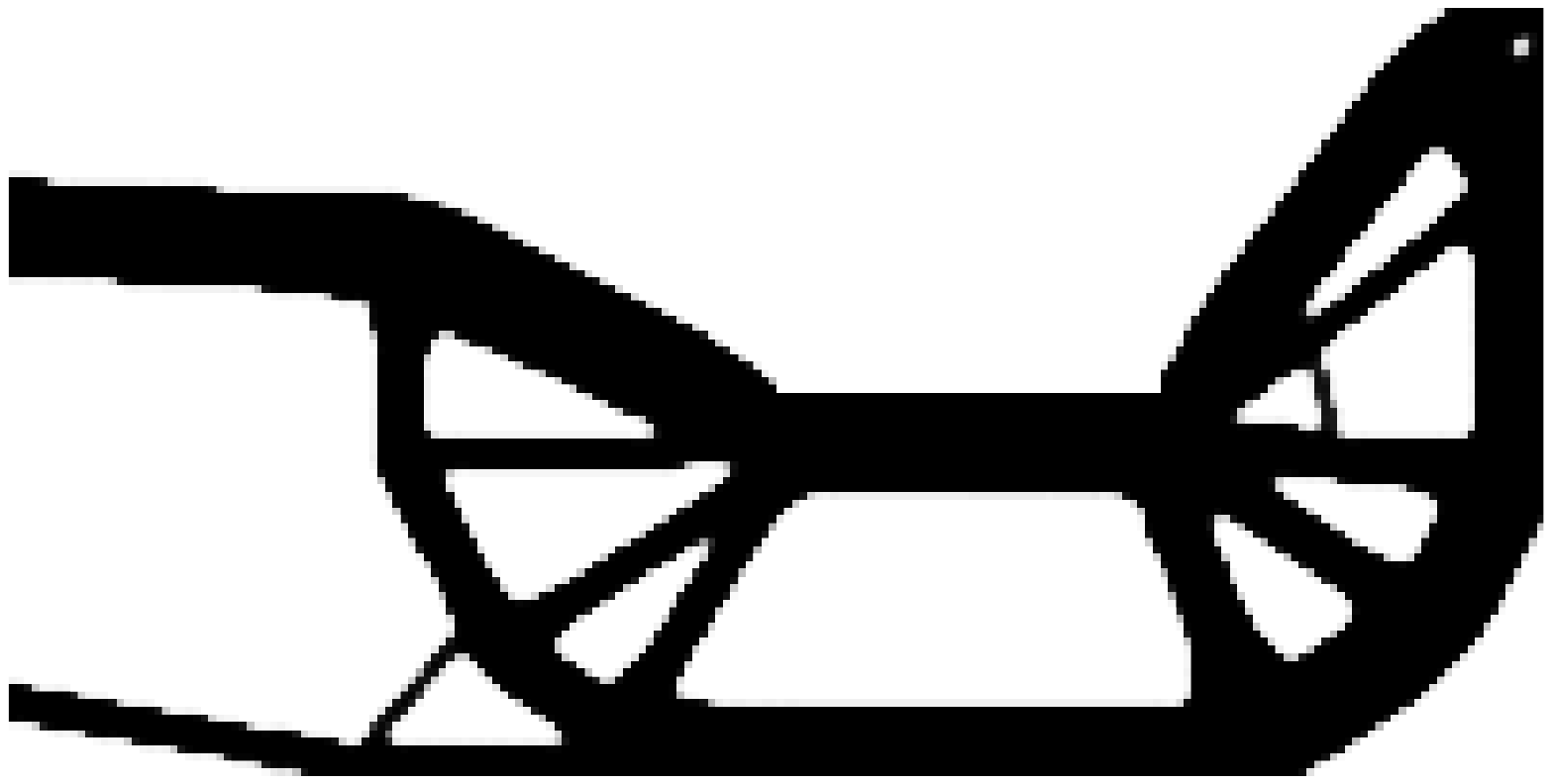}
		\caption{}
		\label{fig:maxcomp200layout}
	\end{subfigure}
	~
	\begin{subfigure}[c]{0.3\textwidth}
		\includegraphics[width=\textwidth,trim={0.5in 2in 1in 3in}]{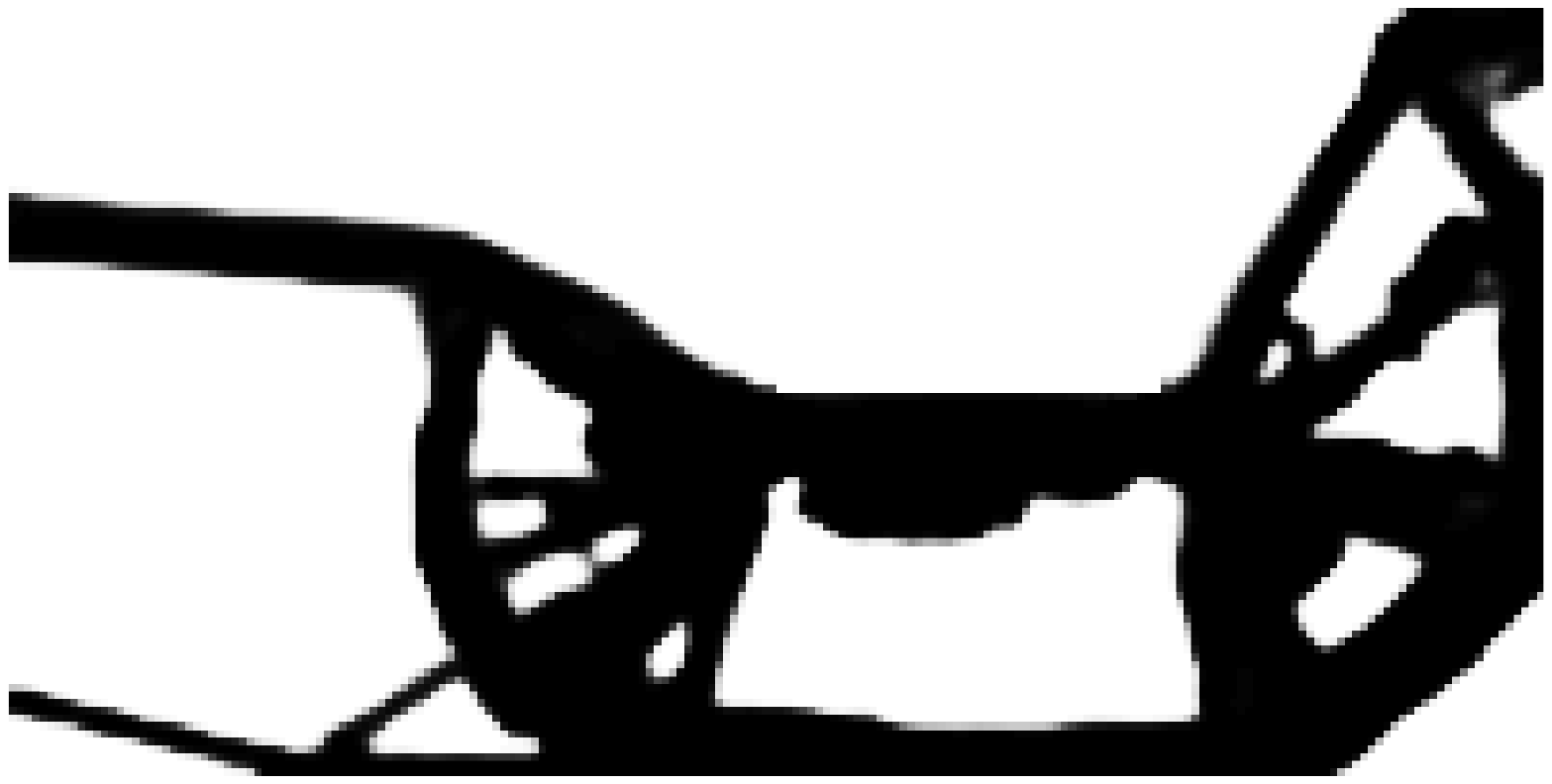}
		\caption{}
		\label{fig:plstr200layout}
	\end{subfigure}
	~
	\begin{subfigure}[c]{0.3\textwidth}
		\includegraphics[width=\textwidth,trim={0.5in 2in 1in 3in}]{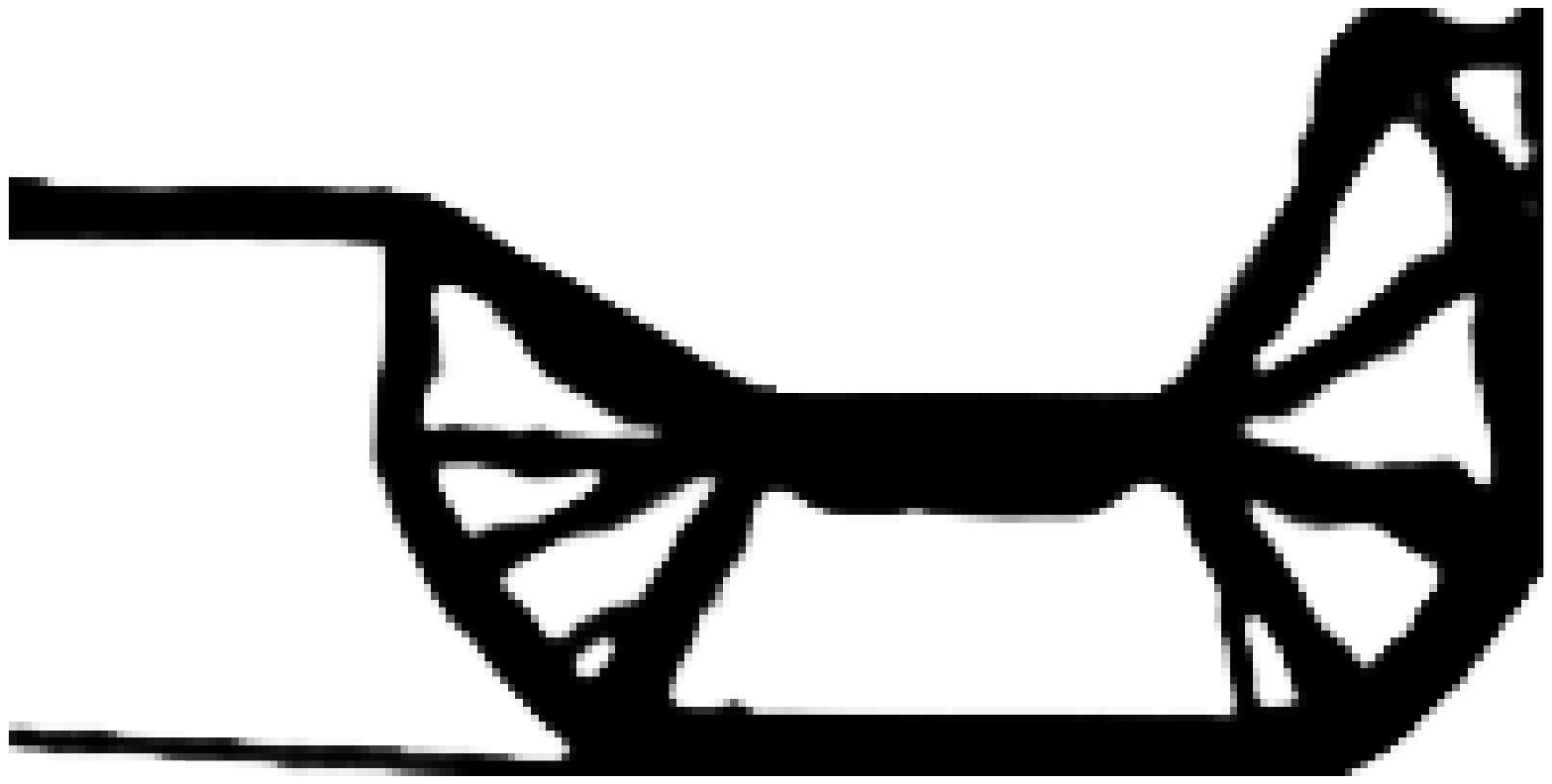}
		\caption{}
		\label{fig:minvol200layout}
	\end{subfigure}
	
	\begin{subfigure}[c]{0.32\textwidth}
		\includegraphics[width=\textwidth]{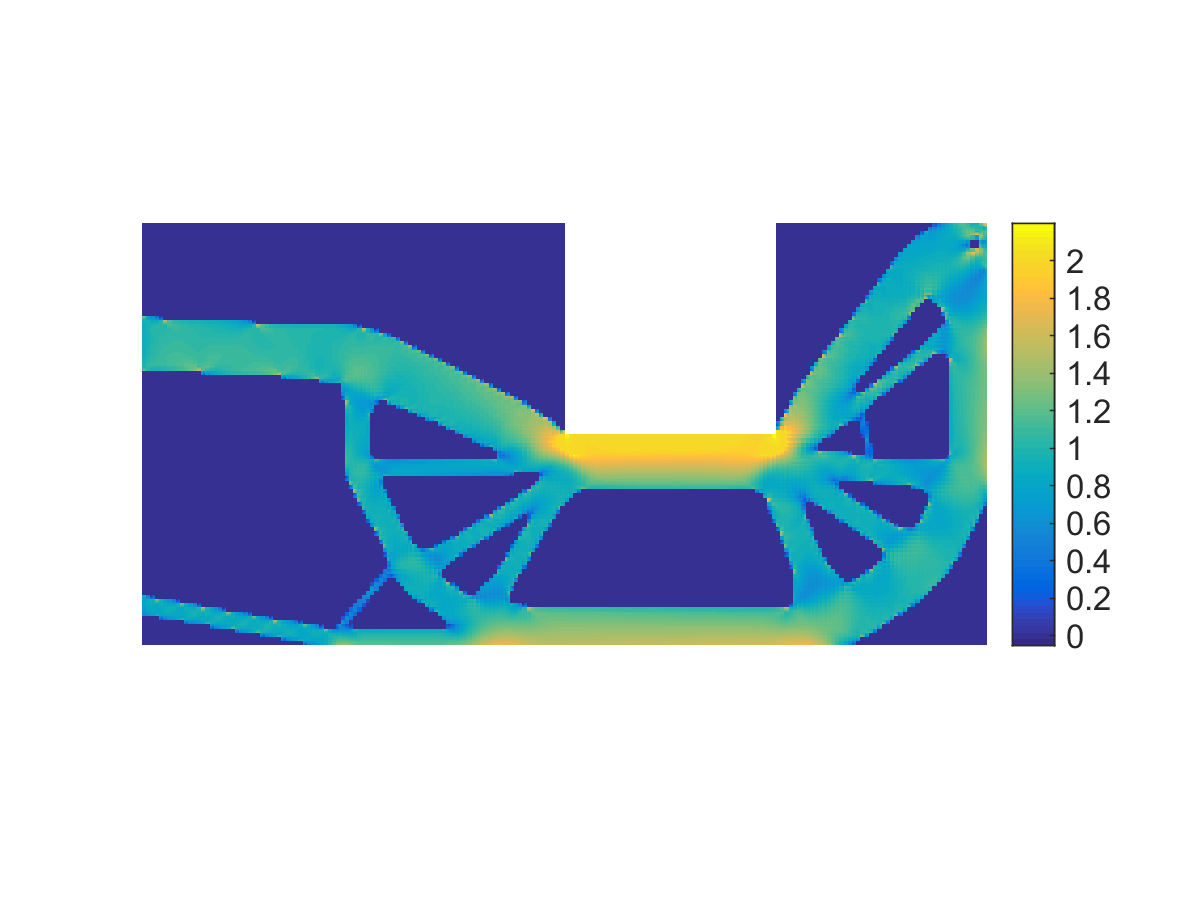}
		\caption{}
		\label{fig:maxcomp200vmstress}
	\end{subfigure}
	~
	\begin{subfigure}[c]{0.32\textwidth}
		\includegraphics[width=\textwidth]{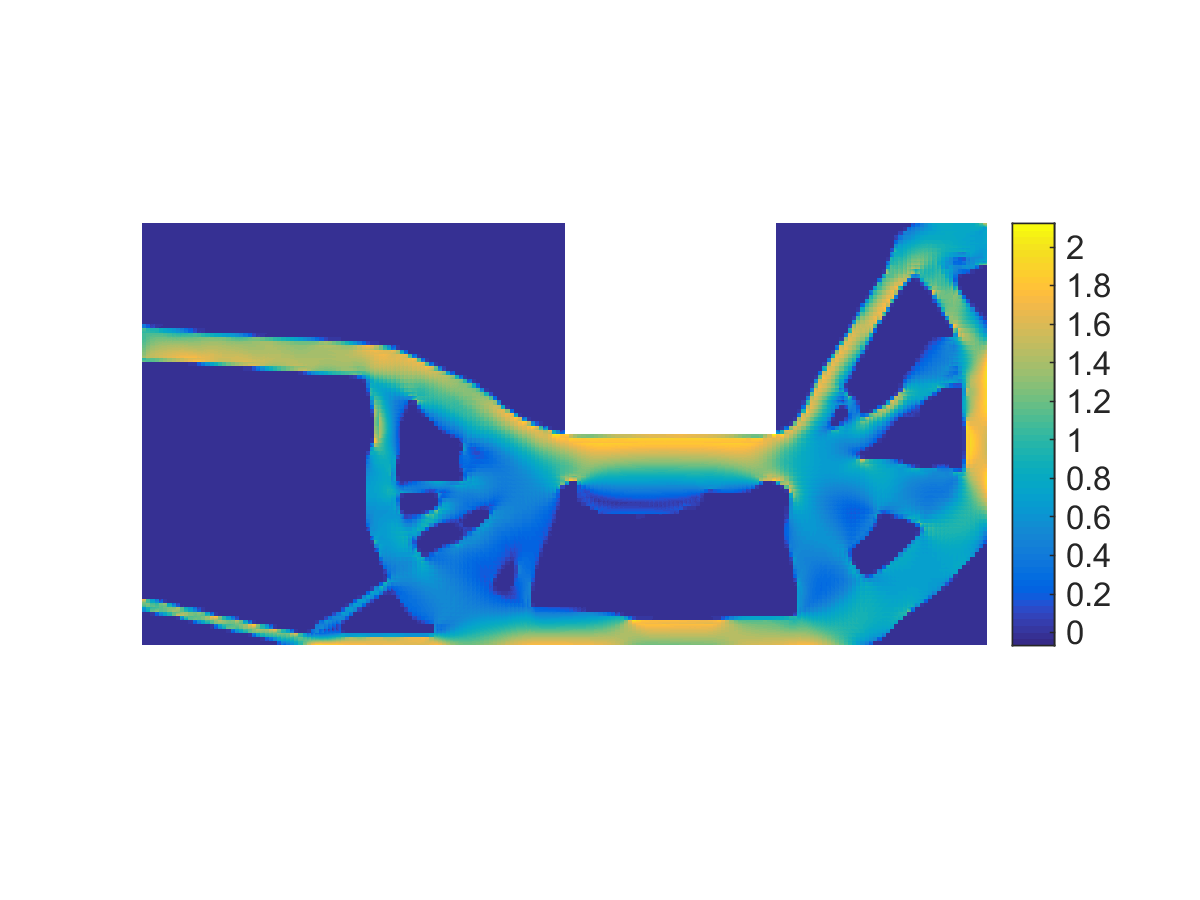}
		\caption{}
		\label{fig:plstr200vmstress}
	\end{subfigure}
	~
	\begin{subfigure}[c]{0.32\textwidth}
		\includegraphics[width=\textwidth]{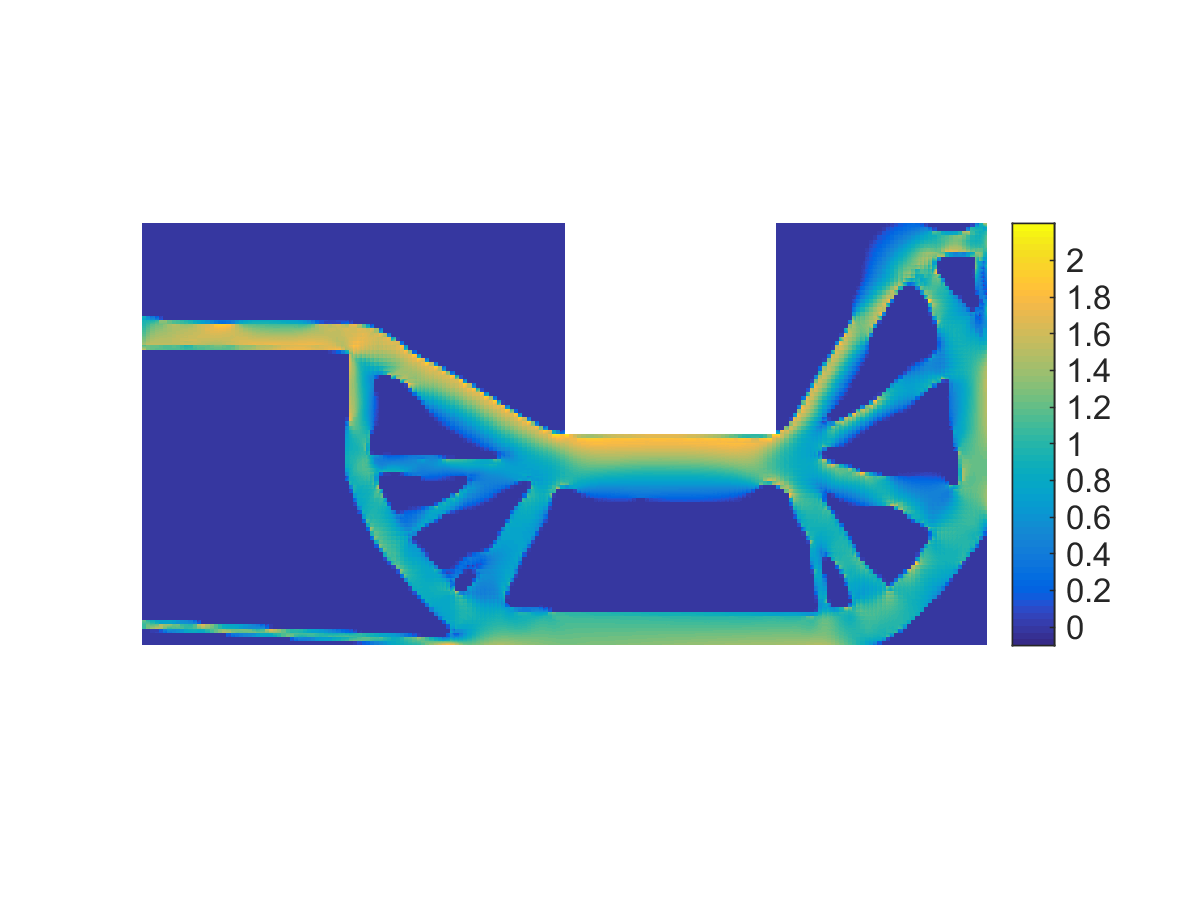}
		\caption{}
		\label{fig:minvol200vmstress}
	\end{subfigure}
	
	\begin{subfigure}[c]{0.32\textwidth}
		\includegraphics[width=\textwidth]{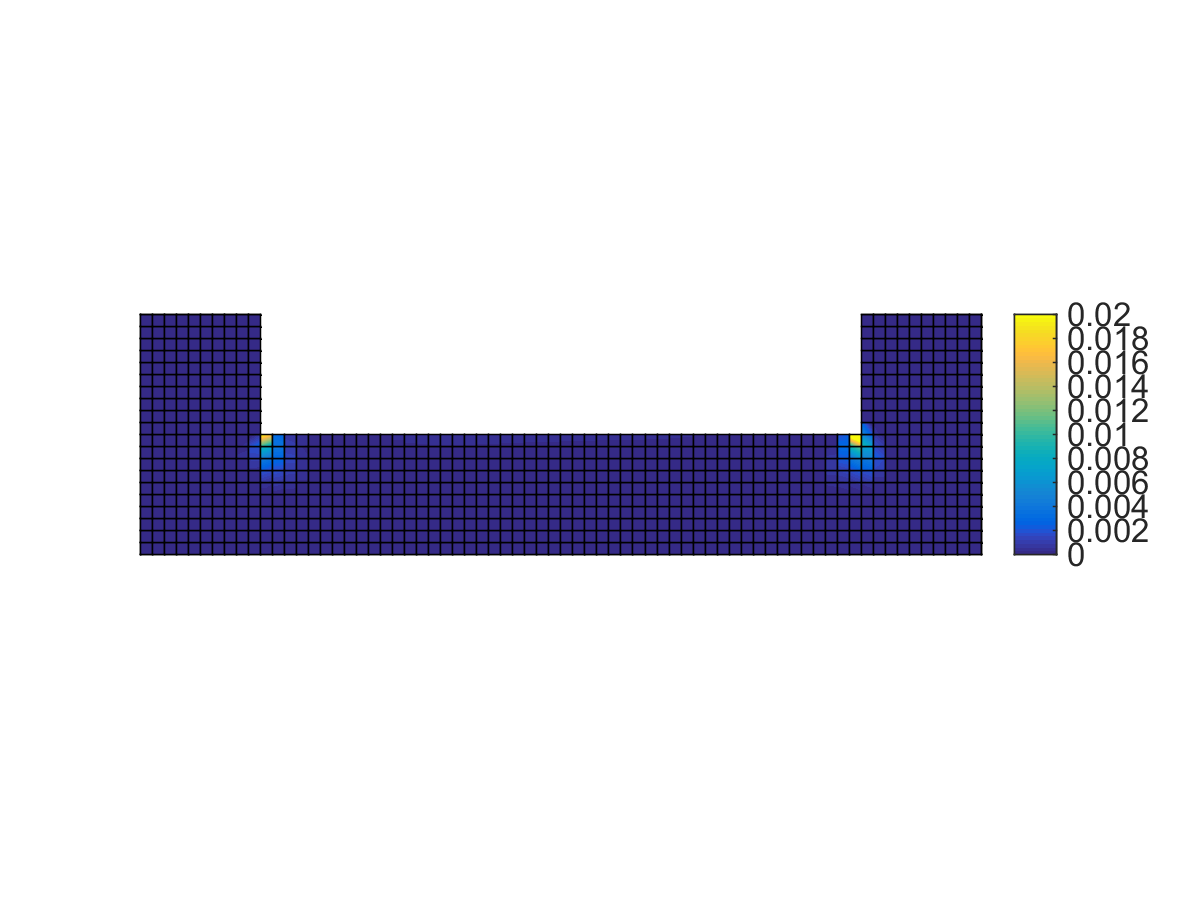}
		\caption{}
		\label{fig:maxcomp200equivps}
	\end{subfigure}
	~
	\begin{subfigure}[c]{0.32\textwidth}
		\includegraphics[width=\textwidth]{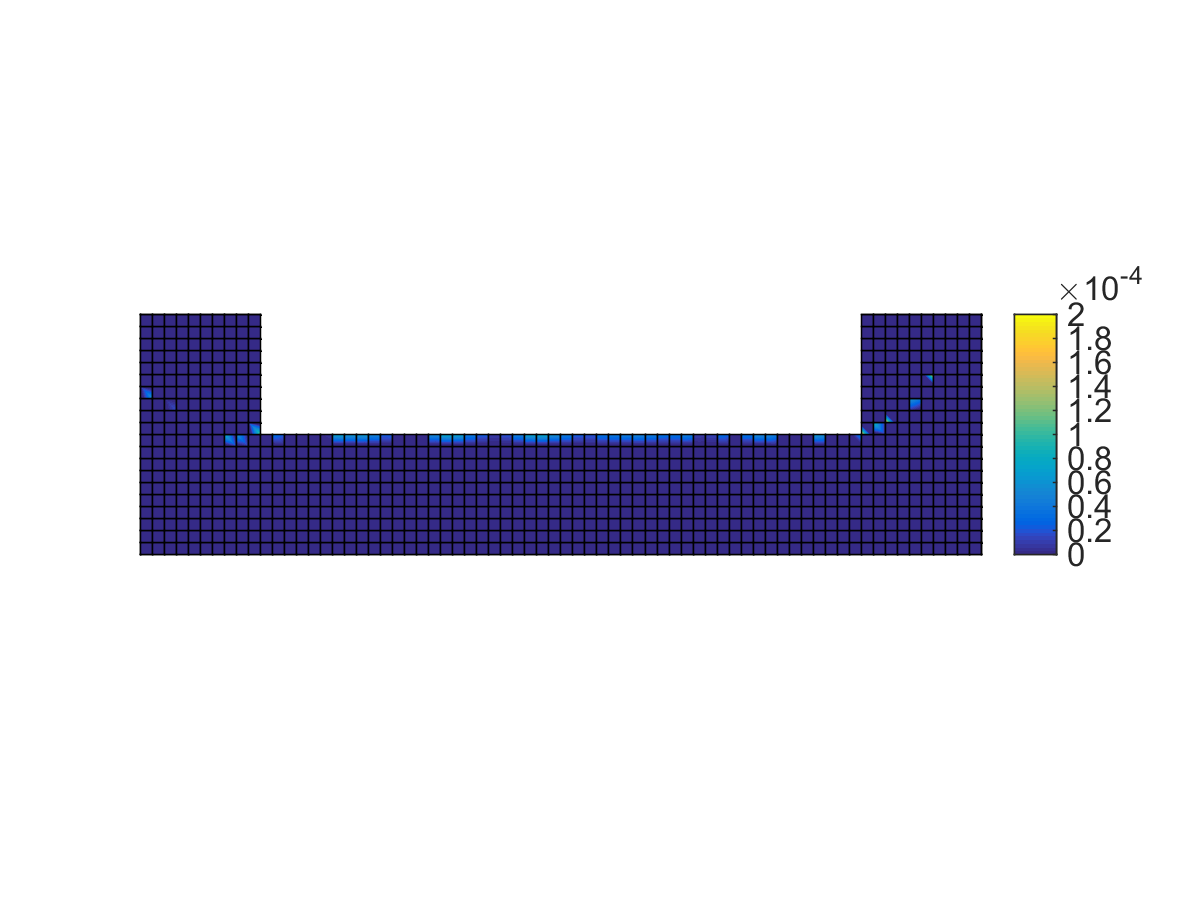}
		\caption{}
		\label{fig:plstr200equivps}
	\end{subfigure}
	~
	\begin{subfigure}[c]{0.32\textwidth}
		\includegraphics[width=\textwidth]{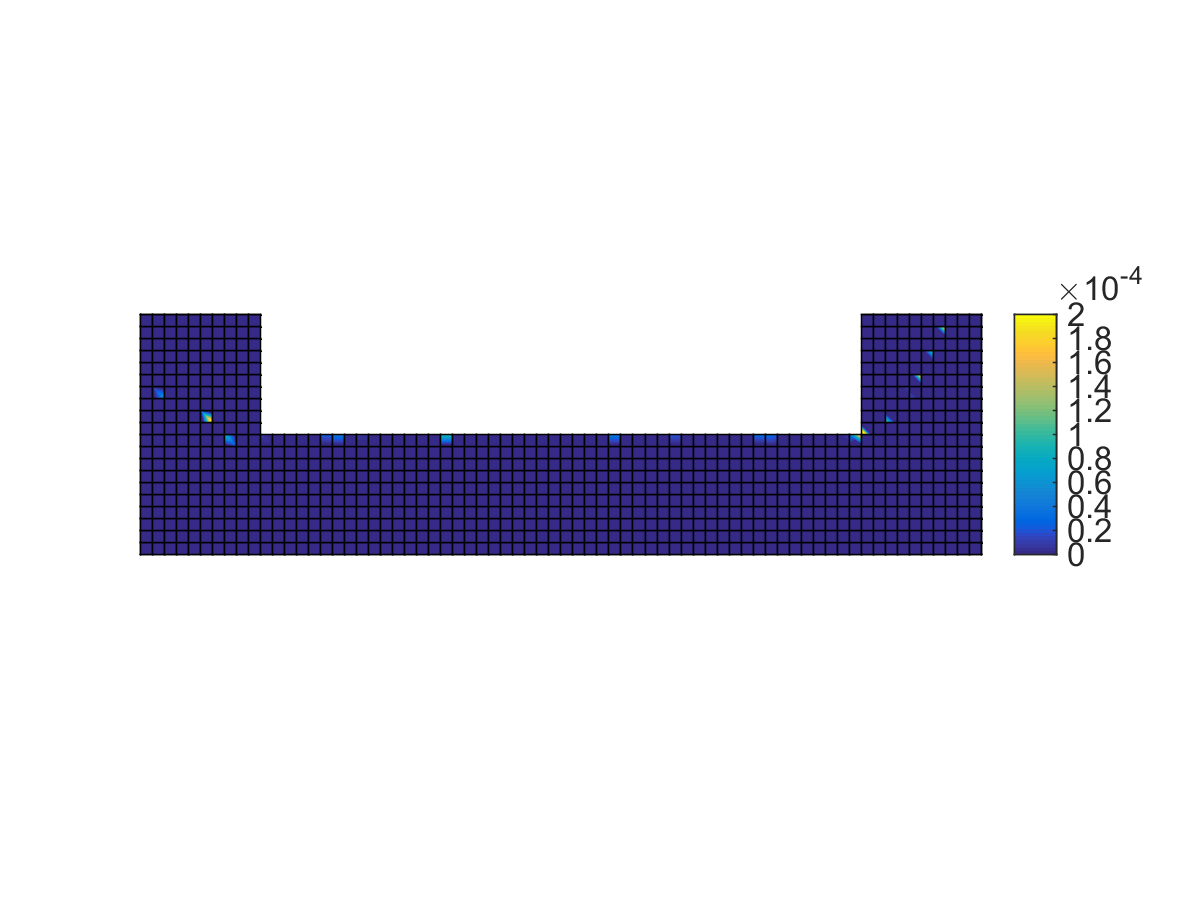}
		\caption{}
		\label{fig:minvol200equivps}
		
	\end{subfigure}
	
	\caption{Topology optimization of a U-bracket with a load and a prescribed displacement at the top right corner. From the left: maximizing the end-compliance s.t. a volume constraint; maximizing the end-compliance s.t. constraints on volume and on the total sum of equivalent plastic strains; minimizing volume s.t. constraints on end-compliance and on the total sum of equivalent plastic strains. From top: optimized layouts; von Mises stress distributions; equivalent plastic strains in the vicinity of the re-entrant corners. Note the two orders of magnitude difference between the scale of sub-figure (g) compared to (h) and (i).}%
	\label{fig:ubracketresults}%
\end{figure*}

As before, we examine three optimization problems: maximizing the end-compliance s.t.~a volume constraint; maximizing the end-compliance s.t.~constraints on volume and on the total sum of equivalent plastic strains; and minimizing volume s.t.~constraints on end-compliance and on the total sum of equivalent plastic strains.
The optimized topologies, the von Mises stress distributions and the distributions of equivalent plastic strains are presented in Figure \ref{fig:ubracketresults}.
The end-compliances, the sums of equivalent plastic strains and the volumes are presented in Table \ref{tab:ubracketresults}.

It can be clearly seen that the suggested approach generates designs that circumvent stress concentrations in the vicinity of the re-entrant corners.
The two latter solutions provide different trade-offs of stiffness to weight while keeping plastic strains at a very low level.
It is interesting to see that the maximization of end-compliance subject to a constraint on plastic strains suggests an alternative load path to that of the reference design.
Considerable forces are transferred via a vertical bar in the right side edge, enabling the reduction of stresses near the re-entrant corners. 
This force transfer appears also in the minimum volume design but to a lesser extent. 
In both designs, the compliance is compromised by roughly 22\% in comparison to the reference solution which exhibits two significant stress concentrations. 
The minimum volume procedure appears to deliver slightly better results as it provides the same compliance, but uses only 36.4\% of the design domain.
Finally, a comparison of the initial yield level of the optimized designs reveals an increase of approximately 89\% in the applied force prior to yielding, compared to the optimized design achieved without stress considerations.
This demonstrates the capability of the proposed approach to deal with multiple stress concentrations without any added complexity.

\begin{table*}
	\footnotesize
	\centering
		\begin{tabular}{|c||c||c||c|}
		\hline
			\multirow{2}{*}{} & Max. end-comp.  & Max. end-comp. s.t. vol. & Min. vol. s.t. end-comp. \\
							                  & s.t. vol.       & and plastic strains      & and plastic strains \\
		\hline
			End-compliance \hfill $\theta_{N} \hat{{f}^p} {{u}_{N}^p}$                                          & $8.6786\cdot10^{-5}$ & $6.7559\cdot10^{-5}$ & $6.7756\cdot10^{-5} $  \\
			Plastic strains \hfill $\sum_{e=1}^{N_e} \sum_{k=1}^{N_{GP}} {\kappa^{ek}_N}$                       & $5.0729\cdot10^{-1} $ & $8.9298\cdot10^{-4}$ & $2.9604\cdot10^{-3} $  \\
			Volume \hfill $\frac{\sum_{e=1}^{N_e} v_e \overline{\mathbf{x}}_e}{0.40\cdot{N_e}\cdot{v_e}} - 1$ & $-4.4559\cdot10^{-7} $ & $-2.0678\cdot10^{-3}$ & $-9.0041\cdot10^{-2}$ \\
			Figures \hfill { } & \ref{fig:maxcomp200layout}, \ref{fig:maxcomp200vmstress}, \ref{fig:maxcomp200equivps} & \ref{fig:plstr200layout}, \ref{fig:plstr200vmstress}, \ref{fig:plstr200equivps} & \ref{fig:minvol200layout}, \ref{fig:minvol200vmstress}, \ref{fig:minvol200equivps} \\ 
		\hline
		\end{tabular}
	\caption{Results of the topology optimization of an U-bracket with a load at the top right corner. For the same volume and under the same prescribed displacement, constraining the sum of equivalent plastic strains leads to nearly zero plastic strains while compromising the end-compliance by 22\%.}
	\label{tab:ubracketresults}
\end{table*}

\section{Conclusion} \label{sec:conclusion}

A new approach for achieving stress-constrained topological designs in continua was presented.
The main novelty is in the use of material nonlinearity, in the form of classical elasto-plasticity models, in order to avoid imposing a very large number of local stress constraints.
Stress constraints are implicitly satisfied by imposing a single global constraint on the spatial sum of the equivalent plastic strains.
Incorporating this constraint into formulations for maximizing the end-compliance (for a given prescribed displacement) subject to a volume constraint, or vice-versa, leads to optimized designs with practically no stress violations.
A critical comparison of the proposed approach to existing techniques, in terms of accuracy and efficiency, is hereby presented.

\subsection{Designed trade-off}
In view of real-world applications, stress-constrained topology optimization should aim at finding the best trade-off between three competing quantities: volume, compliance and stress.
The results presented in this article demonstrate the capability of the plasticity-based formulation to attain high-quality layouts in terms of this trade-off. 
Two types of optimized L-brackets were favorably compared to results obtained with constraint aggregation, using either $p$-norms or an external penalty.   
This highlights one advantage of the current formulation--stress constraints are captured accurately without actually imposing a large number of constraints on local stress values.
There is an evident computational price tag for the improvement in design quantities, as will be discussed in the following.
Nevertheless, the results achieved in this study provide another, unexplored view on stress-constrained topology optimization.
This can motivate and fertilize further development of more efficient, approximate techniques. 

\subsection{Computational cost}
In its present form, the proposed formulation is not as efficient as linear-elastic approaches. 
The results reported in this paper were achieved with a constant number of 500 design iterations for enabling continuation on penalty parameters and on the sharpness of the smoothed Heaviside.
Each design iteration requires a full nonlinear finite element analysis, which typically uses 20-40 solutions of linear equation systems, depending on the automatic incrementation and on the convergence of Newton-Raphson iterations.
The compliance and plastic strains functionals require an adjoint procedure in each design iteration, which typically uses 5-10 solutions of linear equation systems, depending on the automatic incrementation.
Stress-constrained procedures based on linear elasticity and a certain constraint aggregation technique will require only a few solutions of linear equation systems per design iteration, depending on the specific aggregation scheme.  
Therefore, the current implementation of the plasticity-based approach is expected to be slower than existing approaches.
This includes $p$-norm techniques (e.g. \cite{le2010stress}), external penalty (e.g. \cite{amstutz2010topological}) and to some extent also formulations that introduce local constraints (e.g. \cite{bruggi2008mixed}).
Nevertheless, the relatively good design trade-offs encourage further exploration of this approach, focusing on reducing the computational burden.
It may be possible to utilize other material models that may not be suitable for capturing the full elasto-plastic response accurately, but may suffice for the purpose of achieving a no-yield design.
Another path to be explored is the reduction of displacement increments and Newton-Raphson iterations, by exploiting the fact that in most design cycles the response is either linear-elastic or very close to linear.
Then the computational cost of a single design cycle can be reduced to a level similar to that of a linear-elastic analysis.
Another option is to utilize reanalysis techniques, which in fact motivated the investigation of this formulation in the first place \cite{AmirPhDThesis}.
These possibilities will be investigated thoroughly in future work.

\subsection{Further considerations}
In many publications on stress constraints, the local constraints are aggregated into a single, or a few, global approximations of the maximum stress.
This often requires specific tuning of parameters and does not ensure that the actual local stresses will not exceed allowable values. 
This highlights another advantage of the proposed approach, which only requires a well-known elasto-plastic material model that is already incorporated in standard FEA packages.
In the examples it can be observed that the optimized designs involve both topological as well as shape changes.
This means that simply optimizing the shape of an optimized topology generated without stress considerations, may not suffice for achieving the best possible design. 
Furthermore, incorporating stress considerations in the topological design phase can eliminate the need to post-process the optimized topology, create a CAD model, generate a new mesh and then optimize the shape--a process that can be very time consuming in the industrial context, particularly in 3-D.

A potential disadvantage of the proposed formulation lies in the non-differentiability of the plastic strains.
According to the adopted formulation of the elasto-plastic model, the yield stress limit represents a non-differentiable point.
This does not affect the end-compliance functional because the product of forces and displacements (or stresses and strains) retains smoothness when passing the yield point. 
Plastic strains however are strictly zero up to this point and instantaneously increase when passing this point.
As can be seen in the numerical experiments above, the non-differentiability did not appear to hamper the convergence towards a design which does not violate the stress limit. 
This is despite the fact that in the optimized designs, several material points are indeed very close to their yield limit.
A differentiable approximation was introduced in several numerical experiments in order to examine its affect on the quality of the attained solutions.
The results did not show any improvement in comparison to the original implementation, but this important issue will be thoroughly examined in a continuation of the research.


\section{Acknowledgements}
This research was supported by the Israel Science Foundation (grant No. 750/15).  

\section{Appendix}

In this appendix we present a numerical verification of the adjoint sensitivity analysis procedure.
Implementing this procedure can be a somewhat cumbersome task, so we believe this verification can prove useful for readers who are not well-acquainted with such procedures.
Furthermore, accurate and efficient sensitivity analysis for elasto-plastic response is still a rather open issue, as discussed in a recent publication \cite{kato2015analytical}.
In the following, results of the adjoint computations are compared to numerical derivatives computed by forward finite differences.

\begin{figure}%
\includegraphics[width=0.5\columnwidth]{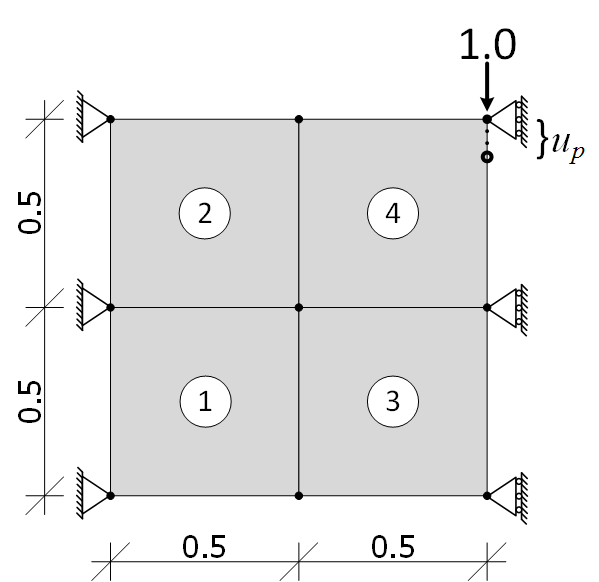}%
\includegraphics[width=0.5\columnwidth]{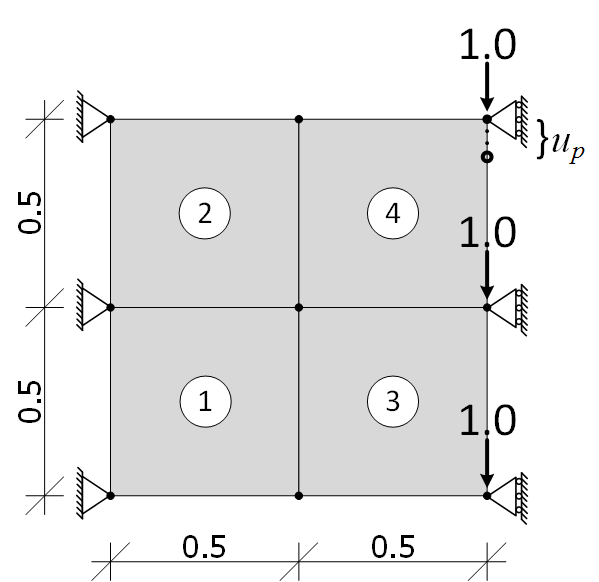}%
\caption{Problem setup for verification of the adjoint sensitivity analysis. Left: point load; right: distributed load.}%
\label{fig:SAvalidation}%
\end{figure}

We consider a small problem of a symmetric clamped beam, where the symmetric half is modeled with a finite element mesh of 2$\times$2 square bi-linear elements.
A downwards vertical displacement is prescribed at the top right corner.
Two separate loading situations are considered, see Figure \ref{fig:SAvalidation} for the problem setup:
1) A point load at the top right corner;
and 2) A distributed load at the right edge.
The first case is easier to implement because the equations for the global adjoint vectors in Eqs.~\eqref{eq:lambdaN} and \eqref{eq:lambdan} take a simple form when the force is applied only at the prescribed DOF.
However, the second case is much more useful, especially in the particular application considered in this article:
It is necessary to distribute the applied load over several adjacent nodes because the numerical solution with a point load will inherently include stress concentrations. 

The material and optimization parameters are given in Table \ref{tab:SAvalidationPar}.
The density $\overline{x}_e$ in all four elements is set to $0.8$.
The prescribed displacement of 0.01 is applied within 10 equal increments.
Convergence of each increment is assumed when the relative norm of the residual forces is below $10^{-6}$. 
For the finite difference check, the perturbation value is set to $\Delta \overline{x}_e = 10^{-6}$.
We compare the design sensitivities of two critical quantities in the context of the current application:
1) The end-compliance at the prescribed DOF, $g_{ec} = -\theta_{N} \hat{{f}^p} {{u}_{N}^p}$, where the superscript $p$ denotes the prescribed DOF;
and 2) The sum of plastic strains in the whole domain at the final equilibrium state, $g_{ps} = \sum_{e=1}^{N_e} \sum_{k=1}^{N_{GP}} {\kappa^{ek}_N}$.

\begin{table*}%
\centering
\begin{tabular}{|c|c|}
\hline 
$E_{min}$ & $1.0\cdot10^{-3}$ \\
\hline 
$E_{max}$ & $1.0\cdot10^3$ \\
\hline 
$\nu$ & 0.3\\
\hline 
$\sigma_{y,min}^0$ & 0.0 \\ 
\hline 
$\sigma_{y,max}^0$ & 2.0 \\
\hline 
${p_E}$ & 3.0\\
\hline 
${p_{\sigma_y}}$ & 3.0 \\
\hline 
$H$ & 0.01 \\
\hline
\end{tabular}
\caption{Material and optimization parameters used in the validation of the adjoint sensitivity analysis.}
\label{tab:SAvalidationPar}
\end{table*}

\begin{figure*}
    \centering
    \begin{subfigure}[b]{0.45\textwidth}
        \includegraphics[width=\textwidth,trim={0.8in 2in 0.8in 3in}]{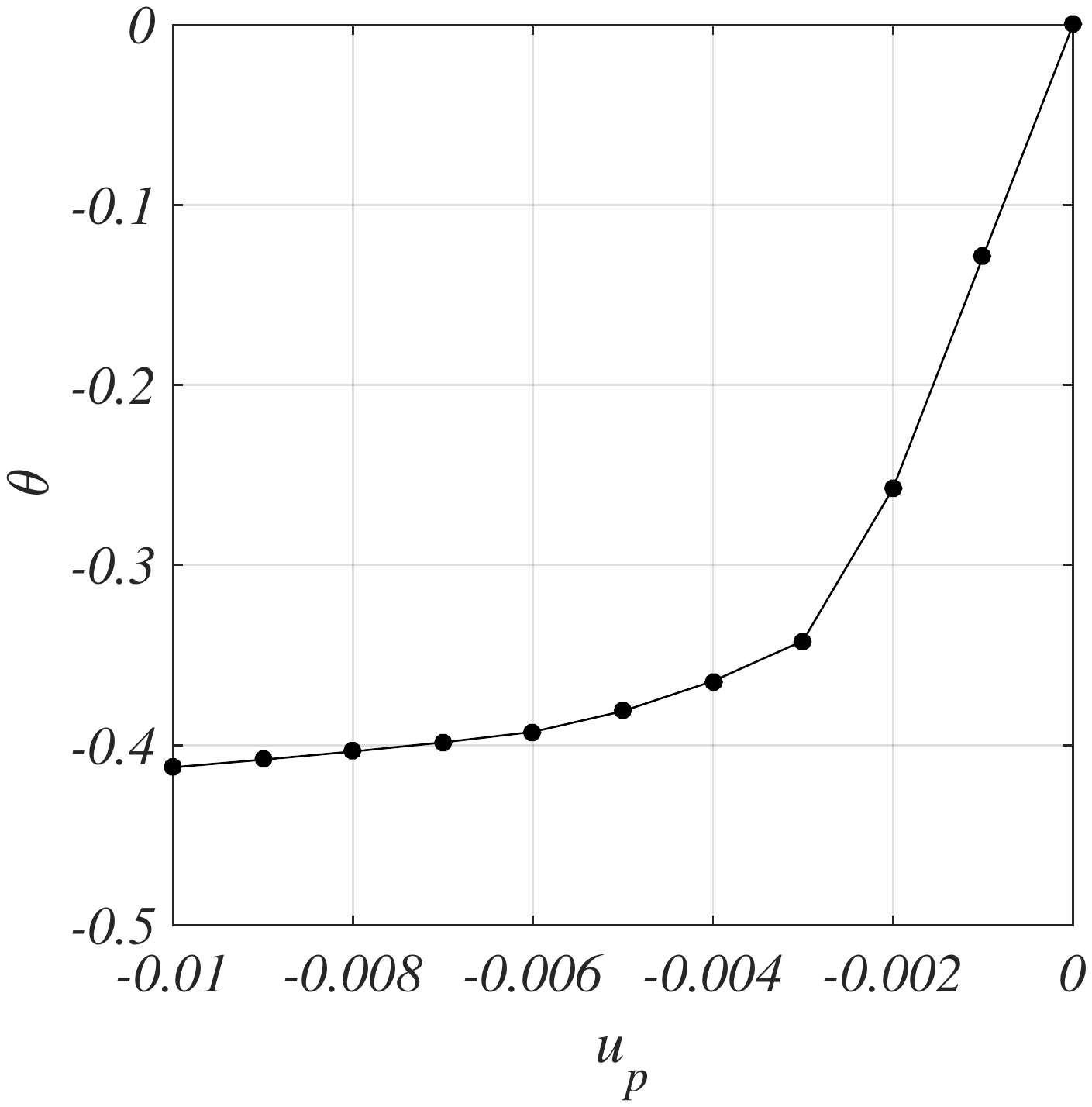}
        \caption{}
        \label{fig:SAPL_UF}
    \end{subfigure}
		~ 
    \begin{subfigure}[b]{0.45\textwidth}
        \includegraphics[width=\textwidth,trim={0.8in 2in 0.8in 3in}]{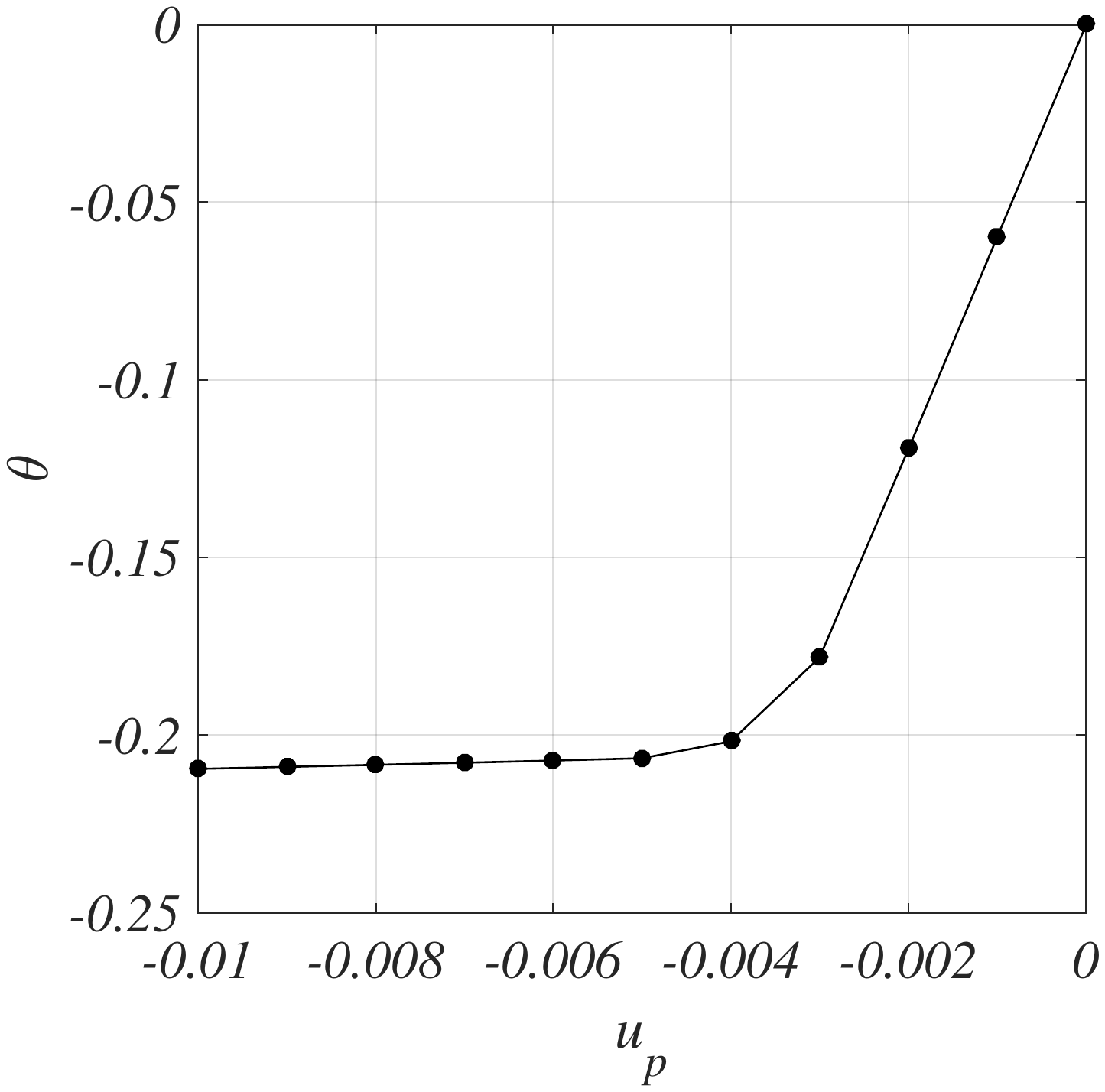}
        \caption{}
        \label{fig:SADL_UF}
    \end{subfigure}
		
    ~ 
    \begin{subfigure}[b]{0.45\textwidth}
        \includegraphics[width=\textwidth,trim={0.5in 2in 0.8in 3in}]{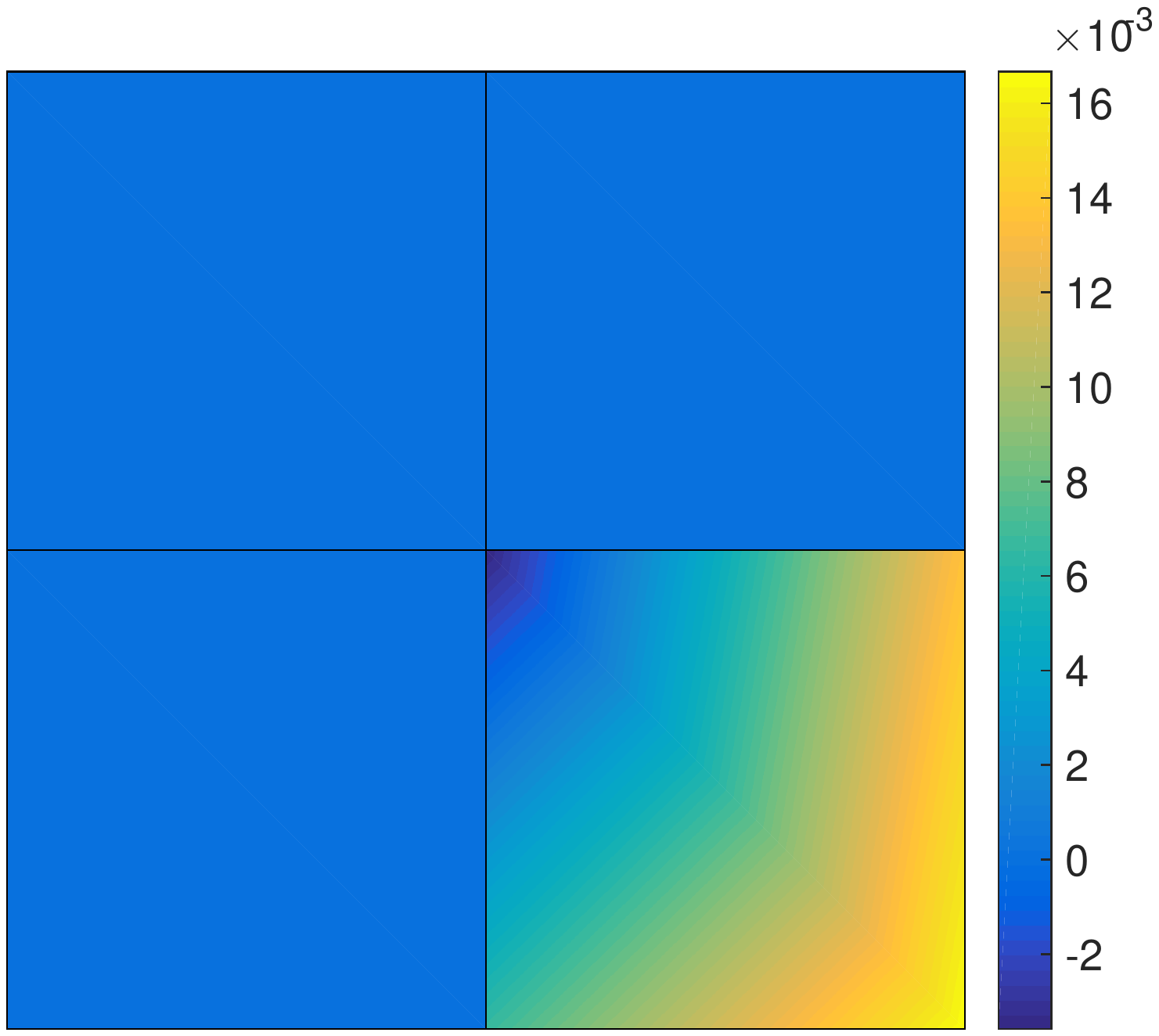}
        \caption{}
        \label{fig:SAPL_VM}
    \end{subfigure}
    ~ 
    \begin{subfigure}[b]{0.45\textwidth}
        \includegraphics[width=\textwidth,trim={0.5in 2in 0.8in 3in}]{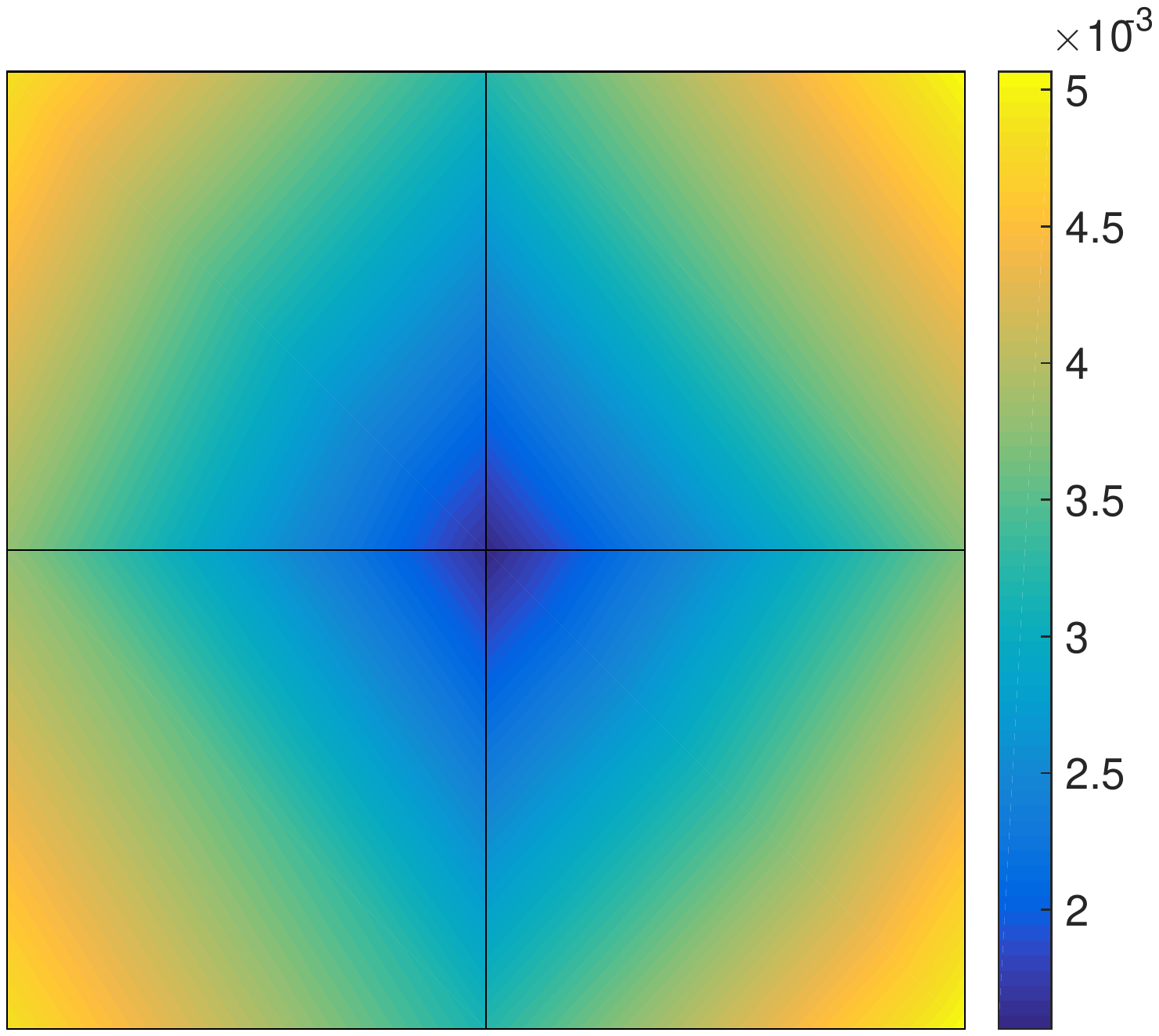}
        \caption{}
        \label{fig:SAPL_PS}
    \end{subfigure}
		\caption{Nonlinear response of the test cases used for verification of the sensitivity analysis. (a) Load factor vs. prescribed displacement, point load; (b) Load factor vs. prescribed displacement, distributed load; (c) Equivalent plastic strains, point load; (d) Equivalent plastic strains, distributed load.}
		\label{fig:SAvaildationplots}
\end{figure*}

The comparisons between the derivatives computed by the adjoint procedure to those obtained by finite differences are presented in Tables \ref{tab:SAvalidationPL} and \ref{tab:SAvalidationDL} for the point load and distributed load, respectively.
It can be seen that the design sensitivities are practically identical, thus verifying the derivation and the implementation of the adjoint procedure.  
The nonlinear response of both test cases is presented in Figure \ref{fig:SAvaildationplots}, in terms of load-displacement curves at the prescribed DOF and equivalent plastic strain.
From the tables it can be seen that even elements that are in the elastic regime contribute to the sum of plastic strains, in two opposite modes--i.e.~the addition of material can either increase or decrease the overall plastic strain, whereas it always has a stiffening effect on compliance.
Finally, the analysis and sensitivity analysis were repeated with 30 and 50 displacement increments. 
Practically identical results were obtained for the nonlinear respones as well as their design sensitivities. 
   
\begin{table*}
	\centering
		\begin{tabular}{|c||ccc||ccc|}
		\hline
			Element & $\frac{\partial g_{ec}}{\partial \overline{x}_e}$ & $\frac{\Delta g_{ec}}{\Delta \overline{x}_e}$ & Rel. error & 
								$\frac{\partial g_{ps}}{\partial \overline{x}_e}$ & $\frac{\Delta g_{ps}}{\Delta \overline{x}_e}$ & Rel. error \\
		\hline
			1 & $-1.6222\cdot10^{-4}$ & $-1.6222\cdot10^{-4}$ & $1.3347\cdot10^{-6}$ & $9.0330\cdot10^{-3}$  & $9.0330\cdot10^{-3}$  & $4.8305\cdot10^{-7}$ \\
			2 & $-1.4139\cdot10^{-4}$ & $-1.4139\cdot10^{-4}$ & $1.7011\cdot10^{-6}$ & $1.5416\cdot10^{-2}$  & $1.5416\cdot10^{-2}$  & $9.9510\cdot10^{-7}$ \\
			3 & $-1.4260\cdot10^{-2}$ & $-1.4260\cdot10^{-2}$ & $1.0736\cdot10^{-6}$ & $-3.3840\cdot10^{-2}$ & $-3.3840\cdot10^{-2}$ & $6.7529\cdot10^{-7}$ \\
			4 & $-2.3066\cdot10^{-4}$ & $-2.3066\cdot10^{-4}$ & $2.2037\cdot10^{-6}$ & $9.3906\cdot10^{-3}$  & $9.3906\cdot10^{-3}$  & $1.2330\cdot10^{-6}$ \\
			\hline
		\end{tabular}
	\caption{Verification of the sensitivity analysis, 4 element domain with a point load.}
	\label{tab:SAvalidationPL}
\end{table*}

\begin{table*}
	\centering
		\begin{tabular}{|c||ccc||ccc|}
		\hline
			Element & $\frac{\partial g_{ec}}{\partial \overline{x}_e}$ & $\frac{\Delta g_{ec}}{\Delta \overline{x}_e}$ & Rel. error & 
								$\frac{\partial g_{ps}}{\partial \overline{x}_e}$ & $\frac{\Delta g_{ps}}{\Delta \overline{x}_e}$ & Rel. error \\
		\hline
			1 & $-1.8691\cdot10^{-3}$ & $-1.8691\cdot10^{-3}$ & $1.5426\cdot10^{-6}$ & $-1.4077\cdot10^{-2}$ & $-1.4076\cdot10^{-2}$ & $1.1225\cdot10^{-5}$ \\
			2 & $-1.8841\cdot10^{-3}$ & $-1.8841\cdot10^{-3}$ & $1.5246\cdot10^{-6}$ & $8.3362\cdot10^{-3}$  & $8.3364\cdot10^{-3}$  & $1.8334\cdot10^{-5}$ \\
			3 & $-1.9104\cdot10^{-3}$ & $-1.9104\cdot10^{-3}$ & $1.8419\cdot10^{-6}$ & $4.4694\cdot10^{-2}$  & $4.4694\cdot10^{-2}$  & $2.9863\cdot10^{-6}$ \\
			4 & $-1.8544\cdot10^{-3}$ & $-1.8544\cdot10^{-3}$ & $1.7176\cdot10^{-6}$ & $-3.8954\cdot10^{-2}$ & $-3.8953\cdot10^{-2}$ & $1.1161\cdot10^{-5}$ \\
		\hline
		\end{tabular}
	\caption{Verification of the sensitivity analysis, 4 element domain with a distributed load.}
	\label{tab:SAvalidationDL}
\end{table*}

\bibliographystyle{basic}
\bibliography{C:/Users/odedamir/OneDrive/WIP/Biblio/references}   

\begin{thebibliography}{}

\bibitem[Allaire and Jouve, 2008]{allaire2008minimum}
Allaire, G. and Jouve, F. (2008).
\newblock Minimum stress optimal design with the level set method.
\newblock {\em Engineering analysis with boundary elements}, 32(11):909--918.

\bibitem[Amir, 2011]{AmirPhDThesis}
Amir, O. (2011).
\newblock {\em Efficient Reanalysis Procedures in Structural Topology
  Optimization}.
\newblock PhD thesis, Technical University of Denmark.

\bibitem[Amir, 2013]{amir2013topology}
Amir, O. (2013).
\newblock A topology optimization procedure for reinforced concrete structures.
\newblock {\em Computers \& Structures}, 114-115:46--58.

\bibitem[Amir and Sigmund, 2013]{amir2013reinforcement}
Amir, O. and Sigmund, O. (2013).
\newblock Reinforcement layout design for concrete structures based on
  continuum damage and truss topology optimization.
\newblock {\em Structural and Multidisciplinary Optimization}, 47(2):157--174.

\bibitem[Amstutz and Novotny, 2010]{amstutz2010topological}
Amstutz, S. and Novotny, A.~A. (2010).
\newblock Topological optimization of structures subject to von mises stress
  constraints.
\newblock {\em Structural and Multidisciplinary Optimization}, 41(3):407--420.

\bibitem[Bends{\o}e, 1989]{bendsoe1989optimal}
Bends{\o}e, M.~P. (1989).
\newblock Optimal shape design as a material distribution problem.
\newblock {\em Structural optimization}, 1(4):193--202.

\bibitem[Bends{\o}e and Kikuchi, 1988]{bendsoe1988generating}
Bends{\o}e, M.~P. and Kikuchi, N. (1988).
\newblock Generating optimal topologies in structural design using a
  homogenization method.
\newblock {\em Computer methods in applied mechanics and engineering},
  71(2):197--224.

\bibitem[Bends{\o}e and Sigmund, 2003]{BendsoeSigmund2003}
Bends{\o}e, M.~P. and Sigmund, O. (2003).
\newblock {\em Topology Optimization - Theory, Methods and Applications}.
\newblock Springer, Berlin.

\bibitem[Bogomolny and Amir, 2012]{bogomolny2012conceptual}
Bogomolny, M. and Amir, O. (2012).
\newblock Conceptual design of reinforced concrete structures using topology
  optimization with elastoplastic material modeling.
\newblock {\em International Journal for Numerical Methods in Engineering},
  90(13):1578--1597.

\bibitem[Bourdin, 2001]{bourdin2001filters}
Bourdin, B. (2001).
\newblock Filters in topology optimization.
\newblock {\em International Journal for Numerical Methods in Engineering},
  50:2143--2158.

\bibitem[Bruggi, 2008]{bruggi2008alternative}
Bruggi, M. (2008).
\newblock On an alternative approach to stress constraints relaxation in
  topology optimization.
\newblock {\em Structural and multidisciplinary optimization}, 36(2):125--141.

\bibitem[Bruggi and Duysinx, 2012]{bruggi2012topology}
Bruggi, M. and Duysinx, P. (2012).
\newblock Topology optimization for minimum weight with compliance and stress
  constraints.
\newblock {\em Structural and Multidisciplinary Optimization}, 46(3):369--384.

\bibitem[Bruggi and Venini, 2008]{bruggi2008mixed}
Bruggi, M. and Venini, P. (2008).
\newblock A mixed fem approach to stress-constrained topology optimization.
\newblock {\em International journal for numerical methods in engineering},
  73(12):1693--1714.

\bibitem[Bruns and Tortorelli, 2001]{bruns2001topology}
Bruns, T.~E. and Tortorelli, D.~A. (2001).
\newblock Topology optimization of non-linear elastic structures and compliant
  mechanisms.
\newblock {\em Computer Methods in Applied Mechanics and Engineering},
  190:3443--3459.

\bibitem[Cheng and Guo, 1997]{cheng1997varepsilon}
Cheng, G. and Guo, X. (1997).
\newblock $\varepsilon$-relaxed approach in structural topology optimization.
\newblock {\em Structural Optimization}, 13(4):258--266.

\bibitem[Cheng and Jiang, 1992]{cheng1992study}
Cheng, G. and Jiang, Z. (1992).
\newblock Study on topology optimization with stress constraints.
\newblock {\em Engineering Optimization}, 20(2):129--148.

\bibitem[Deaton and Grandhi, 2014]{deaton2014survey}
Deaton, J. and Grandhi, R. (2014).
\newblock A survey of structural and multidisciplinary continuum topology
  optimization: post 2000.
\newblock {\em Structural and Multidisciplinary Optimization}, 49(1):1--38.

\bibitem[Duysinx and Bends{\o}e, 1998]{Duysinx1998}
Duysinx, P. and Bends{\o}e, M.~P. (1998).
\newblock Topology optimization of continuum structures with local stress
  constraints.
\newblock {\em International Journal for Numerical Methods in Engineering},
  43:1453--1478.

\bibitem[Duysinx and Sigmund, 1998]{duysinx1998new}
Duysinx, P. and Sigmund, O. (1998).
\newblock New developments in handling stress constraints in optimal material
  distribution.
\newblock In {\em Proceedings of 7th AIAA/USAF/NASA/ISSMO symposium on
  multidisciplinary design optimization, AIAA, Saint Louis, Missouri, AIAA
  Paper}, pages 98--4906.

\bibitem[Eschenauer and Olhoff, 2001]{eschenauer2001topology}
Eschenauer, H.~A. and Olhoff, N. (2001).
\newblock Topology optimization of continuum structures: a review.
\newblock {\em Applied Mechanics Reviews}, 54(4):331--389.

\bibitem[Fancello, 2006]{fancello2006topology}
Fancello, E. (2006).
\newblock Topology optimization for minimum mass design considering local
  failure constraints and contact boundary conditions.
\newblock {\em Structural and Multidisciplinary Optimization}, 32(3):229--240.

\bibitem[Fritzen et~al., 2015]{fritzen2015topology}
Fritzen, F., Xia, L., Leuschner, M., and Breitkopf, P. (2015).
\newblock Topology optimization of multiscale elastoviscoplastic structures.
\newblock {\em International Journal for Numerical Methods in Engineering}.

\bibitem[Guest et~al., 2004]{guest2004achieving}
Guest, J.~K., Pr{\'{e}}vost, J.~H., and Belytschko, T. (2004).
\newblock Achieving minimum length scale in topology optimization using nodal
  design variables and projection functions.
\newblock {\em International Journal for Numerical Methods in Engineering},
  61:238--254.

\bibitem[James and Waisman, 2014]{james2014failure}
James, K.~A. and Waisman, H. (2014).
\newblock Failure mitigation in optimal topology design using a coupled
  nonlinear continuum damage model.
\newblock {\em Computer Methods in Applied Mechanics and Engineering},
  268:614--631.

\bibitem[James and Waisman, 2015]{james2015topology}
James, K.~A. and Waisman, H. (2015).
\newblock Topology optimization of viscoelastic structures using a
  time-dependent adjoint method.
\newblock {\em Computer Methods in Applied Mechanics and Engineering},
  285:166--187.

\bibitem[Kato et~al., 2015]{kato2015analytical}
Kato, J., Hoshiba, H., Takase, S., Terada, K., and Kyoya, T. (2015).
\newblock Analytical sensitivity in topology optimization for elastoplastic
  composites.
\newblock {\em Structural and Multidisciplinary Optimization}, pages 1--20.

\bibitem[Kirsch, 1990]{kirsch1990singular}
Kirsch, U. (1990).
\newblock On singular topologies in optimum structural design.
\newblock {\em Structural Optimization}, 2(3):133--142.

\bibitem[Le et~al., 2010]{le2010stress}
Le, C., Norato, J., Bruns, T., Ha, C., and Tortorelli, D. (2010).
\newblock Stress-based topology optimization for continua.
\newblock {\em Structural and Multidisciplinary Optimization}, 41:605--620.

\bibitem[Maute et~al., 1998]{maute1998adaptive}
Maute, K., Schwarz, S., and Ramm, E. (1998).
\newblock Adaptive topology optimization of elastoplastic structures.
\newblock {\em Structural Optimization}, 15(2):81--91.

\bibitem[Michaleris et~al., 1994]{michaleris1994tangent}
Michaleris, P., Tortorelli, D.~A., and Vidal, C.~A. (1994).
\newblock Tangent operators and design sensitivity formulations for transient
  non-linear coupled problems with applications to elastoplasticity.
\newblock {\em International Journal for Numerical Methods in Engineering},
  37(14):2471--2499.

\bibitem[Nakshatrala and Tortorelli, 2015]{nakshatrala2015topology}
Nakshatrala, P. and Tortorelli, D. (2015).
\newblock Topology optimization for effective energy propagation in
  rate-independent elastoplastic material systems.
\newblock {\em Computer Methods in Applied Mechanics and Engineering}, 295:305
  -- 326.

\bibitem[Par{\'\i}s et~al., 2007]{paris2007block}
Par{\'\i}s, J., Navarrina, F., Colominas, I., and Casteleiro, M. (2007).
\newblock Block aggregation of stress constraints in topology optimization of
  structures.
\newblock In Hern{\'a}ndez, S. and Brebbia, C.~A., editors, {\em Computer Aided
  Optimum Design of Structures X}.

\bibitem[Par{\'\i}s et~al., 2010]{paris2010block}
Par{\'\i}s, J., Navarrina, F., Colominas, I., and Casteleiro, M. (2010).
\newblock Block aggregation of stress constraints in topology optimization of
  structures.
\newblock {\em Advances in Engineering Software}, 41(3):433--441.

\bibitem[Park, 1995]{park1995extensions}
Park, Y.~K. (1995).
\newblock {\em Extensions of optimal layout design using the homogenization
  method}.
\newblock PhD thesis, University of Michigan, Ann Arbor.

\bibitem[Pereira et~al., 2004]{pereira2004topology}
Pereira, J.~T., Fancello, E.~A., and Barcellos, C.~S. (2004).
\newblock Topology optimization of continuum structures with material failure
  constraints.
\newblock {\em Structural and Multidisciplinary Optimization}, 26(1-2):50--66.

\bibitem[Rozvany, 1996]{rozvany1996difficulties}
Rozvany, G. (1996).
\newblock Difficulties in truss topology optimization with stress, local
  buckling and system stability constraints.
\newblock {\em Structural Optimization}, 11(3-4):213--217.

\bibitem[Schwarz et~al., 2001]{schwarz2001topology}
Schwarz, S., Maute, K., and Ramm, E. (2001).
\newblock Topology and shape optimization for elastoplastic structural
  response.
\newblock {\em Computer Methods in Applied Mechanics and Engineering},
  190(15):2135--2155.

\bibitem[Sigmund and Maute, 2013]{sigmund2013topology}
Sigmund, O. and Maute, K. (2013).
\newblock Topology optimization approaches.
\newblock {\em Structural and Multidisciplinary Optimization},
  48(6):1031--1055.

\bibitem[Sigmund and Torquato, 1997]{sigmund1997design}
Sigmund, O. and Torquato, S. (1997).
\newblock Design of materials with extreme thermal expansion using a
  three-phase topology optimization method.
\newblock {\em Journal of the Mechanics and Physics of Solids},
  45(6):1037--1067.

\bibitem[Simo and Taylor, 1986]{Simo1985}
Simo, J. and Taylor, R. (1986).
\newblock A return mapping algorithm for plane stress elastoplasticity.
\newblock {\em International Journal for Numerical Methods in Engineering},
  22:649--670.

\bibitem[Simo and Hughes, 2006]{simo2006computational}
Simo, J.~C. and Hughes, T.~J. (2006).
\newblock {\em Computational inelasticity}, volume~7.
\newblock Springer Science \& Business Media.

\bibitem[Stolpe and Svanberg, 2001]{stolpe2001trajectories}
Stolpe, M. and Svanberg, K. (2001).
\newblock On the trajectories of the epsilon-relaxation approach for
  stress-constrained truss topology optimization.
\newblock {\em Structural and multidisciplinary optimization}, 21(2):140--151.

\bibitem[Svanberg, 1987]{svanberg1987method}
Svanberg, K. (1987).
\newblock The method of moving asymptotes - a new method for structural
  optimization.
\newblock {\em International Journal for Numerical Methods in Engineering},
  24:359--373.

\bibitem[Sved and Ginos, 1968]{sved1968structural}
Sved, G. and Ginos, Z. (1968).
\newblock Structural optimization under multiple loading.
\newblock {\em International Journal of Mechanical Sciences}, 10(10):803--805.

\bibitem[Swan and Kosaka, 1997]{swan1997voigt}
Swan, C.~C. and Kosaka, I. (1997).
\newblock Voigt-reuss topology optimization for structures with nonlinear
  material behaviors.
\newblock {\em International Journal for Numerical Methods in Engineering},
  40(20):3785--3814.

\bibitem[Verbart et~al., 2013]{verbart2013new}
Verbart, A., Langelaar, M., and van Keulen, F. (2013).
\newblock A new approach for stress-based topology optimization: Internal
  stress penalization.
\newblock In {\em 10th World Congress on Structural and Multidisciplinary
  Optimization, Orlando, Florida, USA}.

\bibitem[von Mises, 1928]{vonMises1928}
von Mises, R. (1928).
\newblock Mechanics of the ductile form changes of crystals.
\newblock {\em Zeitschrift f{\"{u}}r Angewandte Mathematik und Mechanik},
  8:161--185.

\bibitem[Xu et~al., 2010]{xu2010volume}
Xu, S., Cai, Y., and Cheng, G. (2010).
\newblock Volume preserving nonlinear density filter based on heaviside
  functions.
\newblock {\em Structural and Multidisciplinary Optimization}, 41(4):495--505.

\bibitem[Yang and Chen, 1996]{yang1996stress}
Yang, R. and Chen, C. (1996).
\newblock Stress-based topology optimization.
\newblock {\em Structural Optimization}, 12(2-3):98--105.

\bibitem[Yoon and Kim, 2007]{yoon2007topology}
Yoon, G.~H. and Kim, Y.~Y. (2007).
\newblock Topology optimization of material-nonlinear continuum structures by
  the element connectivity parameterization.
\newblock {\em International journal for numerical methods in engineering},
  69(10):2196--2218.

\bibitem[Yuge and Kikuchi, 1995]{yuge1995optimization}
Yuge, K. and Kikuchi, N. (1995).
\newblock Optimization of a frame structure subjected to a plastic deformation.
\newblock {\em Structural optimization}, 10(3-4):197--208.

\bibitem[Zienkiewicz and Taylor, 2000]{zienkiewicz2000finite}
Zienkiewicz, O.~C. and Taylor, R.~L. (2000).
\newblock {\em The finite element method: Solid mechanics}, volume~2.
\newblock Butterworth-heinemann.

\end{thebibliography}

\end{document}